\documentclass[msc,10pt,appendixpart,twoside]{ubcthesis}
\usepackage{amsmath, amsthm}

\usepackage[sort&compress]{natbib}
\long\def\symbolfootnote[#1]#2{\begingroup%
\def\thefootnote{\fnsymbol{footnote}}\footnote[#1]{#2}\endgroup}

\usepackage{hyperref}

\usepackage{graphicx}

\usepackage{lscape}

\usepackage{psfrag}

\institution{The University Of British Columbia}
\institutionaddress{Vancouver, Canada}
\department{Department of Physics and Astronomy}

\previousdegree{B.Sc., The University of Waterloo, 2002}
\previousdegree{B.Math., The University of Waterloo, 2001}

\title{Magnetic Vortex Dynamics}
\subtitle{in a 2D easy plane ferromagnet}
\author{Lara Thompson}
\copyrightyear{2004}
\submitdate{\today}
\renewcommand\thepart         {\Roman{part}}
\renewcommand\thechapter      {\arabic{chapter}}
\renewcommand\thesection      {\thechapter.\arabic{section}}
\renewcommand\thesubsection   {\thesection.\arabic{subsection}}
\renewcommand\thesubsubsection{\thesubsection.\arabic{subsubsection}}
\renewcommand\theparagraph    {\thesubsubsection.\arabic{paragraph}}

\begin{document}

\frontmatter

\maketitle
\begin{abstract}
\parskip=0.1in
In this thesis, we consider the dynamics of vortices in the easy
plane insulating ferromagnet in two dimensions. In addition to the
quasiparticle excitations, here spin waves or magnons, this
magnetic system admits a family of vortex solutions carrying two
topological invariants, the winding number or vorticity, and the
polarization.

A vortex is approximately described as a particle moving about the
system, endowed with an effective mass and acted upon by a variety
of forces. Classically, the vortex has an inter-vortex potential
energy giving a Coulomb-like force (attractive or repulsive
depending on the relative vortex vorticity), and a gyrotropic
force, behaving as a self-induced Lorentz force, whose direction
depends on both topological indices.

Expanding semiclassically about a many-vortex solution, the
vortices are quantized by considering the scattered magnon states,
giving a zero point energy correction and a many-vortex mass
tensor. The vortices cannot be described as independent
particles---that is, there are off-diagonal mass terms, such as
$\frac{1}{2} M_{ij}v_iv_j$, that are non-negligible.

This thesis examines the full vortex dynamics in further detail by
evaluating the Feynman-Vernon influence functional, which
describes the evolution of the vortex density matrix after the
magnon modes have been traced out. In addition to the set of
forces already known, we find new damping forces acting both
longitudinally and transversely to the vortex motion. The vortex
motion within a collective cannot be entirely separated: there are
damping forces acting on one vortex due to the motion of another.
The effective damping forces have memory effects: they depend not
only on the current motion of the vortex collection but also on
the motion history.
\end{abstract}

\parskip=0in
\tableofcontents
\listoffigures

\parskip=0.1in

\acknowledgements Thanks to Phil for choosing an excellent masters
research topic. To my mom who read my thesis and corrected it
despite not understanding every third word, although learning that
equations have a grammar all their own! To Talie in Toronto for
housing me in the midst of the crunch and showing me a good time
otherwise to cool off. To Yan for sharing with me the mountains.

``What did the condensed matter theorist say to the soliton? As
long as you aren't empirical, you're all right with me.'' --Lateef
Yang, August 11, 2004

\mainmatter

\chapter{Introduction}
In a wide variety of systems, there exist vortices, high energy
states nonetheless significant in system dynamics at low
temperatures. Despite its high energy, a vortex can nonetheless
form via tunneling processes or at a boundary with only a small
energy barrier. They are exceptionally stable, arguable
topologically, and, in fact, can only be destroyed if one meets
its `anti-vortex' or, equivalently, annihilates at a boundary
(where it has met its image vortex). Cooling a system down
vortex-free is non-trivial, and, in general, we retain a low
density of vortex states down to the lowest temperatures.

Quantum vortices were first proposed in the 1950's in superfluid
helium to explain the decay of persistent currents. Since then,
they have been proposed and measured in, for example,
superconductors and a variety of magnetic systems. The dynamics
are well described phenomenologically as a point-like particle in
2D (or as a line in 3D) endowed with an effective mass and acted
upon by a variety of forces. Microscopic derivations of the
particle properties of a quantum vortex have been plagued by
decades of debate and controversy. A recent resurgence in debates
began in the 1990's concerning the so-called Magnus force, a force
borrowed from classical fluids acting perpendicular to the
velocity. \citet{ao:1993} claimed that in superfluid helium (He
II) there is a universal form of this force, independent of
quasiparticle scattering. Others argue that there should be, in
addition to the bare Magnus force, a tranverse damping force,
reinforcing or opposing the Magnus
force\cite{sonin:1997,volovik:1996c,hall:1998c}.

In this thesis, we consider a relatively simple magnetic system, a
2D insulating ferromagnet with easy plane anisotropy, admitting a
family of topologically stable vortices. We derive microscopically
the vortex effective mass and, in addition to the previously
reported gyrotropic force, the magnetic analogue to the Magnus
force, and inter-vortex Coulomb-like forces, we derive a variety
of vortex damping forces. We find both the usual longitudinal
damping force and a transverse damping that acts in combination
with the gyrotropic force. A transverse damping force has not yet
been considered in a magnetic system. In fact, all treatments of
the dissipative motion of a vortex have been phenomenological,
with the exception of Slonczewski's\cite{sloncz:1984} treatment
with which we compare results in Chapter \ref{chapter:together}. A
collection of vortices cannot be considered as a set of
independent particles---they have mixed inertial terms and damping
force terms.

We first review a few symmetry arguments for the existence and
stability of vortex solutions. Besides revealing the similarity
between vortices from various systems, we find that vortices are
an example of a more general family of topological solitons.

We then briefly discuss the early work on quantizing solitons by
the relativistic field theorists, focussing rather on the
techniques than the various specific contributions. Note that we
will use many of these techniques for quantizing the vortex in the
easy plane magnetic system.

Next, we discuss briefly solitons in condensed matter systems and
the exciting new phenomena found there. For example, by examining
the conducting polymers, fractional charge was first predicted and
observed.

Returning specifically to vortices, we briefly discuss the
controversy in the microscopic derivation of the equations of
motion for a superfluid vortex. This will introduce the variety of
forces we should expect to act on a collection of vortices.
Switching to magnetic systems, we find that despite the ease of
direct experimental observation and simplicity of calculations not
much work has been done here.

Finally, we introduce in detail the magnetic system under
consideration. The symmetry of the system admits topologically
stable vortices and gapless quasiparticles. The purpose of this
thesis is to separate the quantum dynamics of the vortices from
the effects of the perturbative quasiparticles, here magnons.

\section{Symmetry breaking}

Symmetry plays a crucial role in science and we strive to discover
and exploit the symmetries of the laws of nature (Galilean or
Lorentz invariance, gauge invariance, etc.). However, we find that
the symmetry of physical states may be a smaller subset of the
full symmetry in which it resides. For example, in a Heisenberg
ferromagnet, we find a system of spins free to lie in any
direction in 3D, preferring to align parallel to one another,
however, in the absence of any magnetic fields, with no preference
of which direction along which to lie. The ground state then
chooses at random along what direction to align.

\begin{figure}
\centering
\includegraphics[width=6cm]{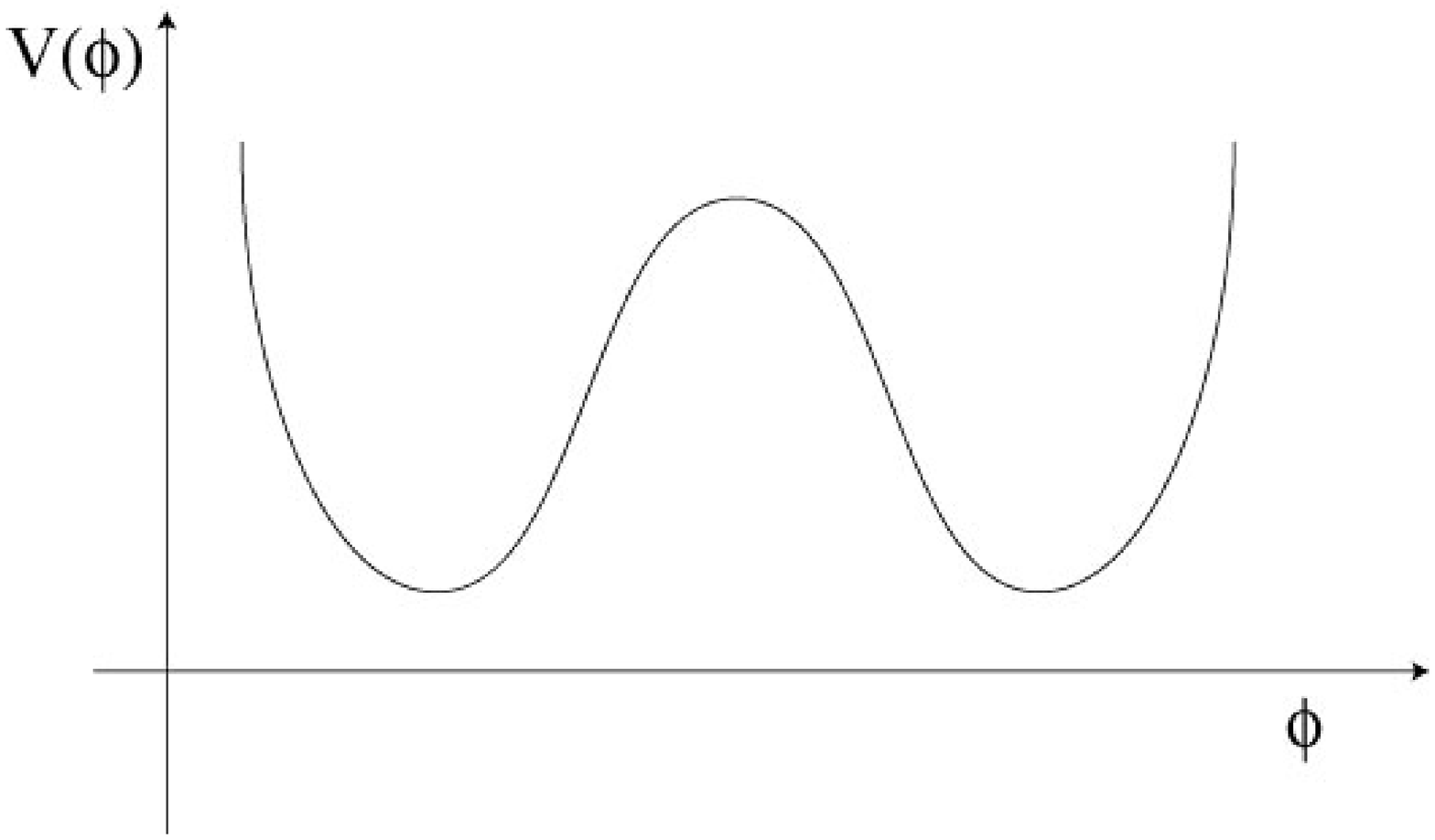}
\includegraphics[width=3.8cm]{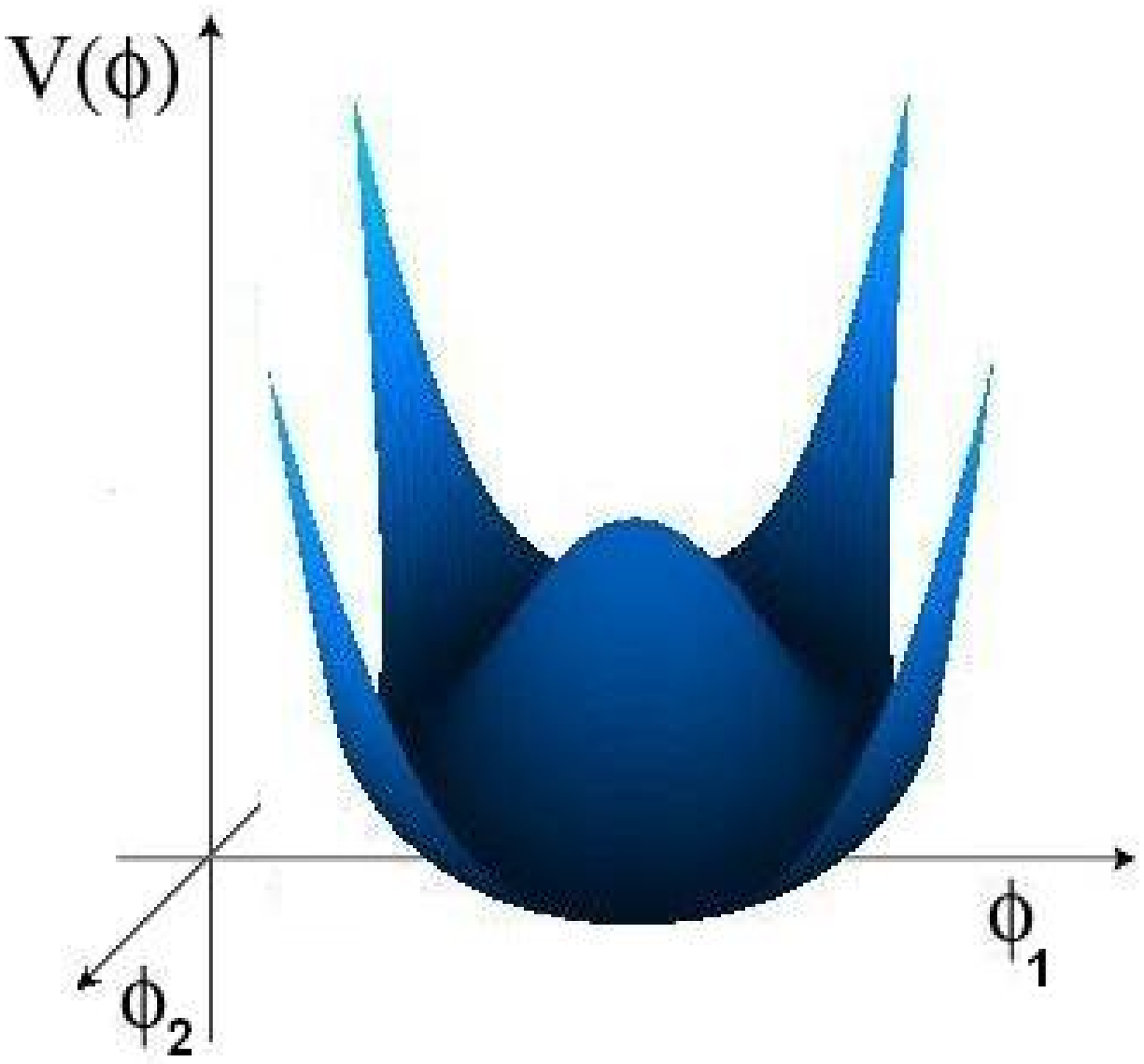}
\caption{Left, an example potential of a 1D field $\phi$ with a
doubly degenerate ground state; right, an example potential of a
2D field with a continuum degeneracy in its ground
state.}\label{figure:potentials}
\end{figure}

A system with a degenerate ground state is forced to spontaneously
choose one state amid the degeneracy, an example of spontaneously
broken symmetry. A discrete degeneracy is found in the problem of
a field residing in a double well potential (as in Figure
\ref{figure:potentials}, left), or, more generally, an $n$-well
potential. A continuous degeneracy in a system has a continuum of
minima in the potential (as, for example, in Figure
\ref{figure:potentials}, right). The ferromagnet is an example of
a system with a continuum of ground states, except that here, the
potential is completely flat: there is no preference at all
between directions.

In general, different regions of a sample may choose different
degenerate states or may even lie in an excited state. A mapping
of the state taken across the sample, in all its available degrees
of freedom, is called the order parameter. In a Heisenberg spin
system, this is simply the spin vector in 3D as a function of
position in the sample. The order parameter here can be mapped
onto a unit sphere---a path along the sample is then traced as a
path on the surface of the sphere. For a spin system confined to
lie in the plane, the so-called XY model, the order parameter is
mapped onto the unit circle.

Incidentally, the order parameter in superfluid helium II can also
be mapped onto the unit circle so that it is topologically
equivalent to the XY model. This does not mean, however, that the
dynamics of the vortices in each system should be the same, but,
rather, only that the topology of vortices is identical in the two
systems.

If a system possesses discrete symmetries, to pass from one ground
state to another there must be some transition region, or domain
wall, separating different states. This domain wall, sometimes
called a kink, is an example of a quasi-1D soliton.

\begin{figure}
\centering
\includegraphics[width=3cm]{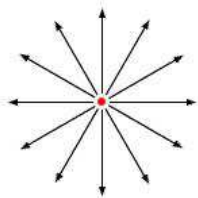}
\caption{A 2D XY-ferromagnet with a vortex connecting the
degeneracy of spin directions. The central red dot signifies the
point of discontinuity.}\label{figure:xyvortex}
\end{figure}

For a continuous symmetry, we can imagine similar cases where
certain regions are forced out of a ground state. As a simple
example, consider the XY spin model. If the spins choose to nearly
align along the boundary, turning very slowly so as to always
radiate outward, as we near some central region the spins are less
and less ferromagnetically aligned and, further, there is a point
discontinuity at the very center (see Figure
\ref{figure:xyvortex}).

If we follow a path surrounding the vortex in order parameter
space, that is along the unit circle, we find we must wrap around
the unit circle once. This vortex is called a topological soliton
with single \emph{wrapping number} or \emph{vorticity}. In this
example, no matter how we smoothly deform the spins, we cannot
continuously deform away this wrapping of the unit circle. We say
that it is homotopically distinct from a zero winding path, or
more simply a point.

There are vortices with higher winding numbers, always integral to
ensure continuity. Each family of solutions corresponding to a
certain winding number is topologically stable. That is, there
exists no homotopy, or continuous mapping, between solutions of
differing winding numbers.

\begin{figure}
\centering
\includegraphics[width=10cm]{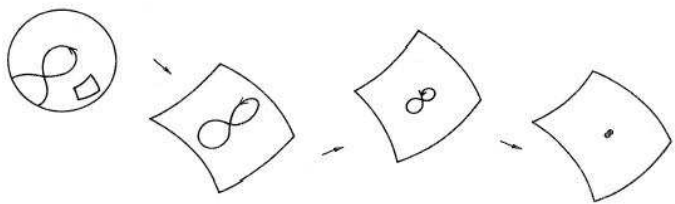}
\caption{A magnetic vortex formed by Heisenberg spins can be
continuously deformed away by expanding about a patch of the unit
sphere not covered by the vortex path, shrinking the vortex to a
point.}\label{figure:removevortex}
\end{figure}

There are systems that admit vortices for which this topological
stability is not guaranteed, and are thus not called topological
solitons. Consider a general vortex residing on a sample for which
the order parameter maps onto a unit sphere (Figure
\ref{figure:removevortex}). The vortex is homotopically equivalent
to a point (that is, a region with constant ground state) since we
can imagine continuously shrinking the vortex away. In real space,
this is equivalent to the ability of the spins to unwind, that is,
all the spins twisting to all lie parallel to one another. Note
that this unwinding is a special feature of the isotropy of the
system. Although such a soliton does not possess topological
stability, the entire plane must unwind, a macroscopic number of
spins in the magnetic vortex case, so that the soliton is still
essentially stable.

\begin{figure}
\centering
\includegraphics[width=3cm]{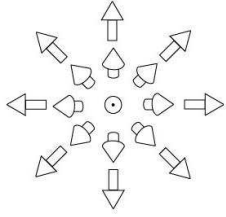}
\caption{A vortex with +1 winding in a 2D Heisenberg ferromagnet
with spins lying preferentially in the
plane.}\label{figure:myvortex}
\end{figure}

The vortices considered in this thesis have an order parameter
lying on the unit sphere, however, with a higher potential at the
north and south poles. They are very similar to the XY vortex
shown in figure \ref{figure:xyvortex}, except that the spins are
not entirely restricted to lie in the plane and, at some energy
expense to restore continuity, the spins twist out of plane at the
vortex center choosing spontaneously between the two possible
perpendicular directions in which to twist. This direction is a
second topological invariant of the vortices and is termed the
\emph{polarization}. An example of a vortex with unit winding
number, or vorticity, and polarization out of the page is shown in
Figure \ref{figure:myvortex}. There exist also zero polarization
vortices lying entirely in the plane.

\section{Classical Solitons}
We found that vortices are examples of a topological solitons.
Generally, a soliton is a finite energy localized solution of a
wave equation, satisfying strict stability conditions under
collisions with other soliton solutions\symbolfootnote[2]{See, for
instance, the excellent book by \citet{raj} on the quantization of
solitons for a rigorous definition of a soliton.}.

\begin{figure}
\centering
\includegraphics[width=11cm]{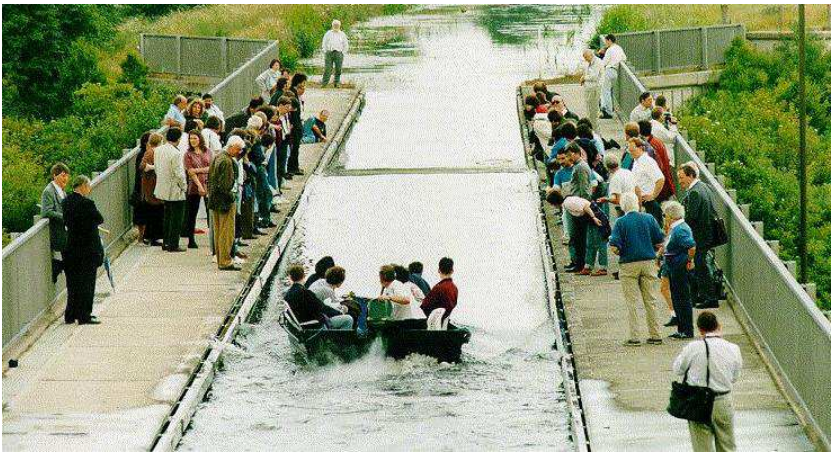}
\caption{Re-enaction of the 1834 `first' soliton sighting on the
Union Canal near Edinburgh by John Scott
Russell.}\label{figure:union}
\end{figure}
The first reported soliton was in 1834 by John Scott Russell
\cite{russell:1845} in the Union Canal near Edinburgh (see Figure
\ref{figure:union}),

\begin{quote}
I was observing the motion of a boat which was rapidly drawn along
a narrow channel by a pair of horses, when the boat suddenly
stopped - not so the mass of water in the channel which it had put
in motion; it accumulated round the prow of the vessel in a state
of violent agitation, then suddenly leaving it behind, rolled
forward with great velocity, assuming the form of a large solitary
elevation, a rounded, smooth and well-defined heap of water, which
continued its course along the channel apparently without change
of form or diminution of speed. I followed it on horseback, and
overtook it still rolling on at a rate of some eight or nine miles
an hour, preserving its original figure some thirty feet long and
a foot to a foot and a half in height. Its height gradually
diminished, and after a chase of one or two miles I lost it in the
windings of the channel. Such, in the month of August 1834, was my
first chance interview with that singular and beautiful phenomenon
which I have called the Wave of Translation.
\end{quote}

He went on to build a 30$'$ wave tank in his back garden in which
to conduct further experiments on his ``waves of translation''.

In physics, there are the familiar optical solitons, with which
demonstrations of long haul, low bit-error-rate transmissions have
been made. In optics, a soliton is a localized EM wave with much
higher power than a traditional optical signal. However, as
opposed to regular low power optical transmissions, an optical
soliton does not suffer dispersion, so that a signal is not
distorted when transmitted over large distances.

A soliton is usually a solution to a partial differential equation
in which competing non-linear terms cooperate to create a
self-reinforcing large amplitude solution. For instance, for a
non-linear dissipative system, ordinarily, wave solutions are
dispersive, that is, different $k$ modes separate, and
dissipative, energy spreads in real space. For these special
soliton solutions the two mechanisms can act in opposition so that
the net result is a non-dispersive, non-dissipative wave.

More specifically, however, a vortex is an example of a
topological soliton. These exist, not because of finely balanced
non-linear terms in the equations of motion, but rather due to a
degenerate freedom in the boundary conditions entailing the
existence of homotopically distinct solutions (that is, solutions
for which there is no continuous deformation from one to another).

\section{Quantum Solitons}
\subsection{The particle theorists}

Solitons resemble extended particles, that is, they are
non-dispersive localized packets of energy, even though they are
solutions of non-linear wave equations. Elementary particles are
localized packets of energy and are also believed to be solutions
of some relativistic field theory. The particle theorists were
thus highly motivated to find some quantum version of these
classical solitons, that is, to quantize the solitons.

It isn't immediately clear how to make the correspondence between
a classical soliton and some extended particle state of a
quantized theory, or between any classical field solution and its
quantum analogue for that matter. To understand the difficulty,
consider first the simple case of a point particle in a potential.
Classically, this particle has some definite position and momentum
with some particular path chosen by its initial conditions.
Quantum mechanically, the picture changes entirely! No longer can
we associate a particle with a definite position and momentum;
instead, we must describe the particle probabilistically via a
wavefunction $\psi(x,t)$ giving the probability $|\psi(x,t)|^2$ to
find the particle at point $x$ and time $t$. How does one go from
the soliton solution to some quantum wavefunction?

Procedures for establishing this correspondence developed in the
mid-70's were essentially a generalization of the semiclassical
expansion of non-relativistic quantum mechanics. It was shown that
not only could we associate a quantum soliton-particle with the
classical solution, but also a series of excited states by
quantizing fluctuations about the
soliton\cite{dashen:1974I,jackiw:1975}.

\begin{figure}
\centering
\includegraphics[width=6cm]{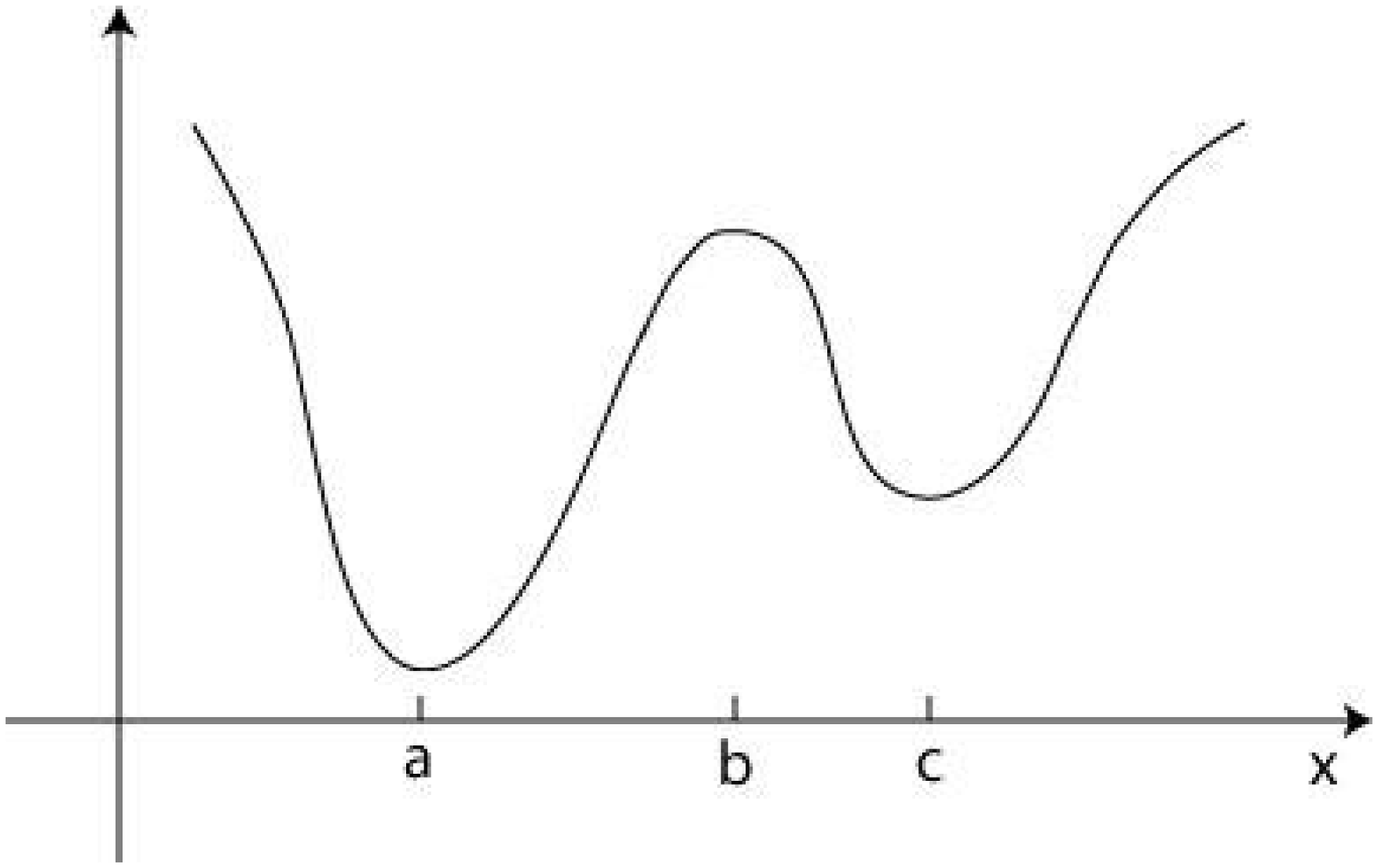}
\caption{An illustrative potential of a one dimensional particle.
A soliton is analogous to the second minimum at
$x=c$.}\label{figure:particlepotential}
\end{figure}

For a soliton, we quantize its motion by defining conjugate
position $\mathbf X$ and momentum $\mathbf P$ operators and
imposing commutation relations. In the original field, however,
there is an entire continuum of degrees of freedom that remain.
These are taken up by the quasiparticle excitations.

The procedure is analogous to the quantization of a particle
residing in a local minimum of the external potential (for
example, $x=c$ in Figure \ref{figure:particlepotential}). This
local minimum is not the global minimum, and hence is not the true
ground state; however, there is a potential barrier blocking it
from decaying to the true ground state. This is the same for a
soliton excitation, or a vortex, which is higher in energy than
the ground state, however, stable against decay.

The quantization of the local minimum begins by assuming to zeroth
order the classical solution, $x = c$. We expand the potential
about this local minimum, finding quadratic behaviour to leading
order, and proceed to quantize the perturbative excitations. Of
course, a quadratic potential has simple harmonic excitations, so
that the quantized solution can be envisioned as a hierarchy of
simple harmonic excitations, centered, of course, about the
classical minimum.

For a soliton in field theory, the procedure is much the same. We
begin by the classical solution, expanding the energy functional
about it and quantizing the leading order corrections. The simple
harmonic analogous solutions are called mesons in quantum field
theory, or quasiparticles in condensed matter. Of course, the
mesons or quasiparticles also exist as excitations in the ground
state, or vacuum state. Thus, quantization of the soliton is
performed by accounting for the spectrum shift in the
quasiparticle excitations and imposing commutation relations for
the soliton position and momentum operators.

For a good introduction on the quantization of solitons from the
quantum field theorist's point of view, see the book of
\citet{raj} or the review articles of \citet{coleman:1975} or
\citet{raj:1975}.

Recall, however, that the soliton is a spontaneously broken
symmetry solution: in has chosen an arbitrary point in space about
which to center. The Goldstone theorem\cite{goldstone:1961}
predicts a gapless boson mode restoring this broken symmetry. This
causes divergences if we consider the next order semiclassical
expansion of the quantized soliton, because of zero energy
denominators that appear.

An analogous situation for a simple particle is when the potential
is completely flat. To all orders we find zero frequencies when
expanding the potential. This is because all points are degenerate
and the particle must randomly choose among them. In the quantum
version, we find that the particle is no longer an eigenvalue of
position at all, but rather of momentum, in the form of a plane
wave.

For the soliton, the Goldstone mode is dealt with in essentially
the same way. For each broken symmetry, the quantized soliton has
an associated momentum which is a good quantum number. For
example, if the soliton exists in a translationally invariant
system, we would find it has a well defined momentum in the
quantized version. This, incidentally, provides a systematic
method for calculating the mass of the soliton.

The general methods for separating the Goldstone mode involve
introducing a collective coordinate for each broken
symmetry\cite{jackiw:1975,sakita:1975,tomboulis:1975}. Since the
original system doesn't depend on these coordinates, the final
expanded energy functional can only depend on their conjugate
momenta.

The magnetic system of this thesis has a two dimensional
translational symmetry broken by the introduction of a vortex.
Thus, we promote the vortex center coordinates to collective
coordinates to we obtain an effective action depending only on the
associated conjugate momentum via a particle-like $\frac{p^2}{2m}$
term.

\subsection{In condensed matter theory}

In condensed matter, we are more specifically interested in the
physical consequences of the quantized solitons, as opposed to
their mere existence and basic properties. Shortly after the
quantum field theorists developed the soliton quantization
methods, Krumhansl and Schrieffer\cite{krumhansl:1975,currie:1980}
showed that one dimensional quantized solitons could be treated
exactly as elementary excitations, in addition to the ever-present
quasiparticles. To explain, suppose we've quantized a soliton in a
translationally invariant system (of length $L$ with minimum
length scale $l$). In the most general case, we would find, in
addition to the regular Goldstone mode, a finite number of
quasiparticle modes localized to the soliton, interpretable as
soliton excited states, followed by the usual continuum of
extended quasiparticle excitations. Krumhansl and Schrieffer show
that the total internal energy of the system can be simplified to
\begin{equation}
U = \left(\frac{L}{l}-N_bN_k^{tot}\right)k_B T +N_k^{tot} \left(
E_k^0 + \frac{1}{2}k_B T + (N_b -1) k_B T \right)
\end{equation}
where $N_b$ is the total number of localized quasiparticle states,
including the translation symmetry-restoring Goldstone mode.

This represents the internal energy of a system with
$\left(\frac{L}{l}-N_bN_k^{tot}\right)$ quasiparticle modes and
$N_k^{tot}$ particles of rest energy $E_k^0$ each having
$\frac{1}{2}k_B T$ translational energy and thermal energy $k_B T$
for each of the $N_b -1$ internal modes. The average number of
particles $N_k^{tot}$ forming the soliton is calculated using
thermodynamic relations once we define a soliton chemical
potential. See \citet{currie:1980} for more details of the
complete thermodynamic description of the soliton as an ideal gas.

Quantum vortices were first considered by condensed matter
theorists as early as the 1940's by \citet{onsager:1949} in
superfluid helium. \citet{feynman:1955} developed further the idea
of these vortex lines to explain the dissipation mechanisms for a
rotating superfluid and conjectured that they may also be
responsible for the superfluid to normal fluid phase transition.
Unfortunately, in 3D the problem is essentially unsolved, so that
no details of a vortex driven phase transition have yet been
developed.

In 2D, the problem is more tractable, and in the 1970's,
\citet{KT:1973} detailed a phase transition due to the
proliferation of dislocations. The theory applies equally to
vortices. Below the transition, the free energy is minimized by
maintaining the vortex-antivortex pairs bound; however, raising
the temperature to the transition, the gain in entropy by
unbinding the pairs balances the increase in energy.

In two dimensions, the energy of a dislocation or vortex diverges
logarithmically in the system surface area,
\begin{equation}\label{logenergy}
E = E_0 \ln\frac{A}{A_0}
\end{equation}
where $A_0\sim a^2$ is the smallest area in the discrete system,
where $a$ denotes the lattice spacing.

The entropy associated with the dislocation also depends
logarithmically on the area since there are approximately $A/A_0$
possible positions for it to center on,
\begin{equation}
\mathcal S = k_B\ln\frac{A}{A_0}
\end{equation}
where $k_B$ is the Boltzmann constant. Since the energy and
entropy depend on the size of the system in the same way, the free
energy, $F=E-TS$, is dominated by the energy term at low
temperatures so that the probability of an isolated dislocation in
a large system is vanishingly small. At high temperatures,
dislocations appear spontaneously as the entropy term takes over.
The phase transition temperature can be roughly estimated as $T_c
= E_0/k_B$.

In the late 1970's, very important new phenomena were discovered
independently by the particle and condensed matter physicists.
\citet{jackiw:1976} in considering the Dirac equation in the
presence of a soliton found it had fermionic $\frac{1}{2}$ states;
while, Su, Schrieffer and Heeger\cite{schrieffer:1979} were
studying kinks in a coupled electron-phonon model for the quasi-1D
conducting polyacetylene and found a neutral spin $\frac{1}{2}$
soliton state.

\begin{figure}
\centering
\includegraphics[width=10cm]{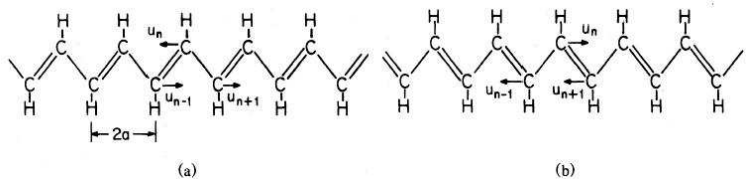}
\caption{The two degenerate dimer states of \emph{trans}-(CH)$_N$,
polyacetylene.}\label{figure:polydimers}
\end{figure}
\begin{figure}
\centering
\includegraphics[width=5cm]{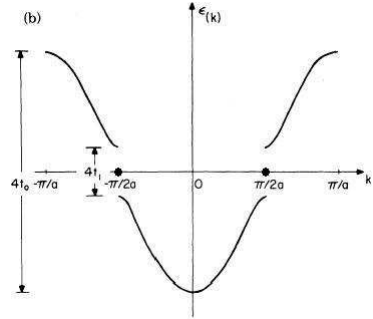}
\caption{The band structure of polyacetylene, gapped due to the
electron-phonon interactions. Note the two isolated electron
states in the gaps are only in the presence of a
kink.}\label{figure:polyspectrum}
\end{figure}
Restricting ourselves to the polyacetylene system, consider a one
dimensional system of electrons in a tight-binding model
interacting linearly with the lattice coordinate displacements
(essentially, coupling the electrons and phonons). The Hamiltonian
of this system is then
\begin{align}
H = & \sum_{n=1}^N \left(\frac{p_n^2}{2m} +
\frac{K}{2}(u_{n+1}-u_n)^2\right) - t_0 \sum_{n=1,
s=\pm\frac{1}{2}}^N \left( c_{n+1,s}^{\dagger} c_{n,s} +
c_{n,s}^{\dagger}c_{n+1,s}\right) \nonumber \\
& + \alpha \sum_{n=1, s=\pm\frac{1}{2}}^N (u_{n+1}-u_n) \left(
c_{n+1,s}^{\dagger} c_{n,s} + c_{n,s}^{\dagger}c_{n+1,s}\right)
\end{align}
where $u_n$ and $p_n$ are the lattice coordinate displacements and
their conjugate momenta, characterized by mass $m$ and stiffness
constant $K$. The electrons are denoted by creation/annihilation
operators $c^{\dagger}_{i,s}$ and $c_{i,s}$ at site $i$ with spin
$s$, with hopping constant $t_0$ and coupling constant $\alpha$
with the lattice displacements.

\begin{figure}
\centering
\includegraphics[width=6cm]{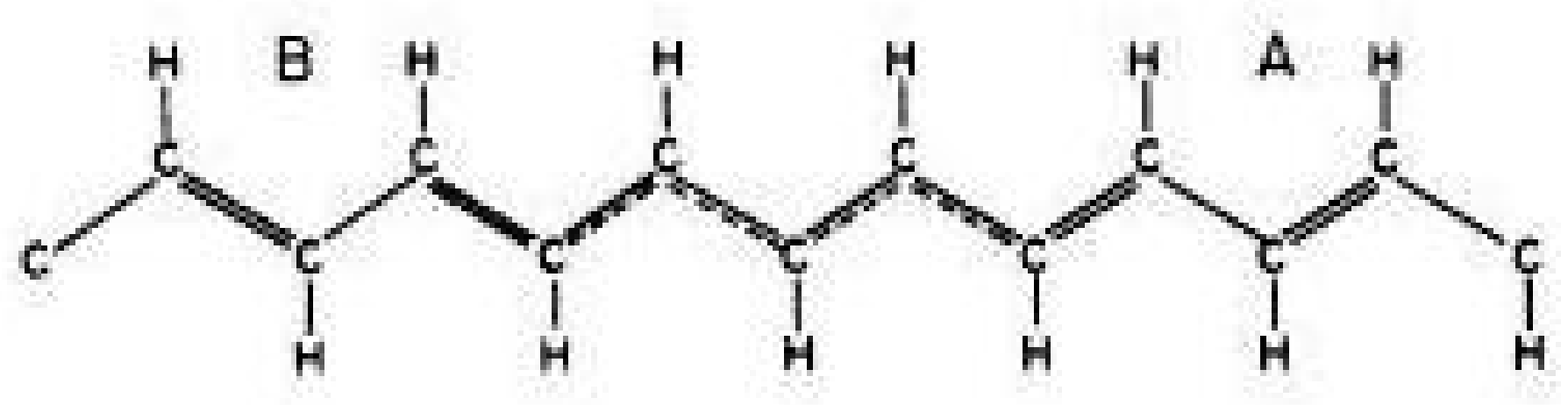}
\includegraphics[width=4cm]{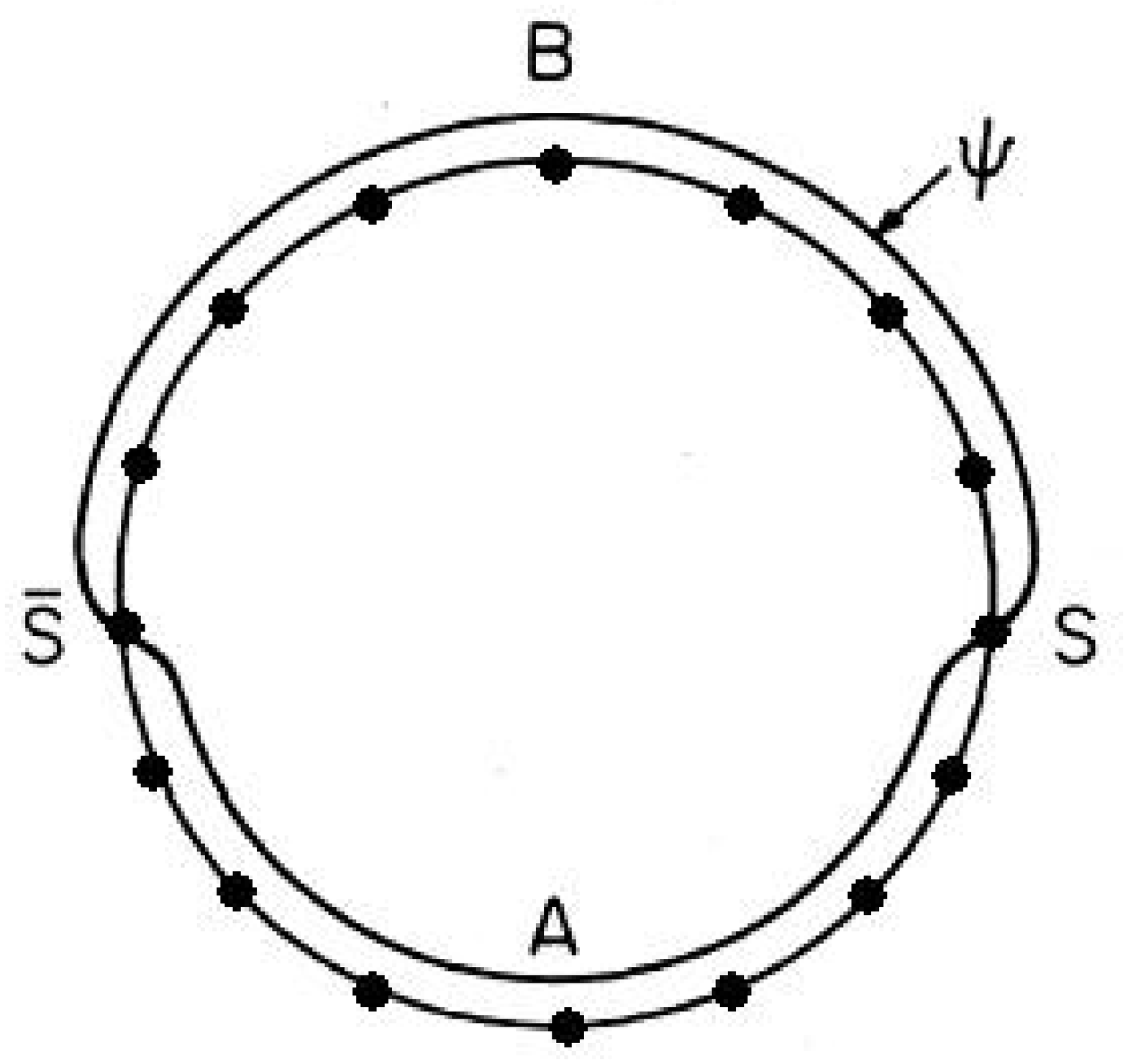}
\caption{A kink solution connecting the two degenerate dimer
ground states, shown, left, on the linear polyacetylene chain,
and, right, on the idealized chain with periodic boundary
conditions.}\label{figure:polykink}
\end{figure}

The ground state of this system is doubly degenerate and
spontaneously breaks reflection symmetry (this was predicted by
\citet{peierls:1955} using mean-field approximation for any
non-zero electron-phonon coupling). Figure \ref{figure:polydimers}
shows the two degenerate dimer states. As a consequence of the
two-fold degeneracy, there exist the kink and antikink topological
solitons connecting the degenerate ground states (see Figure
\ref{figure:polykink}---in actuality, the kink is spread over
$\sim 14a$).

\citet{schrieffer:1979} found that the kink had two states: a
charged state, $Q=\pm e$, with spin $s=0$, and a neutral state
with spin $s=\pm \frac{1}{2}$. In addition, when the kink is in
its neutral state, there is an $s=0$ electron state in the middle
of the gap (see Figure \ref{figure:polyspectrum}, note there are
two states, one localized to the kink, the other to the antikink)
formed by pulling $\frac{1}{2}$ a state per spin out of the Fermi
sea.

The polyacetylene study introduced to condensed matter physics
what the particle theorists independently introduced within a
relativistic field theory: the existence of states with fractional
charge. Although the $\frac 1 2$ charge is obscured by the
doubling of degrees of freedom due to spin, the zero energy state
is still formed by drawing half an electronic state (of each
spin). Furthermore, the spin-charge relations are also unusual:
charged solitons are spinless while neutral solitons carry spin
$\frac 1 2$.

\subsection{Superfluid He$^4$}

Returning our discussion specifically to vortices in condensed
matter, quantum vortices were first proposed by
\citet{onsager:1949} and developed more completely by
\citet{feynman:1955}. A quantum vortex can be imagined as a
regular fluid vortex with a cylindrical core shrunk down to atomic
dimensions. The circulation of the vortex is quantized in units of
$h/m$, where $h$ is the Planck constant and $m$ is the bare $^4$He
mass.

\begin{figure}
\centering
\includegraphics[width=6cm]{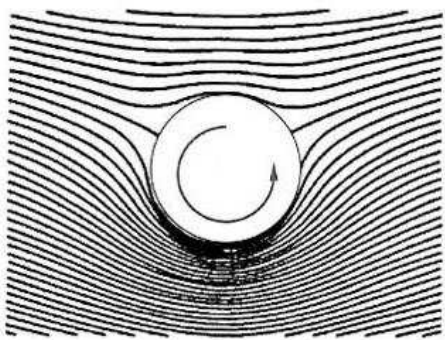}
\caption{The equi-pressure lines of a fluid surrounding a rotating
cylinder. The pressure differential top and bottom creates an
upward force. The fluid flow is to the
left.}\label{figure:rotcylinder}
\end{figure}

Describing the motion of superfluid vortices by making analogy to
the motion of their parent fluid vortices was extremely
successful. Early experiments by Hall and Vinen
\cite{hall:1956I,hall:1956II} found that if they applied an
impulsive force setting a superfluid vortex into motion the vortex
underwent helical motion (resembling that of an electron drifting
in a magnetic field). In general, such a force arises always when
a body with a flow circulation around it moves through a liquid or
gas as in, for example, Figure \ref{figure:rotcylinder}.

First noted in 1852 by Magnus when studying inaccuracies in the
firing of cannon balls, the force responsible, named the Magnus
force after its discoverer, can be explained in terms of the
Bernoulli equation. The speed of the fluid is effectively lower on
one side of the rotating body than the other (perpendicularly to
the flow of the fluid, of course) so that the side with higher
speed has lower pressure---thus the body experiences a force in
that direction (see Figure \ref{figure:rotcylinder}). The Magnus
force in a superfluid is written
\begin{equation}\label{magnus}
\mathbf F_M = \rho_s \kappa \times (\mathbf v - \mathbf v_s)
\end{equation}
where $\rho_s$ is the superfluid density, $\mathbf v$ is the
vortex velocity and $\mathbf v_s$ is the asymptotic superfluid
velocity (affected, of course, by the vortex presence).

Hall and Vinen found the motion of their experimentally observed
vortices could be explained with such a perpendicular Magnus force
and an inertial mass of the order $\rho \xi^2$, where $\rho$ is
the fluid density and $\xi$ is the vortex radius.

In addition, damping forces acting on the vortex were introduced
with phenomenological parameters. The most general damping can act
both longitudinal (as we are most accustomed to) and transverse to
the vortex motion, expressible as
\begin{equation}
\mathbf F_d = D (\mathbf v_n - \mathbf v) + D' \mathbf{\hat
\kappa} \times (\mathbf v_n - \mathbf v)
\end{equation}
where $\mathbf v_n$ denotes the normal fluid velocity, whose exact
definition might vary from one formalism to another. Note that the
transverse damping term has the same behaviour of the Magnus force
(with potentially an additional force $\propto \mathbf v_n -
\mathbf v_s$).

Although this heuristic description is very successful in
explaining observed phenomena, the microscopic derivation of the
various parameters is far less successful. There is considerable
disagreement, especially in calculations of the transverse
dissipation parameter.

An early calculation by Iordanskii \cite{iordanskii:1964,
iordanskii:1966} revealed a transverse damping force, later termed
the Iordanskii force, proportional to the normal fluid density
\begin{equation}
\mathbf F_I = \rho_n \kappa \times (\mathbf v - \mathbf v_n)
\end{equation}
due to the scattering of phonons on the vortex. This entails an
effective Magnus force with the superfluid density replaced by the
total fluid density, $\rho_s \to \rho$, plus additional forces
proportional to $\mathbf v_n - \mathbf v_s$.

In the early 1990's, Thouless, Ao and Niu
\cite{thouless:1996,ao:1993} (TAN) claimed that the transverse
force was exactly the bare Magnus force of equation
(\ref{magnus}), at all temperatures while accounting for the
scattering of phonons. The force on the vortex line due to phonons
is simply the variation of the phonon energy expectation with
vortex position
\begin{equation}
\mathbf F = -\sum_\alpha f_\alpha \langle \psi_\alpha | \nabla_0 H
| \psi_\alpha\rangle
\end{equation}
where $f_\alpha$ denotes the occupation probability of the phonon
state $\alpha$. By expanding the phonon wavefunction to first
order in vortex velocity using time-dependent perturbation theory,
TAN were able to rewrite the force as an integral over the Berry
phase associated with a closed loop around the vortex. Assuming no
circulation in the normal fluid density, this reduces exactly to
the zero temperature Magnus force.

The transverse force on the vortex line can also be expressed as
the commutator of the $x$ and $y$ components of the total momentum
operator
\begin{equation}
[P_x, P_y] = \int\int dx dy \left( \frac{\partial
\psi^\dagger}{\partial x}\frac{\partial \psi}{\partial
y}-\frac{\partial \psi^\dagger}{\partial y}\frac{\partial \psi
}{\partial x}\right)
\end{equation}
Applying Stokes' theorem, the integral over the cross-sectional
area can be expressed instead as a line integral about the
boundary of the one particle density matrix. TAN argue that this
boundary may be extended very far from the vortex core so that
contributions from localized phonon states at the vortex core do
not influence the transverse force \cite{thouless:1996}.

In opposition to Thouless, Sonin \cite{sonin:1997} explained the
transverse damping force via an analogous mechanism to the
Aharonov-Bohm \cite{AB:1959} effect of an electron passing a
double slit in the presence of a magnetic vector potential (though
in regions of no magnetic field). The electrons passing in one
slit relative to the other experience a phase shift due to the
vector potential term, causing a horizontal shift in the observed
interference pattern. However, this entails a momentum transfer
from the magnetic field source, here a conducting coil, to the
electrons, transverse to the double slit screen, and thus a
transverse force acting on the coil.

Similarly, quasiparticles passing above or below a moving vortex
experience a relative Berry's phase shift \cite{Berry:1984}. A
momentum transfer must occur between the vortex and
quasiparticles, again, entailing a transverse damping force.

Sonin calculated the effective transverse force exactly in the
form
\begin{equation}
\mathbf F_t = (\rho_s+\rho_n) \kappa \times (\mathbf v - \mathbf
v_n)
\end{equation}
so that the effective Magnus force is the regular Berry's phase
result plus the Iordanskii force. The normal fluid velocity here
is in the vicinity of the vortex and may differ from the
asymptotic velocity due to viscous dragging of the normal fluid by
the vortex motion\cite{hall:1956I}.

One apparent source of disagreement, first noted by Sonin, is that
the vortex undergoes oscillatory motion due to the passage of
phonon quasiparticles. The scattering calculations of
\citet{fetter:1964} and \citet{demircan:1995}, which supported the
TAN Berry's phase calculation, effectively held the vortex fixed
by an external pinning potential, thereby nullifying the
transverse damping force.

The transverse dissipation is not the only source of controversy.
The effective mass itself of the quantized vortex has not been
agreed upon. Initial estimates are based on the inertial mass of
the circulating fluid, essentially, $\rho r_0^2$, with $r_0$ the
radius of the vortex. In the quantum limit, the vortex radius
shrinks down to atomic dimensions, or zero, so that the vortex
mass tends to zero also.

Alternatively, as suggested by \citet{duan:1992}, the mass of the
vortex must be proportional to
\begin{equation}
M_v \propto \frac{E_v}{v_0^2}
\end{equation}
where $M_v$ is the vortex mass, $E_v$ is the stationary vortex
energy, and $v_0$ is the velocity scale of the superfluid
quasiparticles. This can be explained by purely dimensional
arguments.

For a quasi-2D vortex, however, the stationary vortex energy is
log divergent in the system cross-sectional area, as in
(\ref{logenergy}), suggesting the effective mass is also log
divergent, much larger than the vanishing estimate made earlier.

Clearly, the microscopic derivations of superfluid vortex dynamics
has yet to firmly agreed upon. The variety of conflicting results
suggests we re-examine the different methods used. Doing so in the
simpler magnetic system is an aim of this thesis, though,
unfortunately, a comprehensive study of the various methods could
not entirely be undertaken. Rather, we calculate results here
using regular perturbation theory, expanding in vortex velocity,
and using Feynman-Vernon influence functionals\cite{feynman:1963}.

\subsection{Magnetic vortices}

Magnetic systems have received much attention for their variety of
applications and their lucrative potential\cite{ross:2001}, for
example, in the market of magnetic memory. Vortices in magnetic
systems are very easily observed and manipulated, for example
using Brillouin light scatting \cite{novosad:2002} or magnetic
force microscopy (MFM)\cite{shinjo:2000}.

Despite the ease of experimentally observing magnetic vortices,
there have been relatively few microscopic derivations of the
dynamics of vortices in magnetic systems. In fact, these
derivations should be greatly simplified in a magnetic system;
however, the resulting dynamics still possess many of the same
strange aspects discussed with respect to superfluid vortices.

A magnetic vortex experiences a force transverse to its velocity,
the \emph{gyrotropic} force. This force acts exactly in the same
manner as the Magnus force, however, has a different microscopic
origin. It arises from a self induced Lorentz force, with the
vortex vorticity acting as an analogous charge, while the out of
plane spins create an effective perpendicular magnetic field (this
analogy is more fully developed in Chapter \ref{chapter:vortex}).
Notably, this force is dependent on both topological indices (and
is absent entirely for in-plane vortices for which the
polarization is zero), as compared to the Magnus force dependent
solely on the vortex circulation in a superfluid.

There are interactions with quasiparticles that may alter the
effective gyrotropic force. However, there have been no attempts
to describe a transverse damping force in a magnetic system. In
fact, all descriptions of dissipation in a vortex system have
focussed on calculating an average energy dissipation rate or have
been phenomenological (except for the work of \citet{sloncz:1984}
which we describe in a moment).

The earliest theoretical work on two dimensional magnetic systems
with vortices are adaptations of the work of Thiele
\cite{thiele:1973,thiele:1974}. Thiele first introduced the
gyrotropic force and dissipation dyadic acting on a magnetic
domain wall in a three dimensional system. His dissipative force,
however, was phenomenological employing a Gilbert damping
parameter (the phenomenological damping parameter normally
introduced into the so-called Landau-Lifshitz equations governing
the magnetization dynamics).

In the early 1980's, applying the work of Thiele,
\citet{huber:1982} and \citet{nikiforov:1983} independently
described the basic motion of a magnetic vortex. They calculated
the gyrotropic force and phenomenological damping forces acting on
a single vortex.

\citet{sloncz:1984} shortly thereafter considered perturbations
about a moving vortex, deducing an effective mass tensor. A
collection of vortices behave strongly coupled and the inertial
energy is not diagonal but rather must be expressed as
$\frac{1}{2}M_{ij}v_iv_j$ where there is an implied double sum
over the vortex indices $i$ and $j$. He calculated the vortex
dissipation via a frequency dependent imaginary mass term by
studying the asymptotic behaviour of the lowest order
vortex-magnon coupling. We will compare our dissipation results
with those of Slonczewski in Chapter \ref{chapter:together}.

Scattering phase shifts have been calculated for a variety of
planar magnetic systems
\cite{gouvea:1989,rodriguez:1989,pereira:1996}. They were
primarily interested in the thermodynamic behaviour
\cite{krumhansl:1975} of such systems and searching for a vortex
signature that could be measurable to verify a Kosterlitz-Thouless
\cite{KT:1973} transition. In fact, based on the modified spin
correlations due to the presence of vortices, a central peak found
in neutron-scattering experiments could be
reproduced\cite{mertens:1987}.

In a series of papers\cite{mertens:1997,volkel:1994,
mertens:1999}, Mertens et. al. modeled numerically the motion of a
vortex pair assuming various boundary conditions. The ensuing
motion was best reproduced assuming an non-Newtonian equation of
motion which included a third time derivative of the vortex
position.

We find just such a small third time derivative term in our
influence functional analysis.  We compare our results with
Mertens et. al. in section \ref{section:imagpart}.

However, this is a misapplication of the collective coordinate
formalism: each collective coordinate is meant to replace a
continuous symmetry broken by the vortex. In a planar system, a
vortex breaks the two dimensional translational symmetry allowing
the introduction of a two dimensional center coordinate only.

There has been no work yet to find effective damping forces acting
dynamically on a magnetic vortex. In this thesis, we calculate
these forces assuming an averaged motion of the perturbing
magnons.

\section{Easy plane insulating ferromagnet}

We study an insulating plane of spins, that is, fixed on their
lattice sites, ferromagnetically coupled, lying preferentially in
the plane. The order parameter of the easy plane ferromagnet lies
on the unit sphere but with an energy barrier at both the north
and south poles. There are hence topological solitons
spontaneously breaking the ground state symmetry, the continuous
in-plane symmetry, and, at some energy cost to restore continuity,
twisting out of plane to break the discrete up/down symmetry.
There are also discontinuous vortices lying entirely in the plane
as found in the XY model.

We noted in the symmetry breaking discussion that a vortex lying
in this order parameter space does not have topological stability.
This however is for a completely degenerate sphere. Here, there is
an energy barrier for paths to cross the two poles so that any
homotopy of a vortex to a point would require passing a
macroscopic number of spins through this energy barrier. The
vortex thus has approximate topological stability, unless the
anisotropy becomes vanishingly small.

The energy of a general state $\{\mathbf S_i\}$ of this lattice is
\\
\begin{equation}
E = -\frac{1}{2} \sum_{i,j} J_{ij} \mathbf S_i\cdot \mathbf S_j +
\sum_{i} K S_{iz}^2
\end{equation}
\\
where the indices extend over all lattice points in the 2D
lattice. The first term is the exchange term and is approximated
by including nearest neighbour interactions only, negative to
ensure ferromagnetic coupling,
\\
\begin{equation*}
-\frac{1}{2} \sum_{i,j} J_{ij} \mathbf S_i\cdot \mathbf S_j \to
-\frac{1}{2} \sum_{<i,j>} J \mathbf S_i\cdot \mathbf S_j
\end{equation*}
\\
where $<i,j>$ denotes nearest neighbour pairs. For simplicity,
we've assumed a constant exchange parameter $J$. The second term
enforces the easy plane anisotropy, where $K$ is the anisotropy
parameter (for $S >1/2$).

Since we are interested in the low energy behaviour, we eliminate
the short length scale fluctuations by describing the system in a
continuum approximation. Instead of a spin $\mathbf S_i$ at site
$i$, we now have a spin field $\mathbf S(\mathbf r)$. Sums are
replaced by integrals over space. For instance, the anisotropy
term becomes
\\
\begin{equation*}
\sum_{i} K S_{iz}^2 \to \int d^2r \tilde K S^2_z(\mathbf r)
\end{equation*}
\\
and the exchange term becomes
\begin{equation*}
-\frac{1}{2} \sum_{<i,j>} J \mathbf S_i\cdot \mathbf S_j \sim
\frac{1}{4} \sum_{<i,j>} J \left(\mathbf S_i-\mathbf
S_j\right)\cdot \left(\mathbf S_i-\mathbf S_j \right) \to
\frac{1}{2} \int d^2r \tilde J \left(\nabla \mathbf S \right)^2
\end{equation*}
\\
where adding the constant $S^2$ terms doesn't affect the dynamics.
Note that $\left(\nabla \mathbf S \right)^2 = \left(\nabla S_x
\right)^2+\left(\nabla S_y \right)^2+\left(\nabla S_z \right)^2$.
The redefined constants are given by $\tilde J = J/2$ and $\tilde
K = K/a^2$, noting that we use new dimensions for an anisotropy
density. From here on, we drop the tildes and simply use $J$ and
$K$ for the continuum versions of the exchange and anisotropy
parameters.

The Hamiltonian describing the system is then written
\begin{equation}\label{totHamiltonian}
H = S^2 \int d^{2} r \left(\frac{J}{2} (\mathbf{\nabla}\theta)^{2}
+ \sin^{2}\theta \left(\frac{J}{2}(\mathbf{\nabla}\phi)^2 -
K\right)\right)
\end{equation}
\\
where the spin field is expressed in angular coordinates, $\mathbf
S = S$ $(\sin\theta\cos\phi,$ $\sin\theta\sin\phi,$ $\cos\theta)$.

As explained in Appendix \ref{chapter:mechanics}, $\phi$ and
$-S\cos\theta$ are conjugate variables in the discrete lattice so
that the Lagrangian can be expressed in the continuum limit, $\sum
\to \int d^2r/a^2$ where $a$ is some lattice spacing length scale,
\begin{equation}\label{totLagrangian}
\mathcal L = S \int \frac{d^{2} r}{a^2} \left( -\cos\theta
\dot\phi- \frac{c}{2} \left( (\mathbf{\nabla}\theta)^{2} +
\sin^{2}\theta \left( (\mathbf{\nabla}\phi)^2 - \frac{1}{r_v^2}
\right)\right)\right)
\end{equation}
\\
where we've defined the speed scale $c/r_v$ with $c = SJa^2$ and
the length scale $r_v = \sqrt{J/2K}$.

Using Hamilton's equations (\ref{Hameqns}) or the Euler-Lagrange
equation (\ref{ELeqns}), we find the equations of motion
\begin{align}\label{fulleoms}
\frac{1}{c}\frac{\partial\phi}{\partial t} =&
-\frac{\nabla^2\theta}{\sin\theta}+\cos\theta(\mathbf{\nabla}\phi)^2
- \frac{1}{r_v^2}\cos\theta \nonumber\\
\frac{1}{c}\frac{\partial\theta}{\partial t} =& \sin\theta
\nabla^2\theta + 2 \cos\theta \mathbf{\nabla}\theta \cdot
\mathbf{\nabla} \phi
\end{align}

There are two families of elementary excitations: the perturbative
spin waves, or magnons, and the vortices. The vortices have two
forms: the so-called in-plane solutions with polarization $0$, and
the out-of-plane solutions with polarization $\pm 1$. The
treatment in this thesis considers explicitly the out-of-plane
solutions, however, setting the polarization to $0$ recovers the
results for the in-plane solutions. The out of plane spin
behaviour cannot be solved analytically; however, the core and far
field asymptotic limits suffice for obtaining general results.

The spin waves are small amplitude oscillations about the
ferromagnetic ground state or about a vortex state, in both cases
with an ungapped spectrum. The difference in the two spectra can
be attributed to the vortex presence and yields an effective zero
point energy to the quantized vortex. The equations of motion for
the vacuum magnons are modified to the equations of motion of
magnons in the presence of a vortex. The additional terms are
interpreted as the magnon-vortex interaction terms.

There is a one magnon coupling with the vortex velocity. Normally,
considering a central system coupled to perturbative `bath' modes,
we find to lowest order a one magnon coupling with the vortex
field. There is no such coupling here because the vortex is itself
a minimum action solution of the same system in which the magnons
arise. Thus, there are no first order variational terms. This
assumes, however, that the vortex profile is unchanging in time.
Allowing it to move about the system introduces a first order
coupling between the vortex velocity and the magnons.

There is also a two magnon coupling affecting the magnon energy
with long range effects. This term scatters the magnon modes and
hence alters their zero point energy. We attribute this shift
instead to the quantized vortex state. This two magnon coupling
has other dissipative effects and energy shifts that are not
treated in this thesis.

We first review the basic characteristics of the vacuum magnon
modes and the vortex solutions. The gyrotropic and inter-vortex
forces are found immediately by expanding the Lagrangian about a
many vortex solution.

We then examine the effects of the various couplings between
magnons and vortices. The one magnon coupling can be interpreted
as small vortex deformations when moving at velocity $\mathbf V$
or, alternatively, as a single magnon scattering event. The second
order perturbation energy correction of this one magnon coupling
goes as $V^2$ and is thus interpretable as an inertial energy,
from which we can deduce an effective vortex mass. There is an
additional imaginary energy shift, or a dissipation, from this
coupling.

The two magnon scattering term has a zero point energy shift and
other magnon occupation dependent energy shifts. We do not retain
higher order scattering terms, keeping only one magnon couplings,
although they may indeed contribute more significantly to the
vortex dissipation\cite{stamp:1991,dube:1998}.

The dynamical effect of the one magnon coupling is examined fully
in the Feynman-Vernon influence functional
formalism\cite{feynman:1963}. The two sub-systems are assumed
initially non-interacting with the magnons in thermal equilibrium.
They are thereafter allowed to interact, the magnons generally
shifting out of equilibrium. The effect of the magnons is then
averaged over by tracing out their degrees of freedom. This
yields, in an averaged way, the effect of the magnons on the
vortex motion. As found in perturbation theory, the one magnon
coupling is responsible for two new terms in the vortex effective
action: an inertial energy term and a damping force term.

In addition to the usual longitudinal damping force, we find a
transverse damping force reminiscent of the Iordanskii force in
superfluid helium. Such a term has not before been suggested in a
magnetic system. The damping forces possess memory effects---that
is, they depend on the previous motion of the vortices.

For a collection of vortices, we find that their particle-like
properties are not independent. They have mixed inertial terms
such as $\frac{1}{2} M_{ij} v_iv_j$ and damping forces due to the
motion of one vortex acting on another.

Next, we review the basics of the two elementary excitations,
first the magnons and after the vortices.

\chapter{Magnons}\label{chapter:magnons}

The plane of spins with easy plane anisotropy has a degenerate
ground state. The spins are ferromagnetically coupled and thus
prefer to align, however, they may choose to align along any
direction in the plane---an example of spontaneously broken
symmetry. The Goldstone theorem predicts that there should then
exist boson quasiparticle excitations that are not gapped. In this
system, these Goldstone modes are the small amplitude, or
perturbative, spin waves. When quantized, the excitations are
termed magnons.

The magnon spectrum in the easy plane ferromagnet is ungapped,
however due to the hard axis, the spectrum is not simply the
regular ferromagnet spin wave spectrum $\omega(k) \propto k^2$.
Instead we find a spectrum with reduced density of states near
$\omega = 0$.

We begin by examining the small amplitude equations of motion
satisfied by the magnons; thereby deriving the magnon spectrum and
density of states. We calculate a few old results using spin path
integrals as illustrative examples that we will need in later
calculations. We derive the quantum propagator, a calculation
following closely that of a simple harmonic oscillator. The
quantum propagator is then manipulated to again reveal the magnon
spectrum and, under a simple substitution, to yield the thermal
equilibrium density matrix.

\section{Magnon equations of motion}\label{section:magnons}
The magnons are the quasiparticle excitations of our system. As
such, to describe their motion and properties, we expand in small
deviations about the ferromagnetic in-plane ground state
\begin{align}
\vartheta=& \theta - \pi/2 \nonumber \\
\varphi =& \phi
\end{align}
where we've chosen the ground state $\phi = 0$ amongst the
continuum of ground states without loss of generality. The
complete system Lagrangian in terms of these perturbing variables
$\varphi$ and $\vartheta$ becomes
\\
\begin{equation}\label{linLagrangian}
\mathcal L_m = S\int \frac{d^2r}{a^2} \left( \dot \varphi
\vartheta - \frac{c}{2} \left( -\varphi \nabla^2 \varphi -
\vartheta \nabla^2 \vartheta + \frac{\vartheta^2}{r_v^2}\right)
\right)
\end{equation}
\\
where $J$ is the exchange constant and $K$ is the anisotropy
constant, $a$ is the lattice spacing, $c=SJa^2$ and $r_v^2 =
\frac{J}{2K}$. The conjugate momentum is now $S\vartheta$, the
linearized version of $-S\cos\theta$. We essentially expand the
Lagrangian to second order perturbations to obtain a simple
harmonic-like Lagrangian. Consequently, many calculations to come
here mimic very closely those for a simple harmonic oscillator.

Varying (\ref{linLagrangian}) with respect to $\varphi$ and
$\vartheta$ yields the magnon equations of motion
\\
\begin{align}\label{magnoneom}
    \frac{1}{c}\frac{\partial\varphi}{\partial t} =&
    -\nabla^2\vartheta + \frac{1}{r_v^2}\vartheta \nonumber \\
    \frac{1}{c}\frac{\partial\vartheta}{\partial t} =&
    \nabla^2\varphi
\end{align}
\\
Alternatively, we could have linearized the system equations of
motion, (\ref{fulleoms}), directly with identical results.

\begin{figure}
\centering
\includegraphics[width=6cm]{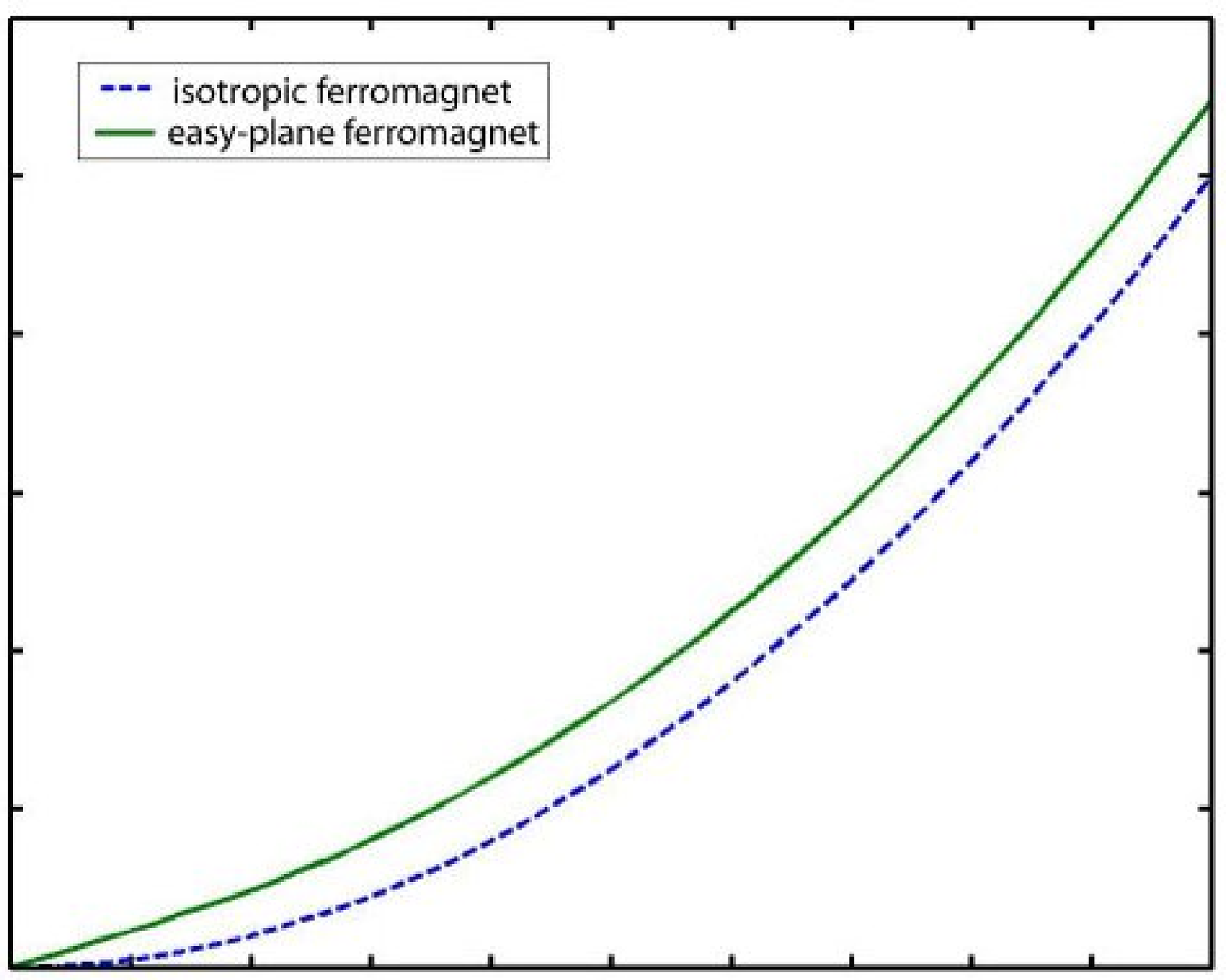}
\includegraphics[width=6cm]{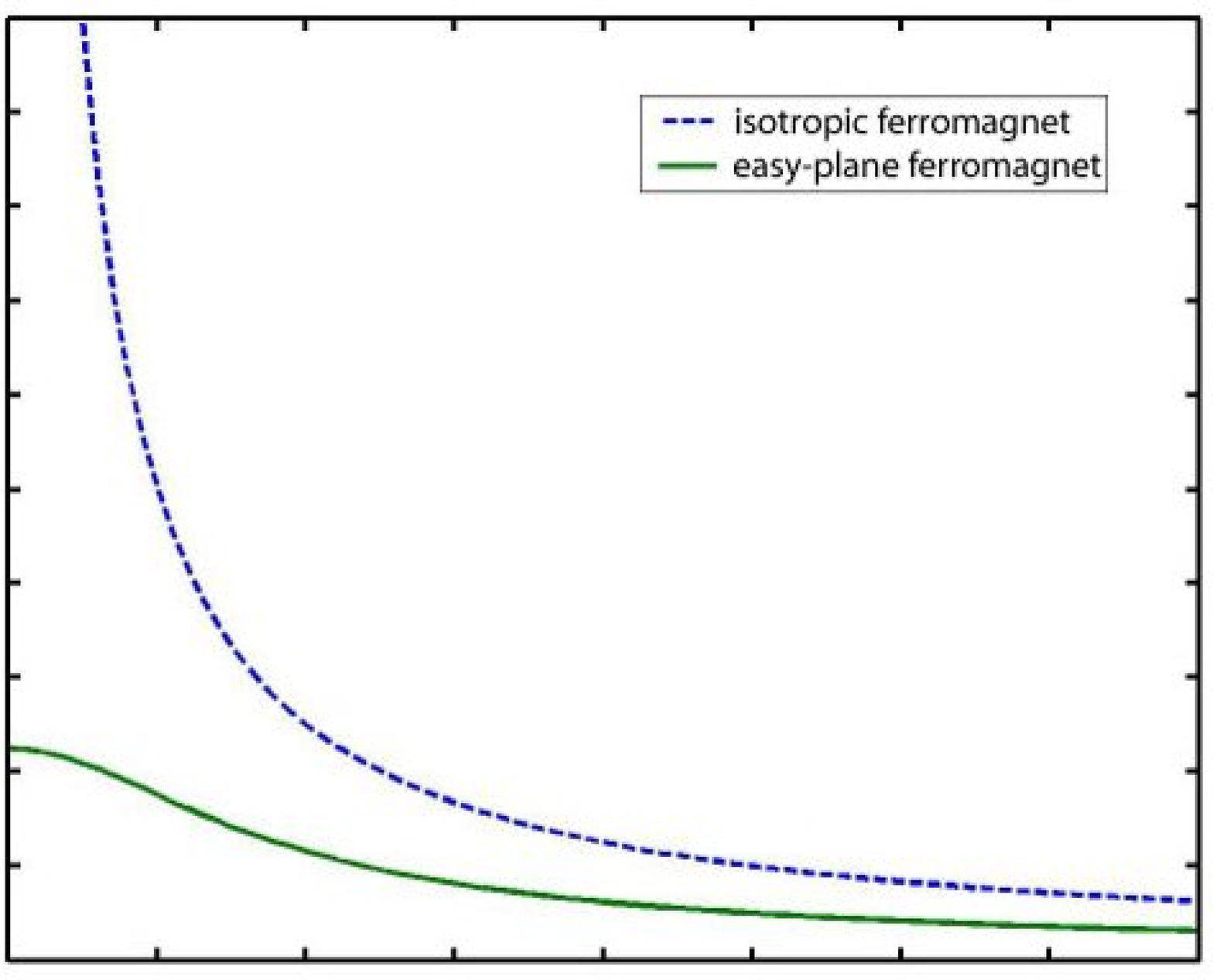}
\caption{A comparison of the easy plane magnon spectrum and
density of states with the regular isotropic
ferromagnet.}\label{figure:spectrums}
\end{figure}

The analysis proceeds in a plane-wave expansion. This system of
equations can be solved by Fourier transforming so that
$\nabla^2\varphi \rightarrow -k^2 \varphi_k$ and $\nabla^2
\vartheta \rightarrow -k^2 \vartheta_k$. Assuming harmonic time
dependence, the eigenvalues of the equations of motion yield the
magnon spectrum
\\
\begin{equation}\label{magnonspectrum}
    \omega(k) = ckQ
\end{equation}
\\
where $Q=\sqrt{k^2+\frac{1}{r_v^2}}$. The spectrum is not gapped
(note the overall factor of $k$), a reflection of the continuous
degeneracy of ground states. However, the density of states goes
as $\frac{Q}{k^2+Q^2}$ remaining finite as $\omega \to 0$ in
comparison to the isotropic ferromagnet with density of states
$\frac{1}{k}$ diverging for zero frequency. The two systems are
compared in Figure \ref{figure:spectrums}.

Alternatively, Fourier transforming the magnon Hamiltonian
directly via
\\
\begin{align}\label{fouriertransform}
    \varphi(\mathbf r) =& \frac{a^2}{(2\pi)^2}\int d^2k
    e^{-i\mathbf k \cdot \mathbf r} \varphi_{\mathbf k} \nonumber \\
    \vartheta(\mathbf r) =& \frac{a^2}{(2\pi)^2}\int d^2k
    e^{-i\mathbf k \cdot \mathbf r} \vartheta_{\mathbf k}
\end{align}
\\
diagonalizes $H_m=\int\frac{dr}{a^2}S\vartheta \dot \varphi
-\mathcal L_m$ to
\\
\begin{equation}
H_m = \frac{Sc}{2} \int \frac{d^2k a^2}{(2\pi)^2} \left( k^2
\varphi_{\mathbf k} \varphi_{-\mathbf k} + Q^2 \vartheta_{\mathbf
k} \vartheta_{-\mathbf k} \right)
\end{equation}

To quantize the magnons, we impose the commutation relations
between the conjugate variables $\varphi_{\mathbf k}$ and
$S\vartheta_{\mathbf k}$
\\
\begin{equation}
    [S \vartheta_k, \varphi_{k'}] = -i\hbar(2\pi)^2\frac{\delta^2
    (\overrightarrow k - \overrightarrow k')}{a^2}
\end{equation}
\\
We diagonalize the system now via the transformation to
creation/anniilation operators
\\
\begin{align}\label{creatannil}
    a_{\mathbf k} =& \sqrt{\frac{Sk}{2\hbar Q}} \left(\varphi_{\mathbf k}+\frac{iQ}{k}
    \vartheta_{-\mathbf k} \right) \nonumber \\
    a_{\mathbf k}^{\dagger} =& \sqrt{\frac{Sk}{2\hbar Q}}
    \left(\varphi_{-\mathbf k}-\frac{iQ}{k} \vartheta_{\mathbf k} \right)
\end{align}
\\
normalized such that $[a_{\mathbf k}, a_{\mathbf k'}^{\dagger}] =
(2\pi)^2\frac{ \delta^2(\mathbf k - \mathbf k')}{a^2}$.
Substituting for $\varphi_{\mathbf k}$ and $\vartheta_{\mathbf k}$
in terms of $a_{\mathbf k}$ and $a_{\mathbf k}^{\dagger}$ into the
Fourier transformed Hamiltonian gives after some manipulation
\\
\begin{align}
H =& \int \frac{a^2d^2k}{(2\pi)^2} \frac{\hbar\omega_k}{2} \left(
a_{\mathbf k}^{\dagger}a_{\mathbf k} +
a_{\mathbf k} a_{\mathbf k}^{\dagger}\right) \nonumber \\
=& \int \frac{a^2d^2k}{(2\pi)^2} \; \hbar\omega_k \left(
a_{\mathbf k}^{\dagger}a_{\mathbf k} + \frac{1}{2} \right)
\end{align}
\\
where $\omega_k$ is again the magnon dispersion relation
(\ref{magnonspectrum}).

We interpret the operators $a_{\mathbf k}^{\dagger}$ and
$a_{\mathbf k}$ exactly as for the simple harmonic oscillator
creation/annihilation operators. The combination $a_{\mathbf
k}^{\dagger}a_{\mathbf k}$ is thus the magnon number operator
$n_{\mathbf k}$ and the spectrum has energy $\hbar\omega_{\mathbf
k}$ for each of the $n_k$ magnons plus an additional
\emph{zero-point energy} $\frac{1}{2}\hbar\omega_{\mathbf k}$ for
each wavevector $\mathbf k$.

Notice throughout that we associate factors of $a^2$ to the
spacial and frequency integration measures to keep them
dimensionless. This is consistent since the integrals replace sums
appearing in the original discrete system.

\section{Quantum propagator}\label{section:magprop}

The quantum propagator is an operator describing the time
evolution of a quantum state. Although the vacuum propagator of
the magnons is not needed for future calculations in this thesis,
its calculation offers a simple application of spin path
integration in our easy-plane ferromagnet. With only slight
modifications to this derivation, that is with the addition of a
perturbing term, or forcing term, we obtain the quantum propagator
for magnons in the presence of a vortex. We must save this
calculation for later after we've derived the appropriate forcing
term.

Suppose initially we know the state of the system of magnons which
can be represented in the $\varphi$ basis. To find the state of
the system at a later time, T,
\\
\begin{equation}
\psi(\varphi,T) = \int d\varphi' K(\varphi,T;\varphi',0)
\psi(\varphi',0)
\end{equation}
where
\\
\begin{equation}
K(\varphi,T;\varphi',0) \equiv \langle \varphi | \exp
-\frac{iHT}{\hbar}| \varphi' \rangle
\end{equation}
\\
is the quantum propagator expressible as a path integral (see
Appendix \ref{section:SPI})
\\
\begin{equation}
K(\varphi,T; \varphi',0) = \int_{\varphi'}^{\varphi} \mathcal
D[\varphi(\mathbf r,t),\vartheta(\mathbf r,t)] \exp
\left(\frac{i}{\hbar} \int_0^T dt \mathcal L_m [\varphi,\vartheta]
\right)
\end{equation}
\\
and where $\mathcal S_m = \int_0^T dt \mathcal L_m$ is the action
with the Lagrangian $\mathcal L_m$ given in (\ref{linLagrangian}).

Before proceeding with the semiclassical approximation---here
exact since we have no terms of higher order than
quadratic---first Fourier transform to diagonalize the problem in
$k$-space. Introducing the Fourier pairs of $\varphi$ and
$\vartheta$, (\ref{fouriertransform}), the Lagrangian becomes
\\
\begin{equation}
\mathcal L_m = S \int \frac{a^2 d^2k}{(2\pi)^2}  \left( \dot
\varphi_{\mathbf k} \vartheta_{-\mathbf k} - \frac{c}{2} \left(
k^2 \varphi_{\mathbf k} \varphi_{-\mathbf k} + Q^2
\vartheta_{\mathbf k} \vartheta_{-\mathbf k} \right) \right)
\end{equation}
\\
The path integration measure is now the product of these Fourier
coefficients
\\
\begin{equation*}
\mathcal D [\varphi(\mathbf r,t),\vartheta(\mathbf r,t)]
\rightarrow \prod_{\mathbf k} d\varphi_{\mathbf k}(t) d
\theta_{\mathbf k}(t)
\end{equation*}

Now to find the classical contribution, the equations of motion
arising from this Lagrangian are
\\
\begin{equation}
\left(%
\begin{array}{cc}
  ck^2 & \frac{\partial}{\partial t} \\
  -\frac{\partial}{\partial t} & cQ^2 \\
\end{array}%
\right)
\left(%
\begin{array}{c}
  \varphi_{\mathbf k}^{cl} \\
  \vartheta_{\mathbf k}^{cl} \\
\end{array}%
\right) = 0
\end{equation}
\\
The general solution with boundary conditions $\varphi_{\mathbf
k}(0) = \varphi_{\mathbf k}'$ and $\varphi_{\mathbf k}(T) =
\varphi_{\mathbf k}$ is
\\
\begin{equation}
\left(%
\begin{array}{c}
  \varphi_{\mathbf k}^{cl} \\
  \vartheta_{\mathbf k}^{cl} \\
\end{array}%
\right) = \frac{\varphi_{\mathbf k}}{\sin \omega_k T}
\left(%
\begin{array}{c}
  \sin\omega_k (T-t) \\
  \frac{k}{Q} \cos\omega_k t \\
\end{array}%
\right) + \frac{\varphi_{\mathbf k}'}{\sin \omega_k T}
\left(%
\begin{array}{c}
  \sin\omega_k (T-t) \\
  -\frac{k}{Q} \cos\omega_k (T-t) \\
\end{array}%
\right)
\end{equation}
where $\omega_{\mathbf k} = ckQ$ as before.

Substituting the classical solution back into the action, after
some simplification, yields the classical contribution to the
action
\\
\begin{equation}\label{action_mcl}
\mathcal S_m^{cl} = \int \frac{a^2 d^2k}{(2\pi)^2}
\frac{Sk}{2Q\sin\omega_kT} \left( \left(\varphi_{\mathbf
k}'\varphi_{-\mathbf k}' + \varphi_{\mathbf k} \varphi_{-\mathbf
k} \right) \cos\omega_kT - 2  \varphi_{\mathbf k}
\varphi_{-\mathbf k}' \right)
\end{equation}

To evaluate the effect of quantum fluctuations, we solve the
relevant Jacobi equation (adapted for a spin path integral as
described in Appendix \ref{section:SPI})
\begin{equation}
\left(%
\begin{array}{cc}
  ck^2 & \frac{\partial}{\partial t} \\
  -\frac{\partial}{\partial t} & cQ^2 \\
\end{array}%
\right)
\left(%
\begin{array}{c}
  \varphi(t) \\
  \vartheta(t) \\
\end{array}%
\right) = 0
\end{equation}
with initial conditions $\varphi(0)=0$ and $S\vartheta(t)=1$. The
determinant of the fluctuations is then given by $ix(T) =
iSQ/k\sin\omega_kT$ for each $k$. Combined with the prefactors in
the path integration measure $S/\hbar$, we find that the Gaussian
integral over fluctuations yields the prefactor
\begin{equation}\label{prefac}
\sqrt{\frac{S k}{2 \pi i \hbar Q \sin \omega_kT}}
\end{equation}

Assembling the various pieces, the propagator of the unperturbed
magnons is
\\
\begin{align}\label{propm}
K(\varphi,T;\varphi',0) = \prod_{\mathbf k}& \sqrt{\frac{S k}{2
\pi i \hbar Q \sin \omega_kT}} \exp \Bigg( \int \frac{a^2 d^2k}{(2\pi)^2}
\frac{iSk}{2\hbar Q\sin\omega_kT} \Bigg. \nonumber \\
&\Bigg. \left( \left(\varphi_{\mathbf k}'\varphi_{-\mathbf k}' +
\varphi_{\mathbf k} \varphi_{-\mathbf k} \right) \cos\omega_kT - 2
\varphi_{\mathbf k} \varphi_{-\mathbf k}' \right) \Bigg)
\end{align}
\\
where $\varphi_{\mathbf k}$ and $\varphi_{\mathbf k}'$ are the
Fourier components of the boundary functions $\varphi(\mathbf r)$
and $\varphi'(\mathbf r)$.

\subsection{Spectrum via tracing over the propagator}
By manipulating the propagator, we can recover the magnon
spectrum. To explain, consider first the propagator of a single
particle starting from position $q_0$ at time $0$ and going to
position $q_T$ at time $T$
\\
\begin{equation}
K(q_T,T;q_0,0) = \langle q_T | \exp -\frac{iHT}{\hbar} | q_0
\rangle
\end{equation}
\\
Taking the trace of this operator, i.e. set $q_T = q_0$ and
integrate over the endpoint $q_0$ of the periodic orbit, we find
\\
\begin{align}\label{part_traceprop}
G(T) =& \int_{-\infty}^{\infty} dq_0 \langle q_0 | \exp
-\frac{iHT}{\hbar} | q_0 \rangle \nonumber \\
=& \int_{-\infty}^{\infty} dq_0 \sum_n \langle q_0 | \xi_n
\rangle \exp -\frac{i E_nT}{\hbar} \langle \xi_n | q_0 \rangle \nonumber \\
=& \sum_n \exp -\frac{i E_nT}{\hbar}
\end{align}
\\
where $\{\xi_n\}$ denote a complete orthonormal set of eigenstates
of $H$. Using the normalization condition of these $\xi_n$ then
yields the excitation spectrum of the Hamiltonian.

The trace of the propagator (\ref{propm}) thus provides another
means to find the excitation spectrum. Set $\varphi_{\mathbf k} =
\varphi_{\mathbf k}'$ and integrate over each Fourier coefficient
\\
\begin{align*}
G_m(T) = \prod_{\mathbf k} \int d\varphi_k & \sqrt{\frac{Sk}{2 \pi
i \hbar Q \sin \omega_kT}} \exp \Bigg( \frac{iSk}{\hbar
Q\sin\omega_kT}  \varphi_{\mathbf k}\varphi_{-\mathbf k}
\left(\cos\omega_kT - 1\right) \Bigg)
\end{align*}
\\
For ease of notation, assume that the product function applies to
everything to the right of it, notably implying an integration
over $\mathbf k$ within the exponential.

Performing the Gaussian integrals
\\
\begin{align*}
G_m(T) = \prod_{\mathbf k}  & \sqrt{\frac{Sk}{2 \pi i \hbar Q \sin
\omega_kT}} \sqrt{\frac{\pi \hbar Q \sin \omega_k T}{-iSk
\left(\cos\omega_kT - 1\right)}}
\end{align*}
\\
and noting that $\cos\omega_kT - 1 = -2 \sin^2
\left(\frac{\omega_kT}{2}\right)$, this reduces to
\\
\begin{align}\label{traceprop}
G_m(T) =& \prod_{\mathbf k}  \frac{1}{2i\sin
\left(\frac{\omega_kT}{2}\right)} \nonumber \\
=&\prod_{\mathbf k} e^{-i\omega_kT/2} \frac{1}{1-e^{-i\omega_kT}} \nonumber \\
=&\prod_{\mathbf k} \sum_{n=0}^{\infty}
e^{-i(n+\frac{1}{2})\omega_kT} \nonumber \\
=&\sum_{\{n_{\mathbf k}\}} e^{-i \sum_{\mathbf k}
(n+\frac{1}{2})\omega_kT}
\end{align}
where $\{n_{\mathbf k}\}$ denotes a set of integers $n_{\mathbf
k}$. Thus, comparing with equation (\ref{part_traceprop}), we find
the excitation spectrum $\sum_{\mathbf k} \hbar \omega_k
(n+\frac{1}{2})$, as expected.

This method of recovering the excitation spectrum is of course
only useful in the special case that $G$ can be cast into this
final form. Nonetheless, by taking the limit $T\to0$, the ground
state term dominates the summation so that we can always at least
find the ground state energy.

\section{Thermal equilibrium density
matrix}\label{section:thermaleq}

\subsection{Magnon density matrix}

A quantum state is represented by a wavefunction $\psi(\mathbf r,
t)$. Generally, this state is a superposition of the system energy
eigenstates, $\{\xi_i\}$. For example
\begin{equation}\label{psi}
\psi(\mathbf r, t)=\sum_i c_i \xi_i
\end{equation}
The probability of finding the $i$th eigenstate upon measurement
is $c_i^2$ and, by conservation of probability, $\sum c_i^2=1$.
This is a pure quantum state. Alternatively, a system may be a
statistical mixture of eigenstates. In that case, the quantum
state isn't expressible as in (\ref{psi}), but, rather, is
described by a set of probabilities $p_i$ of finding the system in
eigenstate $\xi_i$ upon measurement.

We may have a pure quantum state describing the entire interacting
system, which to some extent is the entire universe. Of course, we
may then only be interested in a small subsystem within the whole.
We wish to describe its quantum state only in terms of the
subsystem coordinates.

The density matrix is a notation for describing a quantum state,
necessitated by statistical mixtures such as a thermal equilibrium
state, or entangled states of two sub-systems for which each
individual system must be described by a density matrix even
though the complete system may be in a pure state. As the name
implies, we express the quantum state by a matrix describing the
density of the subsystem or mixture in terms of its eigenstates or
coordinates. More specifically, for a pure state, the density
matrix is
\begin{equation}\label{purestate}
\rho_{ij} = c_i c_j
\end{equation}
where the $c_i$ are the coefficients in (\ref{psi}). For a
mixture,
\begin{equation}
\rho_{ij} = \delta_{ij} p_i
\end{equation}
where $p_i$ are again the probabilities of finding the system is
state $\xi_i$.

Supposing we have a pure quantum state, we can write the density
matrix in the coordinate basis
\begin{align*}
\rho(\mathbf x,\mathbf x') =& \sum_{ij} \xi_i(\mathbf x) \rho_{ij} \xi_j^*(\mathbf x') \\
=& \psi(\mathbf x) \psi^*(\mathbf x')
\end{align*}
The vector $\mathbf x$ is broken into the coordinates of interest
$\tilde{\mathbf x}$ and remaining coordinates $\mathbf q$ such
that $\mathbf x = (\tilde{\mathbf x}, \mathbf q)$. The
\emph{reduced} density matrix for the subsystem of interest is
found by tracing out the uninteresting degrees of freedom
\begin{equation*}
\tilde \rho(\tilde{\mathbf x},\tilde{\mathbf x}')=\int d\mathbf q
\rho(\tilde{\mathbf x},\mathbf q;\tilde{\mathbf x}',\mathbf q)
\end{equation*}
This effectively averages the effects of the external system. For
a pure state, $\textrm{tr}\rho^2=1$. It can be shown that
$\textrm{tr}\rho^2$ is maximal when the ensemble is pure; for a
mixed ensemble $\textrm{tr}\rho^2$ is a positive number less than
one.

A quantum system in thermal equilibrium has its eigenstates
populated with probabilities given by the Boltzmann weighting
factor $e^{-\beta E_i}$. The thermal density matrix in the
coordinate basis is
\begin{equation}
\rho(\mathbf x,\mathbf x') = \sum_i \xi_i(\mathbf
x)\xi_i^*(\mathbf x') e^{-\beta E_i}
\end{equation}
This should be normalized by the partition function $\sum
e^{-\beta E_i}$; however, we will omit it for ease of notation.

But this form is extremely similar to the quantum propagator when
also expressed in this basis
\begin{align*}
K(\mathbf x,T;\mathbf x',0) =& \sum_i \xi_i(\mathbf
x)\xi_i^*(\mathbf x') e^{-i/\hbar H E_i} \\
=& \langle \mathbf x|e^{-i/\hbar H E_i}|\mathbf x' \rangle
\end{align*}
Under the substitution $T\to-i\hbar\beta$, in fact, we recover the
thermal density matrix, though unnormalized. See Appendix
\ref{section:imagtime} for formal details of this substitution.

Noting this imaginary time correspondence between the quantum
propagator and the density matrix in thermal equilibrium, we make
the substitution $T\to-i\hbar\beta$ in the quantum propagator
\\
\begin{align*}
K(\varphi,T;\varphi',0) = \prod_{\mathbf k}& \sqrt{\frac{S k}{2
\pi i \hbar Q \sin \omega_kT}} \exp \Bigg( \int \frac{a^2
d^2k}{(2\pi)^2} \frac{iSk}{2Q\sin\omega_kT} \Bigg. \\
&\Bigg. \left( \left(\varphi_{\mathbf k}'\varphi_{-\mathbf k}' +
\varphi_{\mathbf k} \varphi_{-\mathbf k} \right) \cos\omega_kT - 2
\varphi_{\mathbf k} \varphi_{-\mathbf k}' \right) \Bigg)
\end{align*}
to obtain the thermal equilibrium density matrix
\\
\begin{align}\label{thermaleq}
\rho(\varphi,\varphi') = \prod_{\mathbf k}& \sqrt{\frac{S k}{2 \pi
\hbar Q \sinh \hbar\omega_k\beta}} \exp \Bigg( -\int \frac{a^2
d^2k}{(2\pi)^2} \frac{Sk}{2Q\sinh \hbar \omega_k\beta} \Bigg. \nonumber \\
&\Bigg. \left( \left(\varphi_{\mathbf k}'\varphi_{-\mathbf k}' +
\varphi_{\mathbf k} \varphi_{-\mathbf k} \right) \cosh \hbar
\omega_k\beta - 2 \varphi_{\mathbf k} \varphi_{-\mathbf k}'
\right) \Bigg)
\end{align}
This corresponds to the magnons being excited such that a state
with energy $E_k$ is measured with probability weighting given by
the Boltzmann factor, $e^{-i\beta E_k}$.

\section{Summary}

In summary, the easy-plane magnons perturbing the vacuum ground
state have the spectrum
\\
\begin{equation*}
    \omega(k) = ckQ
\end{equation*}
\\
where $Q=\sqrt{k^2+\frac{1}{r_v^2}}$, $c=SJa^2$ and $r_v^2 =
\frac{J}{2K}$.

We calculated the real time propagator of these magnons and
consequently, making use of the imaginary time path integral of
the density matrix, also the thermal equilibrium density matrix.
The propagator is extremely similar to that of a simple harmonic
oscillator. In fact, under the substitution $\frac{Sk}{Q} \to
m\omega$ the magnon propagator becomes identical to that of the
simple harmonic oscillator.

Next, we examine the vortex excitations.

\chapter{Vortices}\label{chapter:vortex}

The easy plane ferromagnet admits two families of elementary
excitations. In the last chapter, we reviewed the perturbative
excitations, the magnons. Now, we review the other elementary
excitations, the non-perturbative vortices.

Although the out-of-plane spin behaviour cannot be described
analytically, we present the asymptotic behaviour which is
sufficient for getting leading order results. By superposing many
vortex solutions, we expand the action to reveal an inter-vortex
Coulomb-like force. The analogy is complete with the
correspondence of $4\pi\epsilon_0q_i$ with electronic charge in
Coulombs.

The dynamic term ``$p\dot q$" in the action is re-expressed
describing a gyrotropic force (analogous to the Lorentz force) or,
alternatively, as an effective dynamic term in terms of vortex
coordinates, $\mathbf P \cdot \dot{\mathbf X}$, where the momentum
term is a vector potential. This is analogous to a charge in a
magnetic field for which the momentum is modified by the magnetic
field vector potential. In this formalism, the corresponding
vector potential describes an effective perpendicular magnetic
field $\mathbf B = \frac{S^2J}{4\epsilon_0 r_v}p_i \hat z$.

We briefly present different possible two-vortex motions:
depending on the relative sign of $p_iq_i$, the pair execute
parallel motion (for opposite signs) or co-orbital motion (for
like signs). This basic motion is perturbed by introducing an
inertial mass term. Finally, the zero point energy shift of the
two magnon coupling is examined in a Born approximation. This
approximation is found to be sufficient for the continuum of
magnons; however, there exist translation modes localized to the
vortex core. These will be reconsidered in the next chapter using
collective coordinates.

The system has two symmetries: a continuous in-plane symmetry and
a discrete up-down symmetry. The vortices are thus characterized
by two topological indices, the vorticity $q=\pm1,\pm2,\ldots$,
sometimes also called the winding number, and the polarization
$p=0,\pm1$.

The $p=0$ vortices are often separately considered, termed the
in-plane vortices, while the $p\neq0$ solutions are called the
out-of-plane vortices. This separation, however, is unnecessary:
allowing $p=0,\pm1$ in the following treatment recovers the proper
results for both types of solutions.

Being non-perturbative solutions, the vortices satisfy the full,
non-linear, equations of motion of the easy plane ferromagnet.
Derived from the system Lagrangian
\\
\begin{equation}
\mathcal L = S \int \frac{d^{2} r}{a^2} \left( -\dot \phi
\cos\theta- \frac{c}{2} \left( (\mathbf{\nabla}\theta)^{2} +
\sin^{2}\theta \left( (\mathbf{\nabla}\phi)^2 - \frac{1}{r_v^2}
\right)\right)\right)
\end{equation}
\\
the equations of motion are
\\
\begin{align}\label{eom}
\frac{1}{c}\frac{\partial\phi}{\partial t} =&
-\frac{\nabla^2\theta}{\sin\theta}+\cos\theta(\mathbf{\nabla}\phi)^2
- \frac{1}{r_v^2}\cos\theta \nonumber\\
\frac{1}{c}\frac{\partial\theta}{\partial t} =& \sin\theta
\nabla^2\theta + 2 \cos\theta \mathbf{\nabla}\theta \cdot
\mathbf{\nabla} \phi
\end{align}
where $J$ is the exchange constant and $K$ is the anisotropy
constant, $a$ is the lattice spacing, $c=SJa^2$ and $r_v^2 =
\frac{J}{2K}$.

The in-plane vortex can be described analytically. The spin
configuration of this solution has $\phi_v = q\xi +\delta$ and
$\theta_v=0$. The parameter $q$ is called the \emph{vorticity} of
the vortex, and $\delta$ is a phase that has little importance on
the vortex dynamics\symbolfootnote[2]{Note that this broken
continuous symmetry entails the existence of gapless boson modes:
the magnons.}. We can solve for its energy within our continuum
approximation, requiring both an infrared and ultraviolet cutoff,
\\
\begin{equation}\label{energy}
    E = S^2 \int d^2 r \frac{J}{2}(\mathbf \nabla \phi_v)^2 =
    S^2J \pi q^2 \ln \frac{R_s}{a}
\end{equation}
\\
where $R_s$ is the radial size of the system and $a$ is a lower
cutoff, the lattice spacing, required since the system is actually
discrete (making $r \to 0$ unphysical). Note that this energy is
independent of where the vortex center is within the circular
integration region.

The out-of-plane solution is also characterized by its
polarization; that is, the direction (up/down) that the spins
twist out-of-plane. The spin configuration has the same polar
angle dependence, $\phi_v = q\xi +\delta$, while the out-of-plane
spin angle cannot be solved for analytically. The asymptotic
behaviour is
\\
\begin{equation}\label{thetav}
\cos\theta_v = \left\{%
\begin{array}{ll}
  1-c_1\left(\frac{r}{r_v}\right)^2, & \hbox{$r \rightarrow 0$;} \\
  c_2 \sqrt{\frac{r_v}{r}}\exp(-\frac{r}{r_v}), & \hbox{$r \rightarrow\infty$.} \\
\end{array}%
\right.
\end{equation}
\\
where $c_1$ and $c_2$ are free constants that can be set by
imposing appropriate continuity conditions.

\begin{figure}
\centering
\includegraphics[width=10cm]{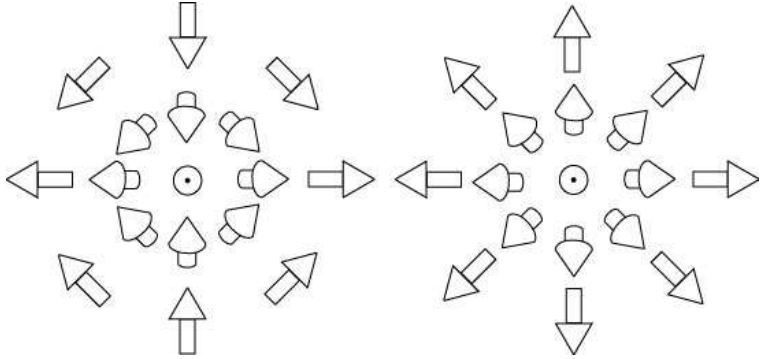}
\caption{Vortex spin configuration: left, a vortex with $q=-1$;
right, a vortex with $q=1$.}\label{figure:vortices}
\end{figure}

Figure \ref{figure:vortices} shows the spin configuration of two
simple out-of-plane vortices. This solution has the same leading
order energy as the in-plane solution
\\
\begin{equation}\label{vortexenergy}
E_v = S^2J\pi q^2 \ln \frac{R_S}{a}
\end{equation}
\\
Core corrections to the energy are finite and hence negligible in
comparison to this log divergent contribution. In fact, in most
that follows, the core will be ignored since it usually offers a
finite contribution next to a log divergent one. A notable
exception is the gyrotropic force that depends on the core
behaviour via the core polarization. This is a differentiating
feature of magnetic vortex dynamics from that of classical fluid
or superfluid vortices where the analogous Magnus force depends
only on the vortex circulation, the fluid analogue to the magnetic
vorticity.

The motion of the in-plane vortex undergoes many of the same
corrections. In fact, with the substitution $p\to0$ the treatment
here reduces to that of an in-plane vortex. The gyrotropic force
disappears, however, all other forces and correction are
polarization independent.

\section{Force between vortices}\label{section:intervortex}

\begin{figure}
\centering
\includegraphics[width=10cm]{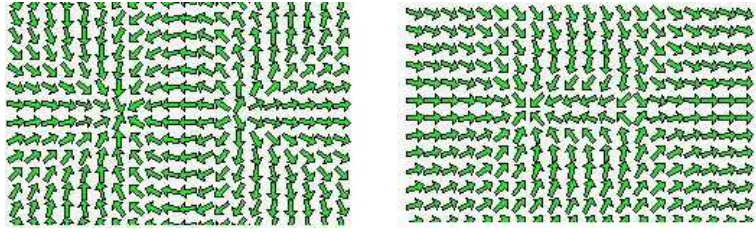}
\caption{Two vortex spin configurations. Left, two vortices with
$q=1$; right, vortices with $q=1$ and $q=-1$; both with no
relative phase shift.}\label{figure:2vortex}
\end{figure}

Consider two vortices of vorticity $q_1$ and $q_2$ and
polarization $p_1$ and $p_2$ well separated so that the only
distortion in their profiles can be assumed to lie in the region
between the two where their profiles are entirely in the plane.
The spins in this middle ground are aligned in the plane with
angle $\phi_{12}$ determined by the sum of spin angles (see Figure
\ref{figure:2vortex}) given by each vortex independently
\\
\begin{equation}
    \phi_{12} = q_1 \chi(X_1) + q_2 \chi(X_2)
\end{equation}
\\
The out-of-plane component of the spin can be neglected here since
we've assumed that the vortex cores are widely separated and each
core gives only a small correction.

\begin{figure}
\centering
\includegraphics[width=8cm]{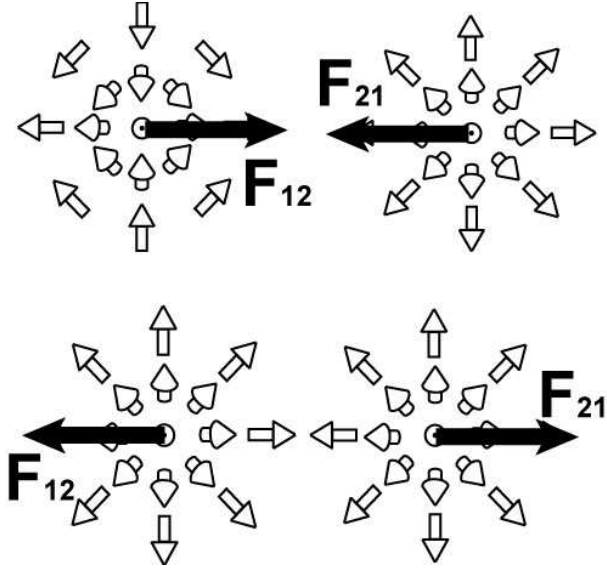}
\caption{Intervortex forces: top, two vortices of opposite
vorticity attract; bottom, two vortices with same sense vorticity
repel.}\label{figure:intervortexforce}
\end{figure}

The energy of the two vortex system is
\\
\begin{equation}
E_{12} = \frac{S^2J}{2} \int d^2r \left( \left( \mathbf{\nabla}
\theta_{12} \right)^2 + \sin^2\theta_v \left( \mathbf{\nabla}
\phi_{12} \right)^2 + \frac{\cos^2 \theta_v}{r_v^2} \right)
\end{equation}
\\
which, except for regions within radius $r_v$ of each vortex core,
is dominated by the $\left( \mathbf{\nabla} \phi_{12} \right)^2$
term. Thus, neglecting core terms, the energy becomes
\\
\begin{align}
E_{12} =& \frac{S^2J}{2} \int d^2r \left( \mathbf{\nabla}\phi_{12} \right)^2 \nonumber \\
=& \frac{S^2J}{2} \int d^2r \left( \frac{q_1 \hat{\mathbf
\phi}_1}{X_1} + \frac{q_2 \hat{\mathbf \phi}_2}{X_2} \right)^2
\end{align}
\\
As an illuminating trick to evaluating this integral, note that
\\
\begin{equation*}
\left( \frac{q_1 \hat{\mathbf \phi}_1}{X_1} + \frac{q_2
\hat{\mathbf \phi}_2}{X_2} \right)^2 = \left( \frac{q_1
\hat{\mathbf X}_1}{X_1} + \frac{q_2 \hat{\mathbf X}_2}{X_2}
\right)^2
\end{equation*}
\\
But $\mathcal E = \frac{q_1 \hat{\mathbf X}_1}{X_1} + \frac{q_2
\hat{\mathbf X}_2}{X_2}$ is just the electric field generated by a
pair of point charges, $4\pi\epsilon_0 q_1$ at $\mathbf X_1$ and
$4\pi\epsilon_0 q_2$ at $\mathbf X_2$, in two-dimensional
electrostatics using SI units. The electrostatic energy, including
the divergent self energies, of this configuration is exactly
\\
\begin{equation}
W = \frac{q_1^2}{2\pi\epsilon_0}\ln\frac{R}{r_v} +
\frac{q_2^2}{2\pi\epsilon_0}\ln\frac{R}{r_v} +
\frac{q_1^2}{\pi\epsilon_0}\ln\frac{X_{12}}{r_v}
\end{equation}
\\
where $\mathbf X_{12}$ is the vector from vortex 1 to vortex 2.
Alternatively \cite{jackson}, we can express the electrostatic
energy as the integral of $\frac{\epsilon_0}{2}\mathcal E^2$.
Thus, upon comparison, the energy of the two vortex system is
\\
\begin{equation}
E_{12} = S^2J\pi \left(q_1^2 \ln\frac{R_S}{r_v} + q_2^2
\ln\frac{R_S}{r_v} + 2 q_1 q_2 \ln\frac{X_{12}}{r_v} \right)
\end{equation}

Similarly, for a collection of $n$ vortices, with cores widely
separated, the spin field pattern is
\\
\begin{align}
\phi_{tot} =& \sum_{i=1}^n q_i \chi(\mathbf X_i) \nonumber \\
\theta_{tot} =& \sum_{i=1}^n \theta_v(\mathbf r - \mathbf X_i)
\approx 0
\end{align}
\\
Following the same analogy to electrostatics as before, we find
the energy of the collection of vortices is now
\\
\begin{equation}
E_{tot} = S^2 J\pi\sum_{i=1}^{n}  q_i^2 \ln\frac{R_S}{r_v} +
2S^2J\pi \sum_{i\neq j}  q_iq_j \ln \frac{X_{ij}}{r_v}
\end{equation}

The force $\mathbf F_{ij}$ acting on vortex $j$ due to vortex $i$,
separated by distance $X_{ij}$
\\
\begin{align}
    \mathbf F_{ij} =& -\mathbf \nabla_{X_{ij}} E_{tot} \nonumber \\
    =& \frac{S^2J 2\pi q_i q_j}{X_{ij}} \hat{\mathbf X}_{ij}
\end{align}
\\
where $\hat{\mathbf X}_{ij}$ is a unit vector pointing from the
center of vortex $i$ to the center of vortex $j$. Thus, if the two
vortices have the same sense, or the same sign vorticities $q_i$
and $q_j$, the force is repulsive, and conversely, for opposite
senses the force is attractive. Note, since in this approximation
there is no interaction between the two vortex cores, the
direction of the spins out of the plane at the cores---the
polarization---is irrelevant.

\section{The gyrotropic force and the vortex momentum}

\subsection{The gyrotropic force}
\begin{figure}
\centering
\includegraphics[width=5cm]{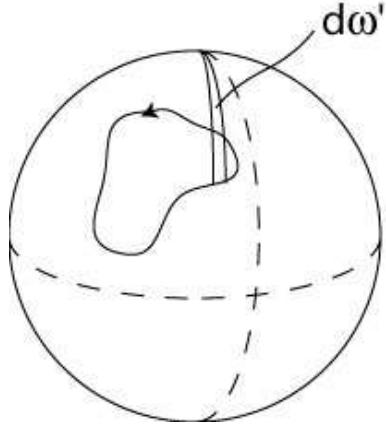}
\caption{The spin path mapped onto the unit sphere. The area
traced out by its motion gives the Berry's
phase.}\label{figure:berry}
\end{figure}

The vortex is a stationary solution of the system. If we assume
that it now moves at a small velocity $\dot{\mathbf X}$, for the
moment with no deformation to the vortex profile, the $p\dot q$
action term, called the Berry's phase in a spin system, is no
longer vanishing. The Berry's phase, $\omega_B$, is a phase
accumulated by the changing spin field
\begin{equation}
\omega_B = \int dt \int \frac{d^2r}{a^2} S\cos\theta\dot\phi
\end{equation}
Considering a single spin, we can interpret this phase
geometrically as the solid angle swept out by the motion mapped
onto the spin sphere. This is clear when we make the change of
variable
\begin{equation*}
\omega_B = S\int d\phi \cos\theta = S\int d\omega'
\end{equation*}
where $d\omega'$ is the area increment on the unit sphere. Refer
to Figure \ref{figure:berry}.

We treat this term as a potential and calculate the corresponding
force acting on the vortex. Let the vortex profile move as a
function of $\mathbf r - \mathbf X(t)$, where $\mathbf X(t)$ is
the center coordinate of the vortex. The Berry's phase term in the
Lagrangian becomes
\begin{equation*}
-S \int \frac{d^2r}{a^2} \dot{\phi_v} \cos\theta_v = S \int
\frac{d^2r}{a^2} \dot{\mathbf X} \cdot \mathbf \nabla \phi_v
\cos\theta_v
\end{equation*}

The gyrotropic force arising from this term is found by varying it
with respect to the center coordinate of the
vortex\cite{sloncz:1984,thiele:1973}, without the usual negative
sign since we take the term from the Lagrangian,
\\
\begin{align}
\mathbf F_{gyro} =& S \partial_{\mathbf X} \int \frac{d^2r}{a^2}
\dot{\mathbf X} \cdot \mathbf \nabla \phi_v
\cos\theta_v \nonumber \\
=& -S \int \frac{d^2r}{a^2} \mathbf \nabla \left(\dot{\mathbf X}
\cdot \mathbf \nabla \phi_v \cos\theta_v \right)
\end{align}
\\
But the integrand is strictly a function of $\mathbf r - \mathbf
X$ so that $\partial_{\mathbf X} \to -\mathbf \nabla$, where
$\mathbf \nabla$ is understood to be with respect to $\mathbf r$.
Note that $\nabla^2 \phi_v = 0$ and thus
\\
\begin{equation*}
\mathbf \nabla \left(\dot{\mathbf X} \cdot \mathbf \nabla \phi_v
\cos\theta_v \right) = \left(\dot{\mathbf X} \cdot \mathbf \nabla
\phi_v\right) \mathbf \nabla \cos\theta_v
\end{equation*}
\\
Using the cross-product relation $\mathbf A \times \left( \mathbf
B \times \mathbf C \right) = \left(\mathbf A \cdot \mathbf C
\right) \mathbf B - \left(\mathbf A \cdot \mathbf B \right)
\mathbf C$, we find
\\
\begin{equation*}
-S\int \frac{d^2r}{a^2} \left( \dot{\mathbf X} \cdot \mathbf
\nabla \phi_v\right) \mathbf \nabla \cos\theta_v = S \int
\frac{d^2r}{a^2} \left( \mathbf \nabla \cos\theta_v \times \mathbf
\nabla \phi_v \right) \times \dot{\mathbf X} - \left( \dot{\mathbf
X}\cdot\mathbf \nabla \cos\theta_v\right) \mathbf \nabla \phi_v
\end{equation*}
\\
where now both terms on the right are integrable. Consider the
first term, noting that
\\
\begin{equation*}
\left(\mathbf \nabla \cos\theta_v \times \mathbf \nabla \phi_v
\right)_z = \frac{\partial \cos\theta_v}{\partial x}
\frac{\partial \phi_v}{\partial y}-\frac{\partial
\cos\theta_v}{\partial y} \frac{\partial \phi_v}{\partial x} =
\frac{\partial(\cos\theta_v,\phi_v)}{\partial(x,y)}
\end{equation*}
\\
Clearly, since $\mathbf \nabla \cos\theta_v$ and $\mathbf \nabla
\phi_v$ both lie entirely in the plane, the $\hat{\mathbf z}$
component is the only non-zero component. The first integral
becomes
\\
\begin{equation*}
S \int \frac{d^2r}{a^2} \mathbf \nabla \cos\theta_v \times \mathbf
\nabla \phi_v = \frac{S}{a^2}\int d\cos\theta_v d\phi_v \;
\hat{\mathbf z} = -\frac{2\pi Spq}{a^2} \hat{\mathbf z}
\end{equation*}
\\
where $p$ is the polarization of the vortex core and $q$ is the
vorticity of the vortex.

For the second integral, consider axes $x_{\|}$ and $x_{\bot}$
parallel and perpendicular to $\dot{\mathbf X}$, where the second
is aligned such that $\hat{\mathbf z} \times \hat{\dot{\mathbf X}}
= \mathbf{x_{\bot}}$. In polar coordinates defined for this frame,
the integral can be written
\\
\begin{equation*}
 - \frac{S}{a^2}\int d^2r \left( \dot{\mathbf X}\cdot\mathbf \nabla
\cos\theta_v\right) \mathbf \nabla \phi_v = -\frac{Sq}{a^2}\int dr
d\chi \dot X \frac{d\cos\theta_v}{dr} \cos \chi (-\sin\chi,
\cos\chi)
\end{equation*}
\\
where we decompose $\hat{\mathbf \chi} = (-\sin\chi,\cos\chi)$
into the $(x_{\|}, x_{\bot})$ basis. Evaluating this gives
\\
\begin{equation}\label{Fgyroterm2}
 - \frac{S}{a^2}\int d^2r \left( \dot{\mathbf X}\cdot\mathbf \nabla
\cos\theta_v\right) \mathbf \nabla \phi_v = \pi p q \dot X
\mathbf{x_{\bot}} = \frac{\pi Spq}{a^2} \hat{\mathbf z} \times
\dot{\mathbf X}
\end{equation}
\\
The gyrotropic force is then
\\
\begin{equation}
\mathbf F_{gyro} = - \frac{\pi Spq}{a^2} \hat{\mathbf z} \times
\dot{\mathbf X}
\end{equation}

\begin{figure}
\centering
\includegraphics[width=8cm]{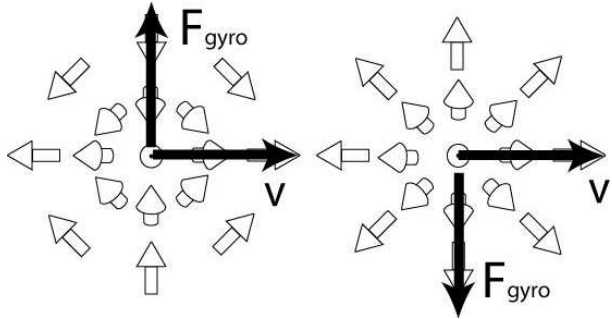}
\caption{The gyrotropic force: left, a vortex with $p=1$ and
$q=-1$ traveling to the right experiences an upward force; right,
a vortex with $p=1$ and $q=1$ traveling to the right experiences a
downward force. Note $\hat{\mathbf z}$ is defined out of the
page.}\label{figure:gyroforce}
\end{figure}

Note, this result differs by a factor of 2 from that of Huber
\cite{huber:1982} using the formalism of Thiele, found for a
magnetic domain wall \cite{thiele:1973}. Thiele's starting point
for the kinetic term was
\begin{equation}
-\cos\theta_v \; \dot\phi_v + \frac{d}{dt}\cos\theta_v \; \phi_v
\end{equation}
which is exactly twice our starting point that includes only the
first of these two terms.

Notice that the gyrotropic force is derivable from the equivalent
Lagrangian term
\\
\begin{equation}\label{gyroLagrange}
\mathcal L_{gyro} = \frac{\pi Sp q}{2a^2} \mathbf X \times
\dot{\mathbf X} \cdot \hat{\mathbf z} =-\frac{\pi Sp q}{2a^2}
\mathbf X \times \hat{\mathbf z} \cdot \dot{\mathbf X}
\end{equation}

\subsection{The vortex momentum}\label{section:vectorpot}
The gyrotropic force can be written in the suggestive form
\begin{equation}
\mathbf F_{gyro} = -\frac{d\mathbf P_{gyro}}{d t}
\end{equation}
where $\mathbf P_{gyro}$ is a momentum term from the equivalent
Lagrangian term (\ref{gyroLagrange}) written in the form $\mathbf
P \cdot \dot{\mathbf X}$
\begin{equation}
\mathbf P_{gyro} = -\frac{\pi Spq}{2a^2} \mathbf X \times
\hat{\mathbf z}
\end{equation}

We now examine a direct evaluation of the vortex momentum as given
in a general field theory by the operator\cite{sakita}
\begin{equation}
\mathbf P=-\int d^2r \tilde\pi (\mathbf{r},t) \nabla \tilde \phi
(\mathbf{r},t)
\end{equation}
\\
where $\tilde\pi$ is the conjugate momentum density to the field
variable $\tilde\phi$ (the tilde's are there to differentiate the
field variable here to the azimuthal angle $\phi$ used
previously). This operator is chosen because it is the
infinitesimal generator of spatial translations, eg.
\\
\begin{align*}
\tilde\phi(\mathbf{r}+\delta\mathbf{r}) = \tilde \phi (\mathbf{r})
+ \delta\tilde\phi(\mathbf{r}) = \tilde\phi(\mathbf{r})+ \nabla
\tilde \phi(\mathbf{r}) \cdot \delta\mathbf{r}\\
\delta\tilde\phi(\mathbf{r})=\{\delta\mathbf{r}\cdot P, \tilde
\phi (\mathbf{r})\}=\nabla\tilde\phi(\mathbf{r}) \cdot \delta
\mathbf{r}
\end{align*}
\\
where we use the Poisson bracket here as defined in equation
\ref{poisson} (note there we used $q$ for $\tilde\phi$ and $p$ for
$\tilde\pi$).

For the magnetic vortex, this gives the momentum expression
\\
\begin{equation}
\mathbf P = \int d^2r \frac{S}{a^2}\cos \theta_v \nabla \phi_v
\end{equation}

Before attempting to evaluate this expression, first note that the
$1/r$ behaviour in $\nabla \phi_v$ is balanced by the $r$ in the
integration measure so that the integrand is nowhere divergent.

If we blindly set the vortex at the origin of the integration
region, the $\hat{\mathbf \chi}_r$ direction of the integrand sums
to 0 by symmetry, there being no other angular dependence. The
integral is non-zero, however, if we displace the vortex by
$\mathbf X$ from the origin.

To evaluate this integral note that
\\
\begin{equation}
\mathbf \nabla \phi_v = -q \hat{\mathbf z} \times \mathbf \nabla
\ln|\mathbf r - \mathbf X|
\end{equation}
\\
Considering the momentum integral one component at a time, first
the $y$ component
\\
\begin{align*}
\int d^2r \cos \theta_v \partial_x \ln|\mathbf r - \mathbf X| =&
-\int dx dy \partial_x \cos \theta_v
\left( \ln r - \frac{\mathbf r \cdot \mathbf X}{r^2} \right) \\
=& \int dr d\chi_r \partial_r \cos \theta_v \cos\chi_r
\hat{\mathbf r} \cdot \mathbf X \\
=& -\pi qp X
\end{align*}
\\
where $X$ is the $x$ component of $\mathbf X$. We expanded the
$\ln$ above and truncated the series to $\mathcal O(1/r)$. This is
in keeping with the $r\to0$ behaviour noted in the original
integral. Of course, for $r\to \infty$ the integrand decays to
zero exponentially as before.

After the analogous treatment for $y$, we find the momentum is
exactly the $\mathbf P_{gyro}$ describing the gyrotropic force
\\
\begin{equation}\label{momvectpotl}
\mathbf P_{gyro} = -\frac{\pi Sqp}{a^2} \mathbf X\times
\hat{\mathbf z}
\end{equation}

What does it mean exactly to have a momentum that is speed
independent and coordinate dependent? Isn't this extremely
bizarre? Recalling the problem of a charged particle in a magnetic
field, the momentum of such a particle is modified by the presence
of the magnetic field according to \cite{LL:fields}
\\
\begin{equation}
\mathbf p \to \mathbf p - \frac{e}{c}\mathbf A
\end{equation}
\\
where $\mathbf A$ is the vector potential describing the magnetic
field $\mathbf B = \nabla \times \mathbf A$, $e$ is the electric
charge and $c$ here is the speed of light.

For the magnetic vortex, this momentum term must also correspond
to a vector potential term. Completing the analogy, using
$4\pi\epsilon_0q$ as charge as in section
\ref{section:intervortex}, replacing the speed of light by the
speed of magnons $SJa^2/r_v$, we find an effective perpendicular
magnetic field $\mathbf B = \frac{S^2J}{4\epsilon_0 r_v}p \hat z$.

To further explore this interpretation, we expect the
gyro-momentum to be gauge dependent. That is, we should be able to
rewrite the vector potential
\begin{equation}
\mathbf A \to \mathbf A + \nabla_r f(r)
\end{equation}
for any continuous function $f(r)$, changing the momentum
expression $\mathbf P_{gyro}$, however, and still describe the
same physical system.

Considering this gauge change in reverse, we use the gauge freedom
of the Berry's phase. The Berry's phase is written in a general
$\hat{\mathbf\Omega}$ basis
\\
\begin{equation}
\omega_B = \int dt d^2r \mathbf A(\hat{\mathbf\Omega})
\dot{\hat{\mathbf\Omega}}
\end{equation}
\\
where $\mathbf A$ is a unit magnetic monopole vector potential. We
change the gauge of this vector potential $\mathbf A$ via
\\
\begin{equation}
\mathbf A \to \mathbf A + \nabla_{\hat{\mathbf \Omega}} f
\end{equation}
\\
where $f$ is a general function of $\hat{\mathbf \Omega}$.

The momentum of the magnetic vortex is altered by noting the
correspondence
\begin{align*}
\omega_B =&\int dt d^2r \mathbf A(\hat{\mathbf\Omega})
\dot{\hat{\mathbf\Omega}}\\
\mathbf P_{j,gyro} = &-\int d^2r A_i \nabla_{r_j}\hat\Omega_i
\end{align*}
The Berry's phase gauge change shifts the momentum definition
according to
\\
\begin{equation}
\mathbf P = \int d^2r \frac{S}{a^2}\cos \theta_v \nabla_{\mathbf
r} \phi_v  + \nabla_{\hat \Omega_i} f \nabla_{\mathbf r}\hat
\Omega_i
\end{equation}
\\
But
\\
\begin{equation*}
\nabla_{\hat \Omega_i} f \nabla_{\mathbf r}\hat \Omega_i =
\nabla_{\mathbf r} f(\hat{\mathbf \Omega}) = -\nabla_{\mathbf X}
f(\hat{\mathbf \Omega})
\end{equation*}
\\
since $\hat{\mathbf \Omega} = \hat{\mathbf \Omega}(\mathbf r
-\mathbf X)$. Thus, the additional term to the vortex momentum
becomes
\\
\begin{equation}
-\int d^2r \nabla_{\hat \Omega_i} f \nabla_{\mathbf r}\hat
\Omega_i = \nabla_{\mathbf X} \int d^2r f = \nabla_{\mathbf X}
F(\mathbf X)
\end{equation}
\\
where $F(\mathbf X) = \int d^2r f$ is now some general function of
$\mathbf X$.

A continuous function $F(\mathbf X)$ can always be expressed as
the integral over another function $f(\mathbf r, \mathbf X)$.
Thus, the gauge freedom in the Berry's phase allows exactly the
necessary gauge freedom in the gyrotropic momentum term,
supporting our vector potential interpretation.

\section{Motion of vortex pairs}
Consider the motion of a pair of vortices, separated enough that
the cores do not significantly interact, with polarization $p_i$
and vorticity $q_i$, $i=1,2$. The motion so far is dictated by the
balance of the inter-vortex and gyrotropic forces acting on each
vortex
\\
\begin{align}
\frac{2\pi S^2Jq_1q_2}{X_{12}^2} (\mathbf X_1 - \mathbf X_2) - \pi
p_1 q_1 \hat{\mathbf z} \times \dot{\mathbf X}_1 =& 0 \nonumber \\
\frac{2\pi S^2Jq_1q_2}{X_{12}^2} (\mathbf X_2 - \mathbf X_1) - \pi
p_2 q_2 \hat{\mathbf z} \times \dot{\mathbf X}_2 =& 0
\end{align}
\\
or taking the cross-product of each equation with $\hat{\mathbf
z}$
\\
\begin{equation*}
\dot{\mathbf X}_1=-\frac{p_1 q_2}{p_2 q_1} \dot{\mathbf X}_2 =
\frac{2S^2Jp_1q_2}{X_{12}^2} \hat{\mathbf z} \times (\mathbf X_2 -
\mathbf X_1)
\end{equation*}
\\
In the case $p_1 q_2=p_2 q_1$, $\dot{\mathbf X}_1=-\dot{\mathbf
X}_2$ and the vortices move on a common circular orbit, keeping
separation $X_{12}$ with angular frequency $\omega = \frac{4 S^2 J
p_1 q_2}{X_{12}^2}$ where $\omega>0$ denotes counter-clockwise
rotation.

For the opposite case, $p_1 q_2=-p_2 q_1$, we have $\dot{\mathbf
X}_1=\dot{\mathbf X}_2$ and the vortex pair move with a common
velocity on parallel lines (upward for $p_2 q_1>0$ and downward
for $p_2 q_1<0$).

\begin{figure}
\centering
\includegraphics[width=2.75cm]{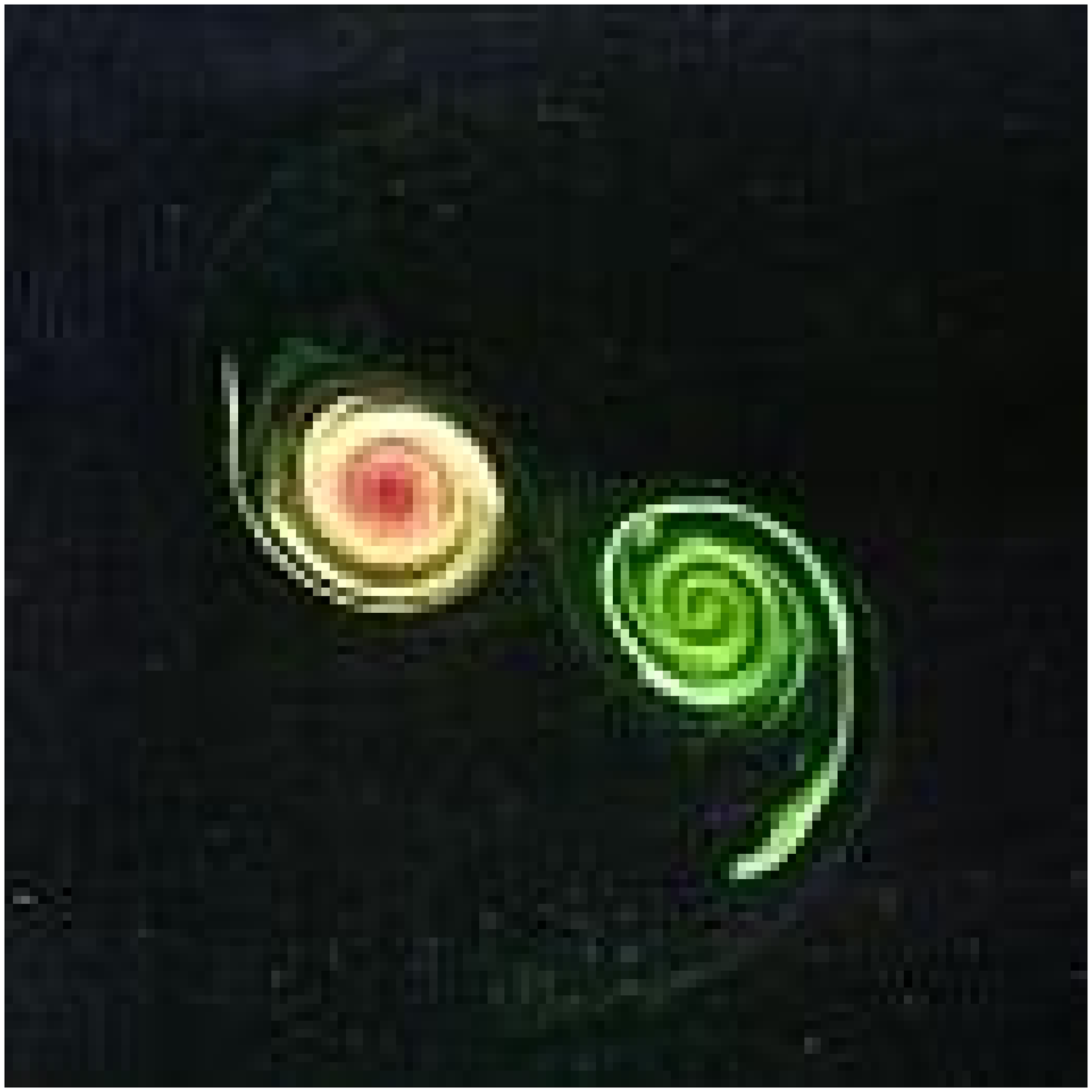}
\includegraphics[width=2.75cm]{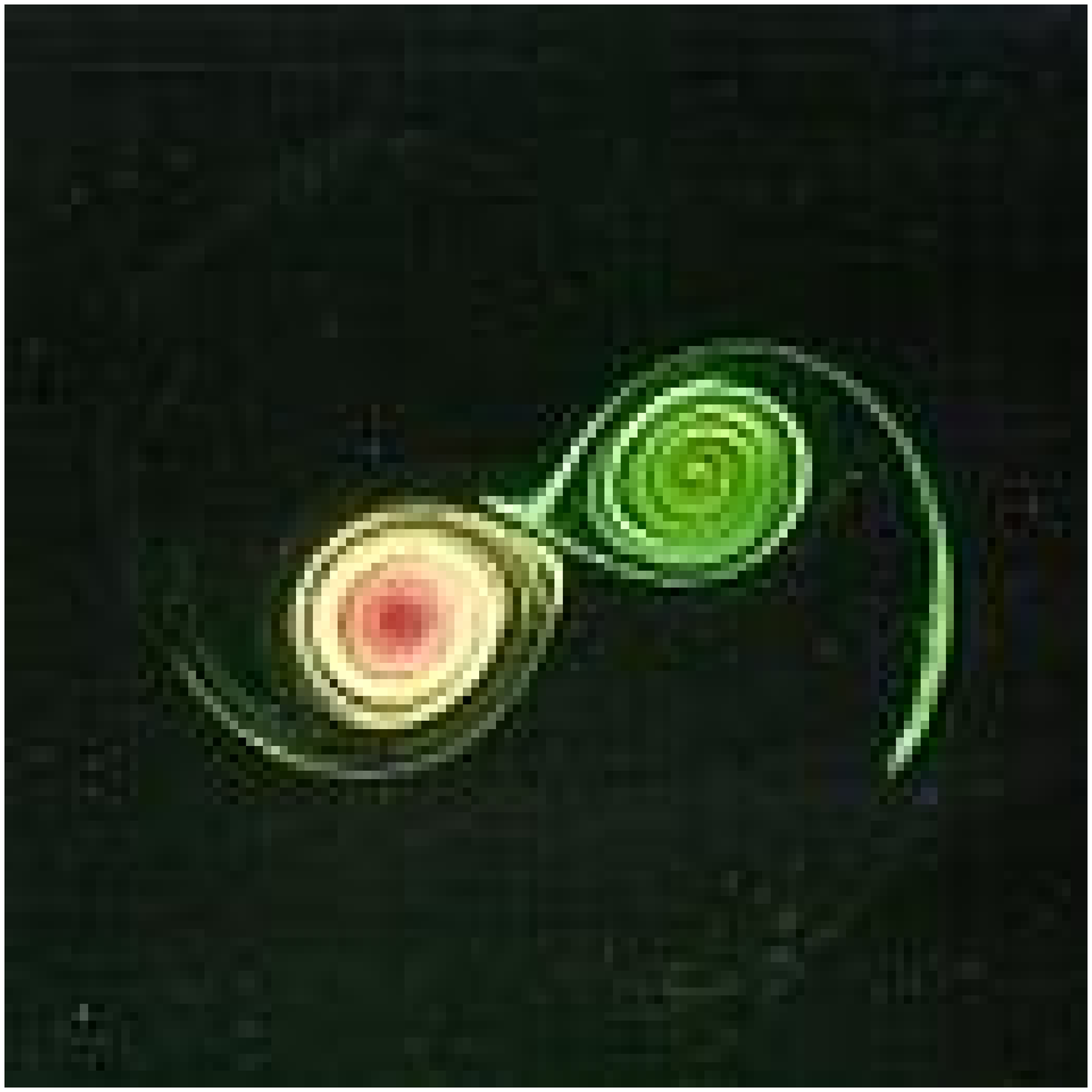}
\includegraphics[width=2.75cm]{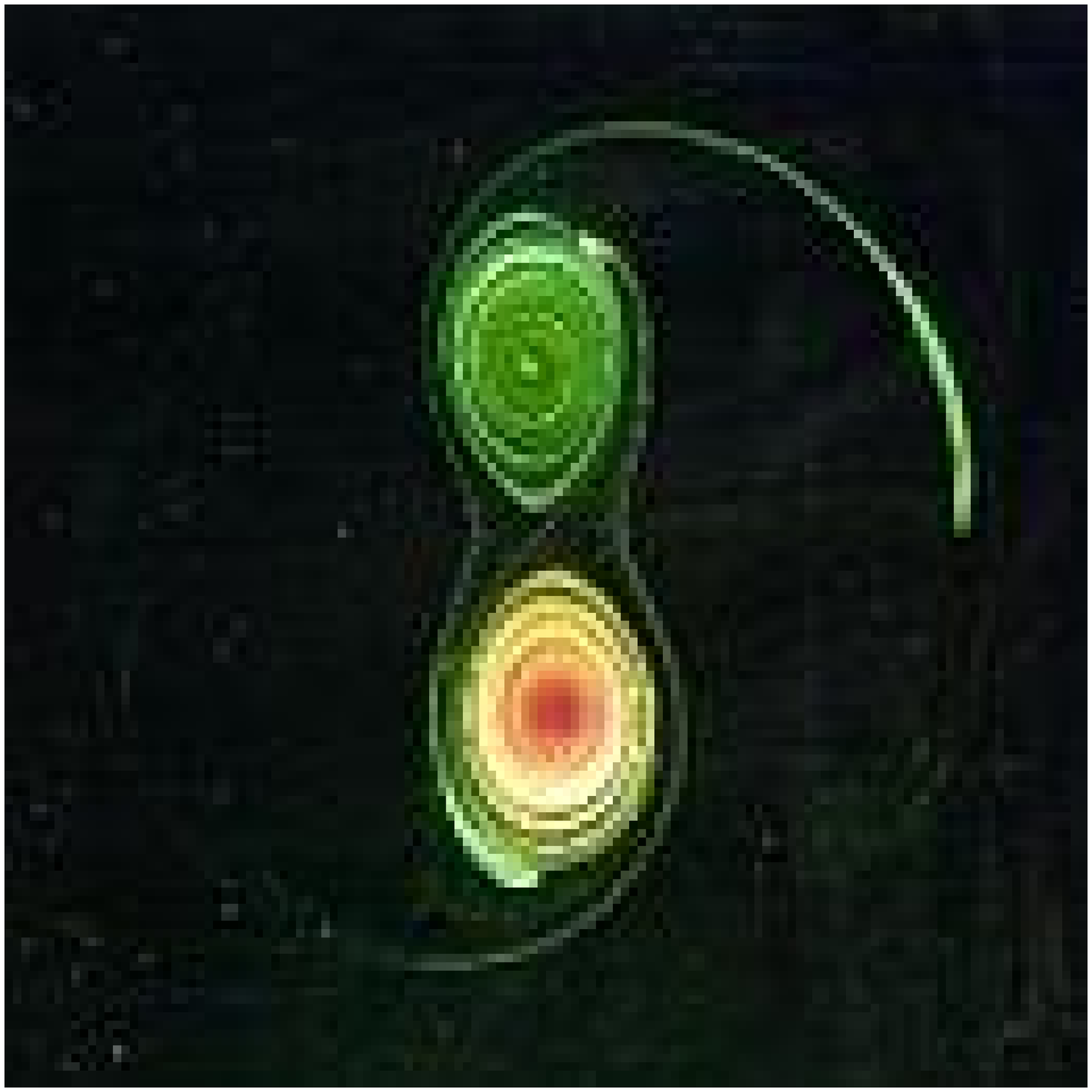}
\includegraphics[width=2.75cm]{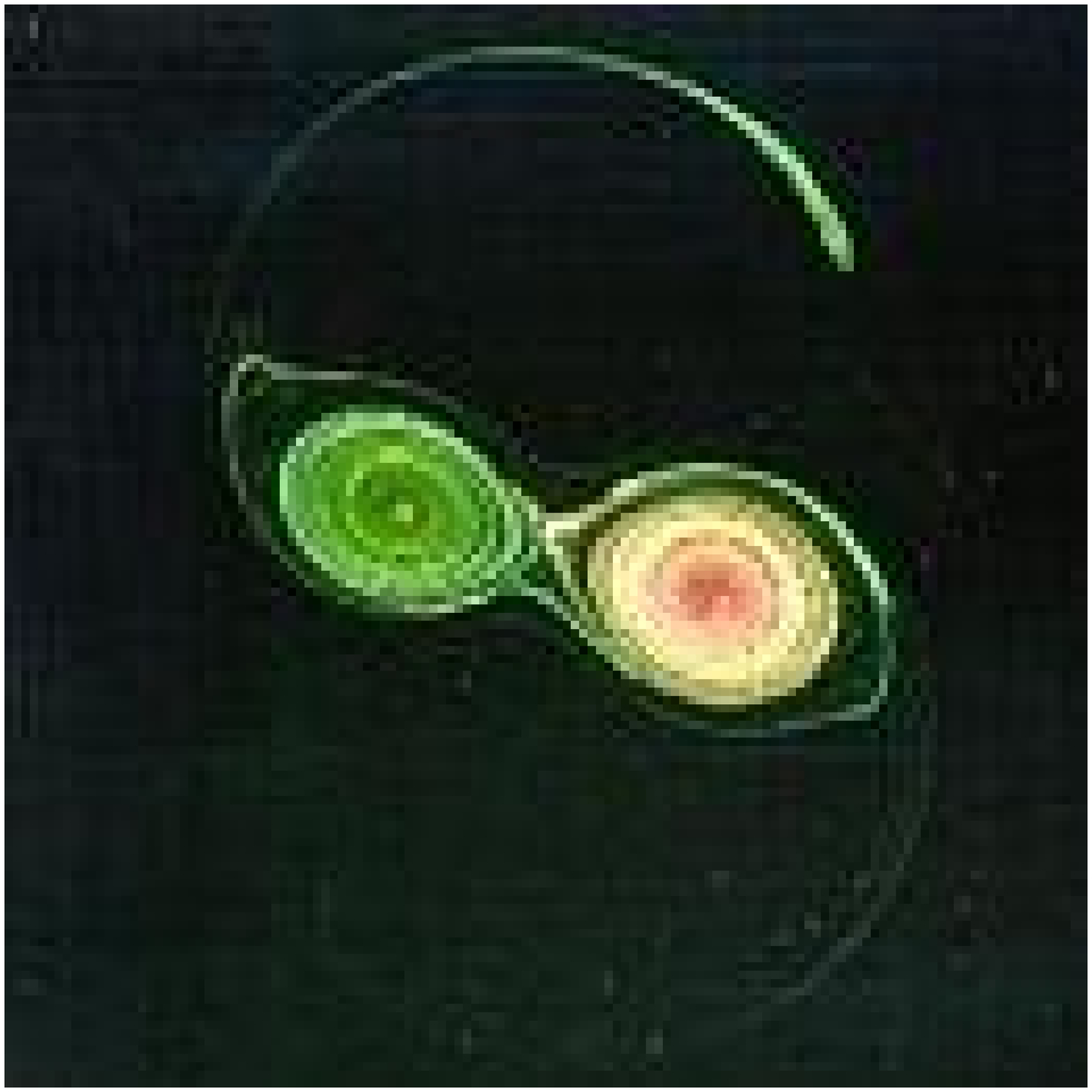}
\caption{Sequenced photographs of a pair of fluid vortices with
same sense vorticity. Photos were taken at 2 second
intervals\cite{fluidpics}.}\label{figure:fluidpairalike}
\end{figure}

\begin{figure}
\centering
\includegraphics[width=2.75cm]{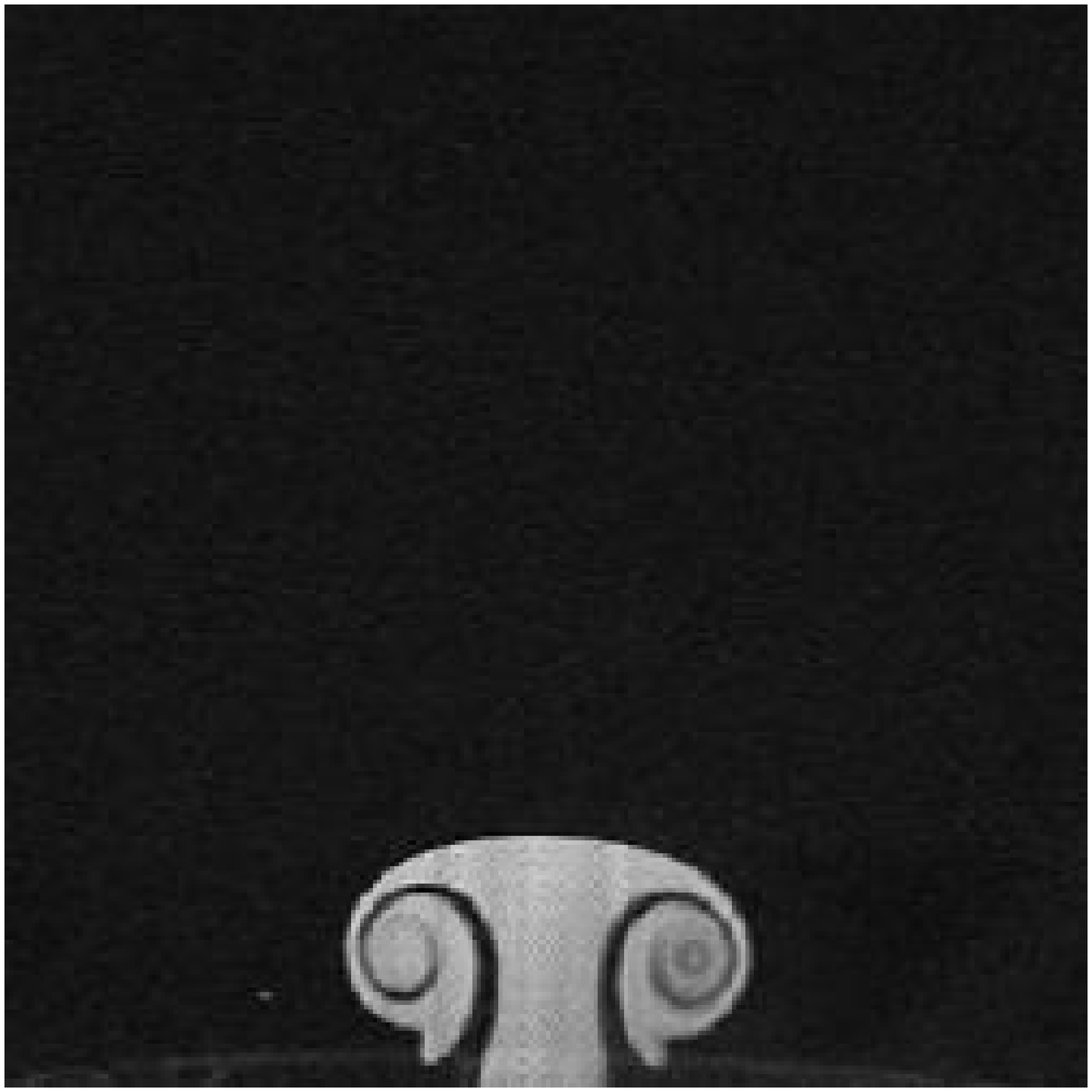}
\includegraphics[width=2.75cm]{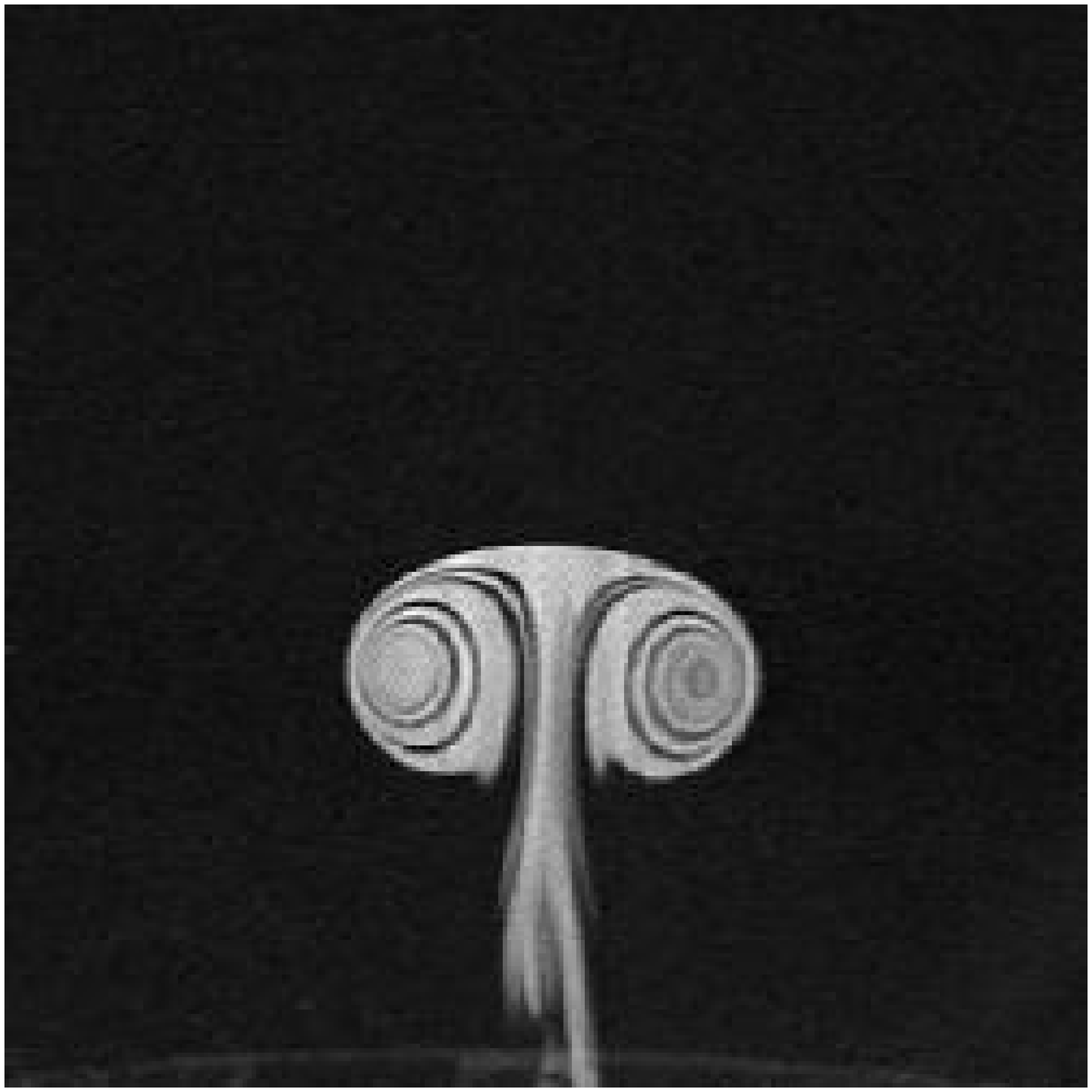}
\includegraphics[width=2.75cm]{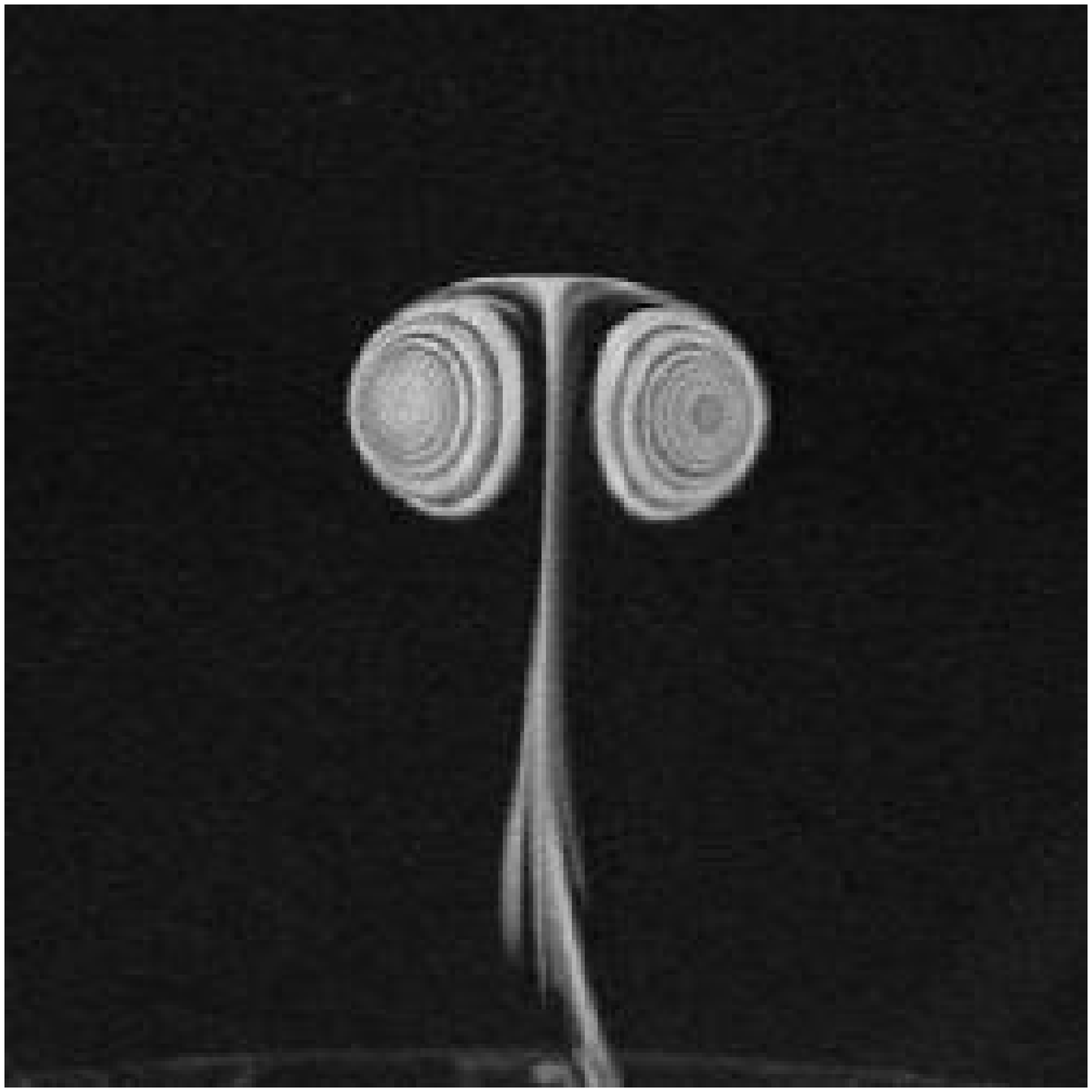}
\includegraphics[width=2.75cm]{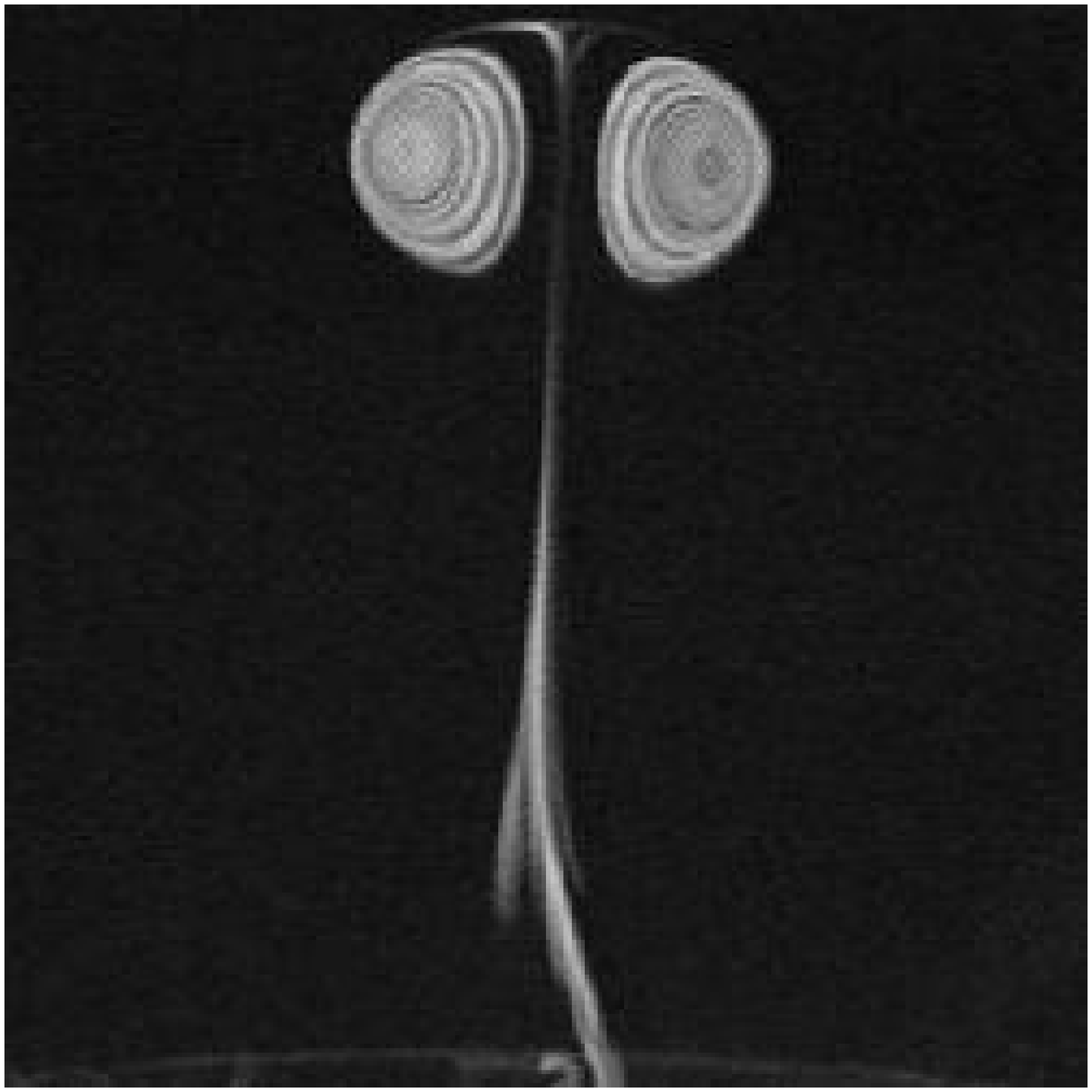}
\caption{Sequenced photographs of a pair of fluid vortices with
opposite sense vorticity. Photos were taken at 4 minute
intervals\cite{fluidpics}.}\label{figure:fluidpairopposite}
\end{figure}

In this approximation, the dynamics of the vortex pair is
identical to the analogous motion of a pair of fluid vortices.
Referring to Figures (\ref{figure:fluidpairalike}) and
(\ref{figure:fluidpairopposite}), a pair of fluid vortices with
the same direction circulation move in a common circular motion
while a pair of opposite circulation move along parallel paths.
There is the notable difference, of course, that here the type of
motion is dictated by the products $pq$ rather than just $q$ as in
regular fluid dynamics.

\section{Vortex mass}\label{section:vortexmass}
Up to now, we've assumed that the vortex profile is rigid when in
motion. In fact, the profile is modified linearly in $\dot X$.

Assuming that the vortex moves at small velocity $\dot X$, expand
about the rigid vortex profile
\\
\begin{align*}
\phi_v =& \phi_v^{(0)}(\mathbf{r}-\mathbf{X}) + \phi_v^{(1)}\\
\theta_v =& \theta_v^{(0)}(\mathbf{r}-\mathbf{X})+ \theta_v^{(1)}
\end{align*}

Substituting this into the equations of motion, (\ref{eom}), to
first order in $\dot X$, making use of the zeroth order equations
of motion, these reduce to
\\
\begin{align}
-\frac{\sin\theta_v^{(0)}}{c}\dot{\mathbf X}\cdot\mathbf{\nabla}
\phi_v^{(0)} = -\nabla^2\theta_v^{(1)} - \cos2\theta_v^{(0)}&
\left(\frac{1}{r_v^2} - \left( \mathbf{\nabla} \phi_v^{(0)}
\right)^2 \right) \theta_v^{(1)} \\
&+\sin2\theta_v^{(0)}
\mathbf{\nabla}\phi_v^{(1)} \cdot \mathbf{\nabla}\phi_v^{(0)} \nonumber\\
-\frac{1}{c}\dot{\mathbf X}\cdot\mathbf{\nabla} \theta_v^{(0)} =
\sin\theta_v^{(0)} \nabla^2\phi_v^{(1)} + 2\cos\theta_v^{(0)}&
\left( \mathbf{\nabla}\theta_v^{(0)} \cdot \mathbf{\nabla}
\phi_v^{(1)} + \mathbf{\nabla}\theta_v^{(1)} \cdot \mathbf{\nabla}
\phi_v^{(0)} \right) \nonumber
\end{align}

Using the asymptotic expansion of the stationary vortex
(\ref{thetav}), the asymptotic forms of the profile perturbation
become, keeping only the dominant terms in the $r\rightarrow0$ and
$r\rightarrow\infty$ limits
\\
\begin{align}\label{profilechange}
\theta_v^{(1)} =& \left\{%
\begin{array}{ll}
    -\frac{qc_1\dot Xr^2}{2c r_v}\cos\chi, & \hbox{$r\to 0$} \\
    -\frac{q\dot Xr_v^2}{c r}\cos\chi, & \hbox{$r\to \infty$} \\
\end{array}%
\right. \nonumber \\
\phi_v^{(1)} =& \left\{%
\begin{array}{ll}
    -\frac{\dot Xr}{2c}\sin\chi, & \hbox{$r\to 0$} \\
    -\frac{\dot Xc_2r_v^{3/2}}{c} \frac{e^{-r/r_v}}{\sqrt{r}}\sin\chi, & \hbox{$r\to \infty$} \\
\end{array}%
\right.
\end{align}
where $c_1$ and $c_2$ are free parameters in the unperturbed
asymptotic form (\ref{thetav}).

Substituting these asymptotic expressions into the energy
integral, we find energy terms that are quadratic to lowest order
in $\dot X$ (the linear terms integrate to zero by symmetry)
interpretable as a $\frac{1}{2}M_v \dot X^2$ kinetic term:
\\
\begin{equation}
    E = E_v^{(0)} + E_{core}^{(1)} +E_{\infty}^{(1)}
\end{equation}
\\
where $E_{core}^{(1)}$ accounts for the $r=0..r_v$ and
$E_{\infty}^{(1)}$ accounts for the remaining $r=r_v..\infty$.
Evaluated,
\\
\begin{align}
E_{core}^{(1)} =& \frac{\pi q^2\dot X^2}{8Ja^4} \left(3c_1^2r_v^2
-
\frac{1}{6}c_1^2r_v^2 - \frac{5}{12}c_1^4r_v^2\right)\nonumber \\
E_{\infty}^{(1)} =& \frac{\pi \dot X^2}{2Ja^4} \left(q^2r_v^2 \ln
\frac{R_S}{r_v} + \frac{r_v^2}{2} + \frac{c_2^2 r_v^2}{2e^2}
\right)
\end{align}

Assuming an energy correction of the form
\\
\begin{equation}
    \Delta E = \frac{1}{2} M_v \dot X^2
\end{equation}
\\
the leading term describing the vortex mass is deduced as
\\
\begin{equation}
    M_v =\frac{\pi q^2 r_v^2}{Ja^4}
    \ln\frac{R_S}{r_v}.
\end{equation}
Note that this mass is, in fact, identical to the mass estimate
suggested by \citet{duan:1992} based on purely dimensional
arguments, $M_v = \frac{E_v}{(c/r_v)^2}$.

\section{Quantization of magnetic vortices}\label{section:zpt}
Quantum fluctuations in a system introduce a zero-point energy. In
the previous chapter, we quantized the magnons finding this
zero-point energy to be $\sum \frac{1}{2}\hbar \omega_k$, summed
over the entire $k$-spectrum. In the presence of a magnetic
vortex, the magnon spectrum is shifted. Since we prefer to have a
consistent definition of the magnons and vortices, the shift in
the zero-point energy of the magnons is associated with the
quantized vortex.

Quantization of a magnetic vortex involves quantizing the small
variations about it and examining how the energy of these modes
shift from the analogous modes in the absence of a
vortex\cite{raj}. See Appendix \ref{section:solitonzpe} for more
details.

Expanding $\theta$ and $\phi$ about a vortex, $\theta = \theta_v +
\vartheta$ and $\phi = \phi_v + \varphi$, in the non-linear
equations of motion (\ref{eom}), yields the linearized equations
in $\vartheta$ and $\varphi$
\\
\begin{align}\label{lineom}
\frac{\sin\theta_v}{c} \frac{\partial \varphi}{\partial t} =&
-\nabla^2\vartheta - \cos2\theta_v \left(\frac{1}{r_v^2} - \left(
\mathbf{\nabla} \phi_v \right)^2 \right) \vartheta +\sin2\theta_v
\mathbf{\nabla}\varphi \cdot \mathbf{\nabla}\phi_v \nonumber\\
\frac{1}{c} \frac{\partial \vartheta}{\partial t} =& \sin\theta_v
\nabla^2\varphi + 2\cos\theta_v \left( \mathbf{\nabla}\theta_v
\cdot \mathbf{\nabla}\varphi + \mathbf{\nabla}\vartheta \cdot
\mathbf{\nabla}\phi_v \right)
\end{align}

These are very similar to the vacuum magnon equations of motion
with the addition of a few perturbing terms. Notably, these
additional terms all decay away the vortex core and will be
treated in a Born approximation (applicability of this
approximation is discussed at the end of the next section).

Alternatively \cite{ivanov:1998,ivanov:2000,ivanov:2003}, we could
expand as
\\
\begin{align}\label{boundmodenorm}
\theta =& \theta_v + \vartheta \nonumber \\
\phi =& \phi_v + \frac{\varphi}{\sin\theta_v}
\end{align}
\\
yielding the linearized equations in $\vartheta$ and $\varphi$
\\
\begin{align*}
\frac{1}{c} \frac{\partial \varphi}{\partial t} =&
-\nabla^2\vartheta - \cos 2\theta_v \left(\frac{1}{r_v^2} - \left(
\mathbf{\nabla} \phi_v \right)^2 \right) \vartheta + 2
\cos\theta_v \mathbf{\nabla}\varphi \cdot \mathbf{\nabla}\phi_v\\
\frac{1}{c} \frac{\partial \vartheta}{\partial t} =&
\nabla^2\varphi + \cos^2\theta_v \left(\frac{1}{r_v^2} - \left(
\mathbf{\nabla} \phi_v \right)^2 \right) \varphi + \left(
\mathbf{\nabla}\theta_v \right)^2 \varphi + 2 \cos\theta_v
\mathbf{\nabla}\vartheta \cdot \mathbf{\nabla}\phi_v
\end{align*}
\\
or, equivalently, in the more symmetric form
\\
\begin{align}\label{linesym}
\frac{r_v}{c}\frac{\partial\varphi}{\partial t}=& \left( -\nabla^2
+ V_1(x) \right) \vartheta + \frac{2q \cos \theta_v}{x^2}
\frac{\partial\varphi}{\partial\xi} \nonumber \\
-\frac{r_v}{c}\frac{\partial\vartheta}{\partial t} =& \left(
-\nabla^2 + V_2(x) \right) \varphi - \frac{2q \cos
\theta_v}{x^2}\frac{\partial\vartheta}{\partial\xi}
\end{align}
\\
where $x = \frac{r}{r_v}$, radial derivatives are now with respect
to $x$, and
\\
\begin{align}\label{Vs}
V_1(x)=&\left( \left(\mathbf{\nabla}\phi_v \right)^2 -1
\right) \cos 2\theta_v \nonumber \\
V_2(x) =& \left( \left(\mathbf{\nabla}\phi_v \right)^2 -1 \right)
\cos^2\theta_v - \left(\mathbf{\nabla}\theta_v \right)^2
\end{align}
This form is particularly suitable for examining the core effects
and searching for possible bound modes.

\subsection{Phase shifts in the Born approximation}\label{section:Born}
The perturbing terms are localized to the vortex and can be
treated in a Born approximation. The dominant scattering term
decays as $\frac{1}{r^2}$, whereas the remaining terms, neglected
in the following treatment, die exponentially. The error
introduced by neglecting these terms will be discussed in the
final analysis at the end of this section. The magnon equations of
motion are modified to
\begin{align}\label{longrangeeom}
\frac{1}{c} \frac{\partial \varphi}{\partial t} =&
-\nabla^2\vartheta + \left(\frac{1}{r_v^2} - \frac{q^2}{r^2} \right) \vartheta \nonumber \\
\frac{1}{c} \frac{\partial \vartheta}{\partial t} =&
\nabla^2\varphi
\end{align}
\\
The perturbation treatment is most straightforward in a single
variable. Eliminating the $\vartheta$ variable, we have
\\
\begin{equation}
\frac{1}{c^2}\frac{\partial^2 \varphi}{\partial^2 t} =
-\nabla^4\varphi + \left(\frac{1}{r_v^2} - \frac{q^2}{r^2} \right)
\nabla^2\vartheta
\end{equation}
\\
Note the additional term $\frac{q^2}{r^2}$ modifying the vacuum
equations of motion of the magnons (\ref{magnoneom}). The Born
approximation is applied using the standard partial-wave analysis
from scattering theory\cite{LL:1977}.

Consider the orthonormal basis functions $\xi_{\mathbf k}$ such
that $\nabla^2 \xi_{\mathbf k} \to -k^2 \xi_{\mathbf k}$ and
assume harmonic time dependence. We expand $\varphi$ in this basis
\\
\begin{equation}\label{assumeform}
\varphi = \sum_{\mathbf k'} c_{\mathbf k'} e^{i\omega_k t}
\xi_{\mathbf k'}
\end{equation}
\\
where to zeroth order we've assumed
\\
\begin{equation}
c_{\mathbf k'}^{(0)} = \left\{%
\begin{array}{ll}
    1, & \hbox{$\mathbf k'=\mathbf k$;} \\
    0, & \hbox{otherwise.} \\
\end{array}%
\right.
\end{equation}
\\
The zeroth order terms simply reduce to the vacuum equations of
motion. The first order terms are
\\
\begin{equation}
- \sum_{\mathbf k'\neq \mathbf k} \frac{\omega_k^2}{c^2}
c_{\mathbf k'}^{(1)} \xi_{\mathbf k'} = - \sum_{\mathbf k'\neq
\mathbf k} k'^2Q'^2 c_{\mathbf k'}^{(1)} \xi_{\mathbf k'} +
\frac{q^2 k^2}{r^2} \xi_{\mathbf k}
\end{equation}
\\
where we've cancelled the common $e^{i\omega_k t}$ factor. Recall
$Q^2 = k^2 + \frac{1}{r_v^2}$. Multiplying by $\xi_{\mathbf
k''}^{\dagger}$ and integrating over space, enforcing
orthonormality of the $\{\xi_{\mathbf k}\}$, we find an expression
for the first order coefficients
\\
\begin{equation}
c_{\mathbf k'}^{(1)} = -\frac{c^2}{\omega_k^2 - c^2 k'^2 Q'^2}
\int d^2r \xi_{\mathbf k'}^{\dagger} \frac{q^2 k^2}{r^2}
\xi_{\mathbf k}
\end{equation}
\\
Substituting for the unperturbed magnon spectrum
(\ref{magnonspectrum}) and using plane waves for the orthonormal
basis, the first order correction to $\varphi$ is
\\
\begin{equation*}
\varphi^{(1)}(\mathbf r) = -\int \frac{d^2k'}{(2\pi)^2}
\frac{e^{i\mathbf k' \cdot \mathbf r}}{k^2Q^2 - k'^2Q'^2} \int
d^2r' e^{-i\mathbf k' \cdot \mathbf r'} \frac{q^2 k^2}{r'^2}
e^{i\mathbf k \cdot \mathbf r'}
\end{equation*}
\\
First, integrating over the polar angle $\phi_{\mathbf k'}$ from
$0$ to $\pi$, we obtain
\\
\begin{equation*}
\varphi^{(1)}(\mathbf r) = \frac{1}{4} \int_{-\infty}^\infty
\frac{dk' \; k'}{2\pi} \frac{H_0^{(1)}(k|\mathbf r- \mathbf r'|) +
H_0^{(2)}(k|\mathbf r- \mathbf r'|)}{k'^2Q'^2 - k^2Q^2 \pm
i\epsilon} \int d^2r' e^{-i\mathbf k' \cdot \mathbf r'} \frac{q^2
k^2}{r'^2} e^{i\mathbf k \cdot \mathbf r'}
\end{equation*}
\\
The $\pm i\epsilon$ are chosen to displace the poles so as to pick
the outgoing wave (the plus is for the $H_0^{(2)}$ integral, the
minus for the $H_0^{(1)}$ integral). Considering the $H_0^{(2)}$
integral, there are poles in the complex $k'$ plane at $k' = \pm (
k + i\epsilon')$. Noting the asymptotic behaviour
\\
\begin{equation}
H_0^{(2)}(kr) \to \sqrt{\frac{2}{\pi kr}}
e^{-i(kr-(l+\frac{1}{2})\frac{\pi}{2})}
\end{equation}
\\
we close the contour about the positive imaginary axis for the
$H_0^{(2)}$  integral, to pick up the $k' = k+i\epsilon$ pole with
residue $\frac{1}{2(k^2+Q^2)} H_0^{(2)}(k|\mathbf r- \mathbf
r'|)$. The integral over $H_0^{(1)}$ is just the complex conjugate
(c.c) of that over $H_0^{(2)}$ and hence follows immediately.

Thus we have
\\
\begin{equation*}
\varphi^{(1)}(\mathbf r) =  \frac{ik^2}{8(k^2+Q^2)} \int d^2r'
H_0^{(2)}(k |\mathbf r- \mathbf r'|)\frac{q^2}{r'^2} e^{i\mathbf k
\cdot \mathbf r'} + c.c.
\end{equation*}

We now expand the Hankel functions according to the identity
\\
\begin{equation}
H_0^{(1,2)}(k |\mathbf r- \mathbf r'|) = \sum_{l=\infty}^{\infty}
J_l(kr')H_l^{(1,2)}(kr) e^{il(\phi - \phi')}
\end{equation}
\\
if $r>r'$, an allowable assumption if we only want the
wavefunction correction for asymptotic $r$, and $\phi$ ($\phi'$)
is the polar angle of $\mathbf r$ ($\mathbf r'$). Expand the plane
wave as
\\
\begin{equation}\label{expandplanewave}
e^{i\mathbf k \cdot \mathbf r'} = \sum_{m=\infty}^{\infty} i^m
J_m(kr') e^{im\phi'}
\end{equation}
\\
After integration over $\phi'$ (giving a factor
$2\pi\delta_{lm}$), the wavefunction correction is
\\
\begin{equation}\label{firstorder}
\varphi^{(1)}(\mathbf r) = \frac{\pi ik^2}{4(k^2+Q^2)}
\sum_{l=\infty}^{\infty} i^l H_l^{(2)}(kr) e^{il\phi} \int dr'
\frac{q^2}{r'} J_l^2(kr') + c.c.
\end{equation}
\\
Recall we assumed the unperturbed $\varphi^{(0)}$ solution was a
plane wave, expandable according to (\ref{expandplanewave}), so
that the entire solution can be written, up to first order,
\\
\begin{equation*}
\varphi(\mathbf r) = \sum_{l=\infty}^{\infty} i^l e^{il\phi}
\left[ J_0(kr) + \left(\frac{\pi ik^2}{4(k^2+Q^2)} H_l^{(2)}(kr)
\int dr' \frac{q^2}{r'} J_l^2(kr') + c.c.\right)\right]
\end{equation*}
\\
Comparing this with the sum of an incoming and outgoing wave
\\
\begin{align}\label{cosinescatt}
\frac{1}{2}& \left( e^{i\Delta_l}H_l^{(1)}(kr) + e^{-i\Delta_l}
H_l^{(2)}(kr)
\right) = J_l(kr) - i\frac{\Delta_l}{2} H_l^{(2)}(kr) +
i\frac{\Delta_l}{2} H_l^{(1)}(kr) \nonumber\\
&  \to \sqrt{\frac{2}{\pi kr}} e^{-i\Delta_l} \cos \left( kr -
(l+\frac{1}{2}) \frac{\pi}{2} - \Delta_l \right) \hbox{as }
r\to\infty
\end{align}
\\
gives for the phase shift of the l$^{th}$ order wave
\\
\begin{equation}\label{phaseshifts}
\Delta_l = -\frac{\pi}{2} \frac{k^2}{k^2+Q^2} \int dr'
\frac{q^2}{r'} J_l^2(kr')
\end{equation}
These phase shifts perturb the magnon wavevector $k = k^{scatt}
-\Delta_l$, and, hence, the magnon spectrum $\omega_k$. For proper
counting of the total energy shift, first discretize $k$ by fixing
the boundary conditions of the wavefunction at $r = R_S$ so that
\\
\begin{equation}
\pi n = k_n R_S = k_n^{scatt} R_S - \Delta_l(k_n)
\end{equation}
\\
Notice that asymptotically we have a cosine wavefunction as
opposed to a plane wave as described by Rajaraman \cite{raj}.
Letting the system size tend to infinity then
\\
\begin{equation*}
\sum_{k_n} \to \frac{R_S}{\pi} \int dk
\end{equation*}
\\
The zero point energy shift, given by the change in the zero point
energy of the small oscillation modes when the vortex is present
as compared to those in vacuum, is then
\\
\begin{equation}
\Delta E = \frac{1}{2} \sum_{k,l} \hbar \delta \omega_k
\end{equation}
\\
But $k=k^{scatt}-\Delta_l$,
\\
\begin{align*}
\delta \omega_k =& \omega(k^{scatt}) - \omega(k) \\
=& \omega(k+\frac{\Delta_l}{R_S}) - \omega(k) \\
=& \frac{\partial \omega(k)}{\partial k} \frac{\Delta_l(k)}{R_S}
\end{align*}
\\
so that
\\
\begin{equation}
\Delta E = \frac{\hbar}{2} \sum_{l=-\infty}^{\infty} \int
\frac{dk}{\pi} \frac{\partial \omega}{\partial k} \Delta_l(k)
\end{equation}
\\
Substituting for $\Delta_l(k)$ from equation (\ref{phaseshifts}),
noting that $\frac{\partial \omega}{\partial k} = c
\frac{k^2+Q^2}{Q}$,
\\
\begin{align*}
\Delta E =& -\frac{\hbar c}{4} \int dk \frac{k^2}{k^2+Q^2}
\frac{k^2+Q^2}{Q} \int dr' \frac{q^2}{r'}
\sum_{l=-\infty}^{\infty} J_l^2(kr')\\
=& -\frac{\hbar c q^2}{4} \int dk \frac{k^2}{Q} \ln
\frac{R_S}{r_v}
\end{align*}
\\
where we've used that $\sum\limits_{l=-\infty}^{\infty}
J_l^2(kr')=1$. Note that the radial integral is cut off by the
vortex core size. This is because the perturbing term changes
behaviour drastically in the core so that our analysis cannot be
extended there. The $k$ integral can be evaluated noting that
\\
\begin{equation*}
\frac{d}{dk}\left(\frac{1}{2}kQ-\frac{1}{2r_v^2} \ln(k+Q)\right) =
\frac{k^2}{Q}
\end{equation*}
\\
The result is ultraviolet divergent so that we must impose a
cutoff of $1/a$, physically reasonable if we recall that $a$ is
the lattice spacing of the discrete lattice. Finally, the energy
shift in the presence of a magnetic vortex is
\\
\begin{equation}
\Delta E = -\frac{\hbar c q^2}{4} \ln \frac{R_S}{r_v} \left(
\frac{\sqrt{r_v^2+a^2}}{2a^2r_v} - \frac{1}{2r_v^2} \ln \frac{r_v
+ \sqrt{r_v^2+a^2}}{a} \right)
\end{equation}
This zero-point energy shift, due to the presence of the vortex,
is associated to the quantized vortex rather than the
magnons\cite{raj}. Thus, $\Delta E$ is the zero-point energy of
the vortex. Note that the interaction actually decreases the
quantum energy of the vortex-magnon system.

We can examine the error in neglecting the exponentially decaying
terms by replacing the $r_v/r'$ behaviour by $\exp(-r/r_v)$.
Essentially, this would replace the log divergence in the final
result with unity. Hence, in comparison with the main
contribution, these exponentially decaying terms are negligible.

The Born approximation amounts to the substitution of
\begin{equation}
\langle \phi_f | U | \psi_i \rangle \to \langle \phi_f | U |
\phi_i \rangle
\end{equation}
where $\phi$ and $\psi$ denote the unperturbed and modified
waveforms, respectively. In general, the validity of the Born
approximation depends on how much the waveforms differ in the
region of the scattering potential \cite{peierls:1979}. In our
case, the Born approximation indicates that the two wavefunctions
in fact differ to first order by equation (\ref{firstorder}) which
is proportional to the predicted phase shifts. This is circular
reasoning; however, in the case of those quasiparticle modes
delocalized over the system, we expect the waveform not to change
significantly. On the other hand, there are quasiparticles that
become trapped by the vortex center. Clearly, for these modes, the
wavefunctions are drastically modified in the vortex core, where
the scattering potential is greatest, so that a Born approximation
is invalid. We examine these bound modes in the next section to
show how they result from the translational symmetry broken by the
vortex solution.

\subsection{Bound modes}\label{section:shortrange}
As pointed out by Ivanov et. al. \cite{ivanov:1998,ivanov:2000,
ivanov:2003}, the short range interactions neglected in
(\ref{longrangeeom}) can drastically alter the behaviour of
certain modes. The symmetric perturbing equations,
(\ref{linesym}), are more suitable for exploring the core region.
Assume a solution of the form
\\
\begin{align}
\vartheta =& f(x) \cos (m\chi + \omega t +\psi) \nonumber\\
\varphi =& g(x) \sin (m\chi + \omega t +\psi)
\end{align}
\\
Substituting this into (\ref{linesym}) yields equations for $f$
and $g$
\\
\begin{align}\label{fg_eqns}
\left( \frac{d^2}{dx^2} + \frac{1}{x}\frac{d}{dx} -
\frac{m^2}{x^2} - V_1(x) \right) f =& \left( \frac{wr_v}{c} +
\frac{2qm\cos\theta_v}{x^2}
\right) g \nonumber\\
\left( \frac{d^2}{dx^2} + \frac{1}{x}\frac{d}{dx} -
\frac{m^2}{x^2} - V_2(x) \right) g =& \left( \frac{wr_v}{c} +
\frac{2qm\cos\theta_v}{x^2} \right) f
\end{align}
\\
recalling that $x=\frac{r}{r_v}$ and that $V_1(x)$ and $V_2(x)$
are defined in (\ref{Vs}).

For $\omega = 0$, there exist exact solutions for $m=0,\pm 1$
\\
\begin{align}\label{w=0solns}
f=&m\theta_v' \nonumber\\
g=&-\frac{q\sin\theta_v}{x^{|m|}}
\end{align}
\\
For $|m|>1$, the asymptotic behaviour of the modes is entirely
unbounded so that the vortex center has not greatly shifted the
magnon wavefunctions and the Born approximation applied in the
previous section should be valid.

Consider first the $m=0$ result. Combining the unperturbed vortex
profile with this result (recall the normalization of the
perturbations as in (\ref{boundmodenorm}))
\\
\begin{align}
\phi =& q \chi - q \delta \chi \nonumber \\
\theta =& \theta_v
\end{align}
\\
where $\delta \chi$ is the coefficient of the linearized solution.
We find that it corresponds simply to the freedom of uniform
rotation in the $xy$-plane.

\begin{figure}
\centering
\includegraphics[width=8cm]{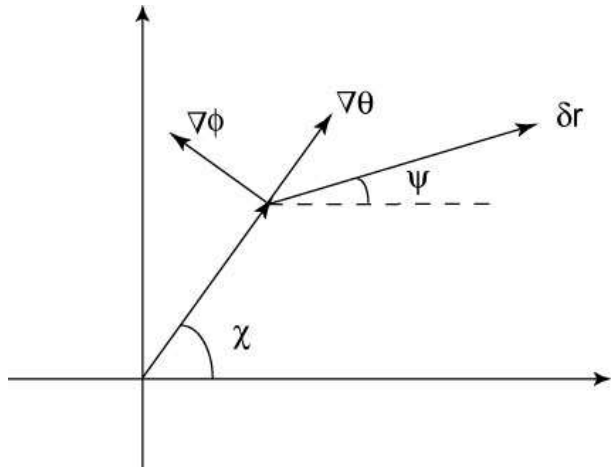}
\caption{The directions relevant to a small translation of the
vortex along $\mathbf{\delta r}$.}\label{figure:transl}
\end{figure}

Similarly, consider the $m=\pm 1$ solutions
\\
\begin{align*}
\phi =& q \chi - \frac{q \delta r}{r} \sin (m\chi -\psi) \\
\theta =& \theta_v + m \delta r \theta_v' \cos (m\chi -\psi)
\end{align*}
\\
with $\delta r$ as the coefficient of the linearized $m=\pm 1$
solution. But note that the additional contributions can be
re-expressed as
\\
\begin{align}
\phi =& q \chi + \mathbf \nabla \phi_v \cdot m \delta{\mathbf r} \nonumber \\
\theta =& \theta_v + \mathbf \nabla \theta_v \cdot m
\delta{\mathbf r}
\end{align}
\\
where $\delta{\mathbf r}$ is now a vector of magnitude $\delta r$
in the direction defined by the polar angle $\psi$ (see Figure
\ref{figure:transl}). Thus, these two modes represent
infinitesimal motion along $\pm\delta{\mathbf r}$ (the sign chosen
by the sign of $m$).

Clearly, these bound modes are inadequately treated using the Born
approximation and must be treated separately somehow. Ivanov et.
al. \cite{ivanov:1998, ivanov:2000, ivanov:2003} attempted to
calculate the phase shifts of these modes separately and to
subsequently use them to describe the angular and translational
motion of the magnetic vortices. Alternatively, however, one can
treat the problem using collective coordinates (see Appendix
\ref{section:collcoord} for more details) conveniently separating
these so-called zero modes and treating the remaining modes in a
Born approximation.

In the next chapter, we expand the interactions of the magnetic
vortex with the environment magnons using collective coordinates.
Using a path integral formalism, we separate the degrees of
freedom of the vortex motion from those of the environment and
proceed to integrate out these modes yielding the effective
dynamics of the vortex.

\chapter{Vortex dynamics}\label{chapter:together}

We now have all the background to interact the vortices and
magnons. Using a variety of techniques, we examine the effects of
couplings between the two systems to the vortex energy and
dynamics. In the previous chapter, we already saw how a
modification in the magnon spectrum can be interpreted as a
quantum energy shift associated with the vortex.

First, using regular perturbation theory, we examine the one
magnon coupling with the vortex velocity giving rise to an
inertial mass and a dissipation rate of a moving vortex. We also
examine the long range two magnon coupling in this language,
finding almost immediately the zero point energy shift that in the
previous chapter required calculating all magnon phase shifts.

The effective vortex dynamics are derived by finding the time
evolution of the vortex-magnon density matrix and tracing over the
magnon modes. We use the Feynman-Vernon formalism, describing the
density matrix with path integrals. We again deduce the vortex
inertial mass, in agreement with perturbation results. The vortex
motion is again dissipative; however, we find the vortex damping
forces explicitly and characterize the associated fluctuating
forces. Generalization to a collection of vortices is carried out.

In addition to the previously derived gyrotropic and inter-vortex
forces, we derive microscopically vortex damping forces. We
introduce for the first time in such a magnetic system a
transverse damping force, analogous to the Iordanskii force acting
on a vortex in a superfluid. These are derived from the action
terms found in the vortex density matrix propagator
(\ref{densmatprop}) and have corresponding fluctuating forces with
correlations given by (\ref{flucforcecorr}).

Alternatively, we consider decomposing the motion in a Bessel
function basis, $\{J_m(kX(t))e^{im\phi_X}\}$, to obtain Brownian
motion for the components with an effective action given by
(\ref{freqeffaction}) and corresponding fluctuating force
correlations (\ref{freqflucforcecorr}).

\section{Vortex-magnon interaction terms}

We work with the complete non-linear Lagrangian for our magnetic
system
\\
\begin{equation}
\mathcal L = S \int \frac{d^{2} r}{a^2} \left( -\dot \phi
\cos\theta- \frac{c}{2} \left( (\mathbf{\nabla}\theta)^{2} +
\sin^{2}\theta \left( (\mathbf{\nabla}\phi)^2 - \frac{1}{r_v^2}
\right)\right)\right)
\end{equation}
\\
Expanding the Lagrangian density about the vortex profile via
$\theta = \theta_v + \vartheta$ and $\phi = \phi_v + \varphi$ we
find the following terms in the integrand
\\
\begin{align*}
&\frac{S}{a^2}\left(\dot\phi_v+\dot\varphi\right)\left(-
\cos\theta_v +\sin\theta_v \; \vartheta \right) - \frac{S^2J}{2}
\left( (\overrightarrow{\nabla}\theta_v)^2 + 2
\overrightarrow{\nabla} \theta_v \cdot
\overrightarrow{\nabla}\vartheta + (\overrightarrow{\nabla}
\vartheta)^2 + \right.\\
&\left. \left( \sin^2\theta_v + \sin 2\theta_v \; \vartheta + \cos
2\theta_v \;\vartheta^2\right) \left( (\overrightarrow{\nabla}
\phi_v)^2 -\frac{1}{r_v^2} + 2 \overrightarrow{\nabla} \phi_v
\cdot \overrightarrow{\nabla}\varphi + (\overrightarrow{\nabla}
\varphi)^2 \right) \right)
\end{align*}
\\
The zeroth order terms in $\varphi$ and $\vartheta$ simply give
the vortex action; the first order terms give $\vartheta$
multiplied by the $\frac{\partial \phi_v}{\partial t}$ equation of
motion and $\varphi$ multiplied by the $\frac{\partial
\theta_v}{\partial t}$ equation of motion and thus are zero,
except, notably, the one magnon dynamic term
\\
\begin{equation}\label{firstorderterm}
\frac{S}{a^2} \dot\phi_v \sin\theta_v \; \vartheta
\end{equation}

Finally, the remaining two magnon terms are
\\
\begin{align}\label{secondorderterms}
\frac{S}{a^2} \dot\varphi \sin\theta_v \; \vartheta -
\frac{S^2J}{2} \left((\overrightarrow\nabla\vartheta)^2 +
\sin^2\theta_v (\overrightarrow\nabla\varphi)^2 + 2 \sin2\theta_v
\overrightarrow\nabla\phi_v \cdot
\overrightarrow\nabla\varphi \; \vartheta \right. & \nonumber \\
\left. + \cos 2\theta_v \left(
(\overrightarrow\nabla\phi_v)^2-\frac{1}{r_v^2}\right)\vartheta^2
\right)&
\end{align}
\\
Minimizing these action terms, we find the perturbed equations of
motion similar to (\ref{lineom})
\\
\begin{align}
\frac{\sin\theta_v}{c} \frac{\partial \varphi}{\partial
t}+\frac{\sin\theta_v}{c}\dot \phi_v =& -\nabla^2\vartheta -
\cos2\theta_v \left(\frac{1}{r_v^2} - \left( \mathbf{\nabla}
\phi_v \right)^2 \right) \vartheta +\sin2\theta_v
\mathbf{\nabla}\varphi \cdot \mathbf{\nabla}\phi_v \nonumber\\
\frac{1}{c} \frac{\partial \vartheta}{\partial t} =& \sin\theta_v
\nabla^2\varphi + 2\cos\theta_v \left( \mathbf{\nabla}\theta_v
\cdot \mathbf{\nabla}\varphi + \mathbf{\nabla}\vartheta \cdot
\mathbf{\nabla}\phi_v \right)
\end{align}

Define the vortex profile relative to the center
coordinate\symbolfootnote[2]{There is no need to add a collective
coordinate reflecting the rotational symmetry of the problem since
this is actually just a restatement of the 2-dimensional
translational freedom, already entirely taken care of in the
2-dimensional center coordinate.} $\mathbf X$
\\
\begin{align}
\phi_v = q\chi(\mathbf r - \mathbf X) \nonumber \\
\theta_v = \theta_v(\mathbf r - \mathbf X)
\end{align}
\\
The center coordinates play the role of the collective coordinates
in this system, introduced to account for the continuous
translational symmetry broken by the vortex. They are elevated to
operators.

Focussing on the one magnon perturbative term,
(\ref{firstorderterm}), expanding in terms of the collective
coordinates, we find
\\
\begin{equation}\label{firstordercoupling}
\frac{S}{a^2} \dot\phi_v \sin\theta_v \; \vartheta =
-\frac{S}{a^2}\dot{\mathbf X}\cdot \nabla \phi_v \sin\theta_v \;
\vartheta
\end{equation}
\\
In the previous chapter, this term perturbed the vortex profile
under motion, introducing a finite vortex mass.

\section{Perturbation theory results}

\subsection{Vortex mass revisited}\label{section:vortexmasspertn}

Consider the one magnon coupling (\ref{firstordercoupling}). This
term can be considered as a perturbing term of the vortex profile
under motion or, alternatively, as a vortex-magnon coupling.
Fourier transforming $\vartheta$ according to
(\ref{fouriertransform}), now with $\vartheta = \vartheta(\mathbf
r - \mathbf X)$, we can rewrite the coupling as
\\
\begin{align*}
-\int \frac{d^2k a^2}{(2\pi)^2} S \int \frac{d^2 r}{a^2}&
\sin\theta_v \dot{\mathbf X}\cdot \nabla \phi_v
e^{-i\mathbf k \cdot \mathbf r} \vartheta_{\mathbf k} = \\
& -\frac{S}{a^2} \int \frac{d^2k a^2}{(2\pi)^2} e^{-i\mathbf k
\cdot \mathbf X} \vartheta_{\mathbf k}  \int d^2 r e^{-ikr
\cos\chi_{kr}} \sin\theta_v \frac{q\dot X \sin\chi_{vr}}{r}
\end{align*}
\\
where the $\mathbf r$ integration has been shifted to move the
vortex center coordinates into the exponential. Expanding
$\chi_{vr} = \chi_{vk} + \chi_{kr}$ and noting that
\\
\begin{equation*}
\int d\chi_{kr} \sin\chi_{kr}e^{-ikr \cos\chi_{kr}}=0
\end{equation*}
\\
the coupling term becomes
\\
\begin{align*}
-\frac{S}{a^2} \int \frac{d^2k a^2}{(2\pi)^2} e^{-i\mathbf k \cdot
\mathbf X}\vartheta_{\mathbf k} &\int d^2 r e^{-ikr \cos\chi_{kr}}
\sin\theta_v \frac{q\dot X \sin\chi_{vk}
\cos\chi_{kr}}{r} \\
=& \frac{2\pi iSq}{a^2} \int \frac{d^2k a^2}{(2\pi)^2}
e^{-i\mathbf k \cdot \mathbf X}\vartheta_{\mathbf k} \int dr \dot
X \sin\chi_{vk}
J_1(kr) \sin\theta_v\\
=& \frac{2\pi iSq}{a^2} \int \frac{d^2k a^2}{(2\pi)^2}
e^{-i\mathbf k \cdot \mathbf X}\vartheta_{\mathbf k}
\frac{\dot{\mathbf X} \cdot \hat{\mathbf \chi}_k}{kQr_v}
\end{align*}
\\
where we approximate $\sin\theta_v \approx 1-e^{-r/r_v}$, which
has the right asymptotic behaviour for $r\to0$ and $r\to\infty$.
Expressing $\vartheta_{\mathbf k}$ in terms of creation and
annihilation operators given in (\ref{creatannil}), we finally
obtain
\\
\begin{equation}
\int \frac{d^2k a^2}{(2\pi)^2}e^{-i\mathbf k \cdot \mathbf X}
\frac{2\pi q}{a^2} \sqrt{\frac{\hbar S}{2kQ}}\frac{\dot{\mathbf X}
\cdot \hat{\mathbf \chi}_k}{Qr_v} (a_{\mathbf k}^{\dagger} -
a_{-\mathbf k})
\end{equation}
\\
or expressed a little more symmetrically
\\
\begin{equation}
\int \frac{d^2k a^2}{(2\pi)^2} \frac{2\pi q}{a^2}
\sqrt{\frac{\hbar S}{2kQ}}\frac{\dot{\mathbf X} \cdot \hat{\mathbf
\chi}_k}{Qr_v} (e^{-i\mathbf k \cdot \mathbf X}a_{\mathbf
k}^{\dagger} - e^{i\mathbf k \cdot \mathbf X}a_{-\mathbf k})
\end{equation}
\\
Note the similarity of the coupling here to that of the polaron
problem discussed, for example, by Feynman\cite{feynman:1972}.

In first order perturbation theory, this coupling provides no
energy shift since it necessarily changes the number of magnons
between the initial and final state.

\begin{figure}
\centering
\includegraphics[width=6cm]{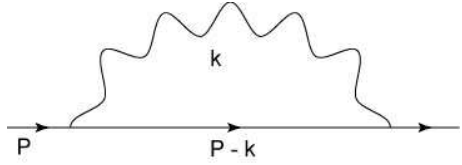}
\caption{Lowest order contributing diagram for the first order
vortex-magnon coupling term.}\label{figure:feynmandiag}
\end{figure}

In second order perturbation theory, we consider the diagram shown
in Figure \ref{figure:feynmandiag} corresponding to the emission
and re-absorption of a virtual magnon. The energy shift provided
by this diagram, which with foresight we call $E_{mass}$,
\\
\begin{equation*}
E_{mass} = \int \frac{d^2k a^2}{(2\pi)^2}  \frac{2\pi^2 \hbar
Sq^2}{a^4kQ^3r_v^2} \left(\dot{\mathbf X} \cdot \hat{\mathbf
\chi}_k\right)^2 \frac{1}{(\mathbf P - \hbar \mathbf k)^2/2M_v +
\hbar ckQ-P^2/2M_v}
\end{equation*}
\\
However, so far the vortex has no inertial energy, $P^2/2M\to0$,
and
\\
\begin{equation}
E_{mass} = \int \frac{d^2k a^2}{(2\pi)^2} \frac{1}{\hbar ckQ}
\frac{2\pi^2 \hbar Sq^2}{a^4kQ^3r_v^2} \left(\dot{\mathbf X} \cdot
\hat{\mathbf \chi}_k\right)^2
\end{equation}

The integration over the polar angle of $\mathbf k$ contributes a
factor $\pi$. We expand the $k$ and $Q$ dependence in partial
fractions
\\
\begin{equation*}
\frac{1}{kQ^4r_v^2} = \frac{r_v^2}{k} - \frac{kr_v^2}{Q^2} -
\frac{k}{Q^4}
\end{equation*}
\\
The radial integral is evaluated as
\\
\begin{align}
E_{mass} =& \frac{\pi q^2 \dot X^2}{2Ja^4} \int dk
\frac{k}{ k^2 Q^3 r_v^2} \nonumber \\
=& \frac{\pi q^2 \dot X^2}{2Ja^4} \int dk \left(
\frac{r_v^2}{k} - \frac{kr_v^2}{Q^2} - \frac{k}{Q^4}\right) \nonumber \\
=& \frac{\pi q^2r_v^2 \dot X^2}{2Ja^4} \left( \ln\frac{R_S}{a} -
\frac{1}{2}\ln\frac{a^2+r_v^2}{a^2} - \frac{1}{2}
\frac{r_v^2}{a^2+r_v^2} \right)
\end{align}
where we've imposed both an upper and lower cutoff, with $a$ the
lattice spacing and $R_S$ the system size.
\\
Thus, identifying this as a $\frac{1}{2}M_v\dot X^2$ inertial
term, we find a vortex mass of
\\
\begin{equation}\label{vortexmass}
M_v = \frac{\pi q^2r_v^2}{Ja^4} \left(
\ln\frac{R_S}{\sqrt{a^2+r_v^2}}  - \frac{1}{2}
\frac{r_v^2}{a^2+r_v^2} \right)
\end{equation}
\\
in agreement to leading order with the analysis of section
\ref{section:vortexmass}. The $r_v \to \sqrt{a^2+r_v^2}$
replacement corrects the $r_v\to0$ limiting behaviour.

\subsubsection{Mass tensor of a collection of vortices}\label{section:manyvortexmass}

We can easily generalize this result to a collection of vortices
in this formalism. Recall that the $n$-vortex superposed solution
is given by
\\
\begin{align}
\phi_{tot} =& \sum_{i=1}^n q_i \chi(X_i) \nonumber \\
\theta_{tot} =& \sum_{i=1}^n \theta_v(\mathbf r - \mathbf X_i)
\end{align}
\\
so that the one magnon coupling becomes
\\
\begin{equation}
\sum_{i=1}^n \int \frac{d^2k a^2}{(2\pi)^2}e^{-i\mathbf k \cdot
\mathbf X_i} \frac{2\pi q_i}{a^2} \sqrt{\frac{\hbar
S}{2kQ}}\frac{\dot{\mathbf X}_i \cdot \hat{\mathbf \chi}_k}{Qr_v}
(a_{\mathbf k}^{\dagger} - a_{-\mathbf k})
\end{equation}

The second order energy correction is now
\\
\begin{equation}
E_{mass} = \sum_{i,j=1}^n \int d^2k
\frac{q_iq_j}{2Ja^4k^2Q^4r_v^2} \left(\dot{\mathbf X}_i \cdot
\hat{\mathbf \chi}_k e^{-i\mathbf k \cdot \mathbf
X_i}\right)\cdot\left(\dot{\mathbf X}_j \cdot \hat{\mathbf \chi}_k
e^{i\mathbf k \cdot \mathbf X_j}\right)
\end{equation}

\begin{figure}
\centering
\includegraphics[width=6cm]{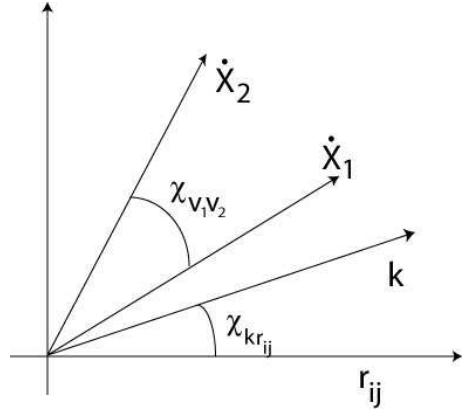}
\caption{Definition of angles for evaluation of off-diagonal mass
terms.}\label{figure:axesmanymass}
\end{figure}

The diagonal terms evaluate exactly as above. The off-diagonal
terms can be evaluated noting that
\\
\begin{align*}
\int & d\chi_{kr_{ij}} \sin\chi_{kv_1} \sin\chi_{kv_2}
\exp(ikr_{ij}\cos\chi_{kr_{ij}}) \\
&= \frac{1}{2} \int d\chi_{kr_{ij}} \left(
\cos(\chi_{kv_1}-\chi_{kv_2}) + \cos(\chi_{kv_1}+\chi_{kv_2})
\right) \exp(ikr_{ij}\cos\chi_{kr_{ij}})
\end{align*}
\\
where $r_{ij} = |\mathbf X_i - \mathbf X_j|$. If we assume the
various angles are defined as in Figure \ref{figure:axesmanymass},
then $\chi_{kv_1}-\chi_{kv_2} = \chi_{v_1v_2}$ and
$\chi_{kv_1}+\chi_{kv_2} = \chi_{v_1r_{ij}}+\chi_{v_2r_{ij}} +
2\chi_{kr_{ij}}$.

Noting by symmetry that the $\sin\chi_{kr_{ij}}$ terms integrate
to zero, we have
\\
\begin{align*}
\int & d\chi_{kr_{ij}} \sin\chi_{kv_1} \sin\chi_{kv_2}
\exp(ikr_{ij}\cos\chi_{kr_{ij}}) \\
&= \frac{1}{2} \int d\chi_{kr_{ij}} \left( \cos\chi_{v_1v_2} +
\cos(\chi_{v_1r_{ij}}+ \chi_{v_2r_{ij}})
\cos2\chi_{kr_{ij}}\right)\exp(ikr_{ij}\cos\chi_{kr_{ij}}) \\
&= \pi\cos\chi_{v_1v_2}J_0(kr_{ij}) + \pi \cos(\chi_{v_1r_{ij}}+
\chi_{v_2r_{ij}})J_2(kr_{ij})
\end{align*}
\\

Next we perform the integrals over $k$, noting that
\\
\begin{align*}
\int_{1/R_S}^{1/a}dk \frac{J_0(kr_{ij})}{k} =& \ln
\frac{R_S}{r_{ij}} \\
\int_{0}^{\infty}dk \frac{J_2(kr_{ij})}{k} = \frac{1}{2}
\end{align*}
\\
and rewrite
\\
\begin{equation*}
\cos(\chi_{v_1r_{ij}}+ \chi_{v_2r_{ij}}) = (\hat{\mathbf
X}_i\cdot\hat{\mathbf e}_{\Delta}) (\hat{\mathbf
X}_j\cdot\hat{\mathbf e}_{\Delta}) -(\hat{\mathbf
X}_i\times\hat{\mathbf e}_{\Delta})\cdot (\hat{\mathbf
X}_j\times\hat{\mathbf e}_{\Delta})
\end{equation*}
\\
Finally the energy correction term becomes
\\
\begin{equation}
E_{mass} = \sum_{i,j=1}^{n} \frac{1}{2} \dot{X}_i M_{ij} \dot{X}_j
\end{equation}
\\
where $M_{ij}$ is the $n$-vortex mass tensor given by
\\
\begin{align}\label{manyvortexmass}
M_{ij} =& \int d^2k  \frac{q_iq_j}{Ja^4 k^2 Q^4 r_v^2}
\left(\hat{\dot{\mathbf X}}_i \cdot \hat{\hat \chi}_k\right)
\left(\hat{\dot{\mathbf X}}_j \cdot \hat{\hat \chi}_k \right)
e^{i\mathbf k \cdot (\mathbf X_i-\mathbf X_j)} \nonumber \\
=& \frac{\pi q_iq_j r_v^2}{Ja^4} \left\{%
\begin{array}{ll}
    \ln \frac{R_S}{r_{ij}} + \frac{1}{2}
    \left( (\hat{\mathbf X}_i\cdot\hat{\mathbf e}_{ij}) (\hat{\mathbf
    X}_j\cdot\hat{\mathbf e}_{ij})   \right. & \\
    \left. -(\hat{\mathbf X}_i\times
    \hat{\mathbf e}_{ij}) \cdot (\hat{\mathbf
    X}_j\times\hat{\mathbf e}_{ij}) \right), & \hbox{$i\neq j$;} \\
    \ln \frac{R_S}{\sqrt{a^2+r_v^2}}, & \hbox{$i=j$.} \\
\end{array}%
\right.
\end{align}
\\
where $\hat{\mathbf e}_{ij} = \frac{\mathbf X_i-\mathbf
X_j}{|\mathbf X_i-\mathbf X_j|}$. This result is in agreement with
that of Slonczewski \cite{sloncz:1984}. Slonczewski's calculation
follows very closely that of section \ref{section:vortexmass}.

\subsection{Radiation of magnons}\label{section:vortmagdiss}

In the previous section, we calculated the vortex inertial energy
using second order perturbation theory. However, we only used the
principle part of the integral. When evaluating the integral
giving the second order perturbative energy
shift\cite{feynman:1972}, to be careful in the divergent region
$E_f \to E_i$, symbolically, we should write
\\
\begin{equation}
\Delta E_i = \sum_f \frac{H_{if} H_{fi}}{E_i-E_f +i\epsilon}
\end{equation}
\\
and then take the limit $\epsilon \to 0$. But
\\
\begin{equation*}
\frac{1}{x+i\epsilon} = \frac{x}{x^2+\epsilon^2} -
\frac{i\epsilon}{x^2+\epsilon^2}
\end{equation*}
\\
The imaginary part approaches a $\delta$-function as $\epsilon \to
0$ since
\\
\begin{equation}
\int_{-\infty}^{\infty} dx \frac{\epsilon}{x^2+\epsilon^2} = \pi
\end{equation}
So then
\\
\begin{equation*}
\frac{1}{x+i\epsilon} = \mbox{principle value} \left( \frac{1}{x}
\right) - i\pi\delta(x)
\end{equation*}

An imaginary part to the energy shift creates a decaying
exponential factor in the time-dependent wave function
\\
\begin{equation*}
e^{-i(E/\hbar-i\gamma/2)t} = e^{iEt/\hbar}e^{-\gamma t/2}
\end{equation*}
\\
and is hence interpreted as dissipation. The factor of 2 is there
so that the probability $|\psi|^2$ decays as $e^{-\gamma t}$.

The rate of decay due to magnon emission is thus given
symbolically by
\\
\begin{equation}
\gamma = \sum_f \frac{2\pi}{\hbar} |H_{fi}|^2 \delta(E_f-E_i)
\end{equation}
\\
If we assume an initial state of no magnons the rate equation
becomes
\\
\begin{equation}
\gamma = 2\pi \int \frac{d^2k a^2}{(2\pi)^2} \frac{2\pi^2
Sq^2}{a^4kQ^3r_v^2} \left(\dot{\mathbf X} \cdot \hat{\mathbf
\chi}_k \right)^2 \delta\left( \frac{(\mathbf P-\hbar \mathbf
k)^2}{2M_v} + \hbar ckQ- \frac{P^2}{2M_v} \right)
\end{equation}
\\
Again, initially we have no inertial term, simplifying $\gamma$ to
\\
\begin{equation}\label{pertndissiK}
\gamma = 2\pi \int \frac{d^2k a^2}{(2\pi)^2} \frac{2\pi^2
Sq^2}{a^4kQ^3r_v^2} \left(\dot{\mathbf X} \cdot \hat{\mathbf
\chi}_k \right)^2 \delta\left(\hbar ckQ\right)
\end{equation}

Rewriting the $\delta$-function as
\\
\begin{equation*}
\delta(\hbar ckQ) = \frac{Q}{\hbar c(k^2+Q^2)}\delta(k)
\end{equation*}
\\
and the rate of dissipation becomes
\\
\begin{align}\label{pertndissi}
\gamma =& \int \frac{d \chi_k }{2\pi} \frac{2\pi^2 q^2r_v^2}{\hbar
Ja^4} \left(\dot{\mathbf X} \cdot \hat{\mathbf
\chi}_k \right)^2 \nonumber \\
=& \frac{\pi^2 q^2 r_v^2 \dot X^2}{\hbar Ja^4}
\end{align}

\citet{sloncz:1984} calculated microscopically a dissipation rate
by extending the simple results of section
\ref{section:vortexmass} to include retardation effects of spin
waves. However, to evaluate the far region perturbations, he
assumes small frequencies which give a log divergent frequency
dependent dissipation.

In our analysis, we used the same vortex-magnon coupling,
although, accounting only for the $k=0$ contribution. We find the
full spectrum contributions in section \ref{section:infldiss} and
will return to this comparison then.

\subsubsection{Second order radiative corrections}

Interestingly, the rate of emission using the finite mass of the
vortex calculated in section \ref{section:vortexmasspertn} is
actually considerably more important. This is reasonable since
when we assume a finite mass, the vortex can lose kinetic energy
by emitting a finite energy magnon.

Thus, with a finite $M_v$ this time,
\\
\begin{equation}
\gamma = 2\pi \int \frac{d^2k a^2}{(2\pi)^2} \frac{2\pi^2
Sq^2}{a^4kQ^3r_v^2} \left(\dot{\mathbf X} \cdot \hat{\mathbf
\chi}_{\mathbf k} \right)^2 \delta\left( \frac{(\mathbf P-\hbar
\mathbf k)^2}{2M_v} + \hbar ckQ- \frac{P^2}{2M_v} \right)
\end{equation}
\\
Note that the vector potential component of the vortex momentum
won't be included because in perturbation theory we assume that
the vortex momentum is changed by changing speed, not position.

Let $k_{\chi}$ be a solution of the delta function condition as a
function of emission angle, $\chi_{\mathbf k}$. The delta function
can then be rewritten
\\
\begin{equation}\label{2ndpertdelta}
\delta\left(\frac{\hbar^2}{2M_v} \left(k - \left(\frac{P}{\hbar}
\cos\chi_{\mathbf k} - \frac{cM_v}{r_v\hbar} \right)\right)^2 -
\frac{\hbar^2}{2M_v} \left(k_{\chi} - \left(\frac{P}{\hbar}
\cos\chi_{\mathbf k} - \frac{cM_v}{r_v\hbar}
\right)\right)^2\right)
\end{equation}
\\
where we've approximated $Q\approx 1/r_v$, a reasonable
approximation assuming small vortex velocities. Changing variables
within the delta function to express it as $\delta(k-k_{\chi})$,
the integral becomes
\\
\begin{equation}
\gamma = \frac{S\pi q^2r_v}{a^2} \int d^2k \frac{
\left(\dot{\mathbf X} \cdot \hat{\mathbf \chi}_{\mathbf k}
\right)^2}{k} \frac{M_v\delta(k-k_{\chi})}{\hbar^2 \left|k_{\chi}
- \left( \frac{P}{\hbar} \cos\chi_{\mathbf k} -
\frac{cM_v}{r_v\hbar} \right)\right|}
\end{equation}
\\
Substituting for $k_{\chi}$,
\\
\begin{align}
\gamma =& \frac{S\pi q^2r_v}{a^2} \int d^2k \frac{
\left(\dot{\mathbf X} \cdot \hat{\mathbf \chi}_{\mathbf
k}\right)^2}{k} \frac{M_v\delta(k-k_{\chi})}{\hbar^2
\left|\frac{P}{\hbar} \cos\chi_{\mathbf k} -
\frac{cM_v}{r_v\hbar}\right|} \nonumber \\
=& \frac{ S\pi q^2r_v M_v\dot X^2}{\hbar P a^2} \int
d\chi_{\mathbf k}
\frac{\sin^2\chi_{\mathbf k}}{\left|\cos\chi_{\mathbf k} - \frac{cM_v}{Pr_v}\right|}\\
=& \frac{2 \pi q^2S^2J}{\hbar} \left( \frac{\pi}{2} + \sqrt{\tilde
P^2-1} - \sin^{-1}\frac{1}{\tilde P} + \sqrt{\tilde P^2-1} \ln
\left( \frac{\hbar \pi r_v }{R_ScM_v} \frac{\tilde P}{\tilde
P^2-1} \right) \right) \nonumber
\end{align}
\\
where $\tilde P = \frac{Pr_v}{cM_v}$. To evaluate this last
integral, an infrared cut-off had to be imposed: $k_{min} =
\frac{2\pi}{R_S}$.

The discontinuity at $\tilde P = 1$ occurs when the vortex attains
the minimum energy to overcome the ``semi-gap'' formed by the $Q =
\sqrt{k^2+1/r_v^2}$ factor in the energy spectrum.

\begin{figure}
\centering
\includegraphics[width=10cm]{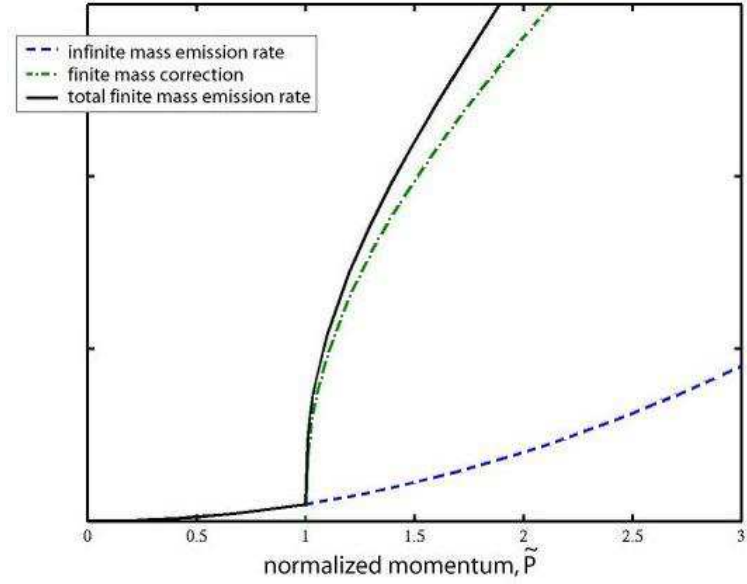}
\caption{The dissipation rate from perturbation theory; first
assuming infinite mass and then adding corrections due to finite
mass.}\label{figure:emissionrate}
\end{figure}

Note that this dissipation is in addition to that calculated in
the previous section. We didn't get both contributions here
because we left out the $k=0$ solution of the $\delta$-function
(\ref{2ndpertdelta}). See Figure \ref{figure:emissionrate} for a
plot of these two contributions.

\subsection{Zero point energy}\label{section:zptpertn}

Consider next the two magnon couplings, (\ref{secondorderterms}),
arising from expanding the Hamiltonian about a stable vortex.
Separate out the terms corresponding to the magnon Lagrangian
expanded about a vacuum solution
\\
\begin{equation*}
\frac{S}{a^2} \dot\varphi \; \vartheta - \frac{S^2J}{2}
\left((\mathbf\nabla\vartheta)^2 + (\mathbf\nabla\varphi)^2 +
\frac{1}{r_v^2} \vartheta^2 \right)
\end{equation*}
\\
and interpret those remaining as an interaction Hamiltonian
\\
\begin{align}
H_{int} =& S \int \frac{d^2r}{a^2}\Bigg( -\dot\varphi
(1-\sin\theta_v) \vartheta + \frac{c}{2}\Big( -\cos^2 \theta_v \;
(\mathbf \nabla\varphi)^2 + 2\sin 2\theta_v \; \mathbf
\nabla\phi_v\cdot\mathbf \nabla\varphi \; \vartheta \Big.\Bigg. \nonumber \\
&\Bigg.\Big. + \cos 2 \theta_v\; (\mathbf\nabla\phi_v)^2
\vartheta^2 - 2\frac{\cos^2\theta_v}{r_v^2} \vartheta^2 \Big)
\Bigg)
\end{align}
\\
The only long range interaction term in $H_{int}$ (i.e. that
doesn't decay exponentially) is the $\sin^2\theta_v$ portion of
the fourth term. Fourier transforming the $\vartheta$ factors
according to (\ref{fouriertransform}), this long range term
becomes
\\
\begin{align}
H_{int} =&- \frac{Sca^2}{2} \int d^2r (\mathbf\nabla\phi_v)^2 \int
\frac{d^2k}{(2\pi)^2} e^{-i\mathbf k \cdot \mathbf r} e^{-i\mathbf
k \cdot \mathbf X}\vartheta_{\mathbf k} \int \frac{d^2k}{(2\pi)^2}
e^{i\mathbf k' \cdot \mathbf r} e^{i\mathbf k' \cdot \mathbf
X}\vartheta_{-\mathbf k'}
\nonumber \\
=& - Sca^2q^2\pi \int dr \frac{J_0(| \mathbf k - \mathbf k'|r)}{r}
\int \frac{d^2k}{(2\pi)^2} \int \frac{d^2k'}{(2\pi)^2}
e^{-i(\mathbf k -\mathbf k')\cdot \mathbf X} \vartheta_{\mathbf k}
\vartheta_{-\mathbf k'}
\end{align}
\\
This integral over $r$ diverges in the short range. However, the
original term actually changes sign as $r\to 0$ so that the
analysis is invalid into the core region anyway and must be
cut-off. We define a form factor $\mathcal F(\mathbf \kappa) =
e^{-i\mathbf \kappa\cdot \mathbf{X}}\int \frac{J_0(\kappa r)}{r}
dr$ where $\mathbf \kappa = \mathbf k - \mathbf k'$.

Expressing this term in the language of quantized magnons (see
section \ref{section:magnons}) gives a Fourier transformed version
\\
\begin{align}
H_{int}^{(1)} =& -S^2Jq^2\pi  \int \frac{a^2d^2k}{(2\pi)^2}
\frac{a^2d^2\kappa}{(2\pi)^2}
\mathcal F(\mathbf \kappa) \vartheta_{\mathbf k+\mathbf
\kappa} \vartheta_{-\mathbf k} \nonumber \\
=& \frac{\hbar SJq^2 a^4\pi}{2}  \int \frac{d^2k}{(2\pi)^2}
\frac{d^2\kappa}{(2\pi)^2} \mathcal F(\mathbf \kappa)
\left(\frac{k\kappa}{Q\;\sqrt{\kappa^2
 + 1/r_v^2}}\right)^{\frac{1}{2}} \nonumber \\
&\times \left( a_{\mathbf k+\mathbf \kappa}^{\dagger}
-a_{-(\mathbf k+\mathbf \kappa)}\right)
\left(a_{-\mathbf k}^{\dagger} -a_{\mathbf k} \right)\\
=& \frac{\hbar SJq^2 a^4\pi}{2} \int \frac{d^2k}{(2\pi)^2}
\frac{d^2\kappa}{(2\pi)^2} \mathcal F(\mathbf \kappa)
\left(\frac{k\kappa}{Q\;\sqrt{
\kappa^2 + 1/r_v^2}}\right)^{\frac{1}{2}} \nonumber \\
&\times \left(a_{\mathbf k+\mathbf \kappa}^{\dagger} a_{-\mathbf
k}^{\dagger}+a_{-(\mathbf k+\mathbf \kappa)} a_{\mathbf k} -
a_{\mathbf k+\mathbf \kappa}^{\dagger}a_{\mathbf k} - a_{\mathbf
k}^{\dagger}a_{\mathbf k+\mathbf \kappa} - (2\pi)^2 \frac{\delta^2
(\kappa)}{a^2} \right)\nonumber
\end{align}
\\
The $a$ and $a^{\dagger}$ terms above correspond, respectively, to
the case of two magnons being created in opposite directions, two
magnons incoming from opposite directions being annihilated, a
magnon given a momentum boost of $\mathbf \kappa$, and the last
combination removes momentum $\mathbf \kappa$ from an existing
magnon. The last term gives the zero point energy shift
\\
\begin{align}
\Delta E =& -\frac{\hbar cq^2\pi}{2} \ln\frac{R_s}{r_v} \int
\frac{d^2k}{(2\pi)^2} \frac{k}{Q}\nonumber \\
=& -\frac{\hbar cq^2}{4} \ln\frac{R_s}{r_v} \int dk \frac{k^2}{Q}\\
=& -\frac{\hbar cq^2}{4} \ln\frac{R_s}{r_v} \left( \frac{{\sqrt
{r_v^2  + a^2 } }}{{2a^2 r_v }} - \frac{1}{{2r_v^2 }}\ln \left(
{\frac{{r_v  + \sqrt {r_v^2  + a^2 } }}{a}} \right)
\right)\nonumber
\end{align}
\\
as found before in section \ref{section:zpt}.

\section{Vortex influence functional}\label{section:vortexinfluence}

In this section, we develop the effective dynamics of the magnetic
vortex using path integration. The temperature is introduced by
assuming as an initial condition that the magnons are in thermal
equilibrium. They are of course allowed to evolve out of
equilibrium when interactions with the vortex are considered.

Populating the magnons at a temperature $\tau$, we have a density
matrix describing them given by equation (\ref{thermaleq}). To
describe the effective dynamics of the reduced density matrix for
the vortex, we trace out the magnon degrees of freedom from the
full vortex-magnon density matrix using the Feynman-Vernon
influence functional formalism \cite{feynman:1963}. Before
proceeding, consider the simple case of a central coordinate
$x(t)$ coupled to a bath of simple harmonic oscillators $r_i(t)$
with frequencies $\omega_i$. This introduces the influence
functional formalism and the interpretation of results for our own
magnetic system.

Separate the Lagrangian describing a coordinate $x(t)$ coupled
linearly to a set of harmonic oscillators $r_i$ as
\\
\begin{equation}
\mathcal L = \mathcal L_x[x(t)] + \mathcal L_{r}[r_i] + \mathcal
L_{int}[x(t),r_i(t)]
\end{equation}
\\
where $\mathcal L_x[x(t)]$ describes subsystem $x(t)$, $\mathcal
L_{r}[r_i]$ describes the environmental modes and $\mathcal
L_{int}[x(t),r_i(t)]$ describes the couplings between the two
systems.

Assume a general Lagrangian $\mathcal L_x[x(t)]$ for the central
coordinate, a simple harmonic Lagrangian in $r_i$
\begin{equation}
\mathcal L_{r}[r_i] = \sum_i \frac{1}{2}\dot r_i^2 +
\frac{\omega_i^2}{2} r_i^2
\end{equation}
and for the interacting Lagrangian, assume linear couplings
\begin{equation}
\mathcal L_{int}[x(t),r_i(t)] = \sum_i C_i x(t) r_i(t)
\end{equation}

Generally, the dynamics of the two subsystems become entangled
which is conveniently described within the density matrix
formalism.

The density matrix of the complete system in operator form evolves
from initial state $\rho(0)$ according to
\\
\begin{equation}
\rho(T) = \exp{-\frac{i H T}{\hbar}}\rho(0)\exp{\frac{i H
T}{\hbar}}
\end{equation}
\\
Alternatively, in the coordinate representation,
\\
\begin{align}
\rho(x, r_i, T; y, q_i, 0) =&
\langle x, r_i | \rho(T) | y, q_i \rangle \nonumber \\
=& \int dx' dy' dr_i' dq_i' \langle x, r_i |\exp{-\frac{i H
T}{\hbar}}|x', r_i'\rangle \\
& \times \langle x', r_i'|\rho(0)| y', q_i'\rangle \langle y',
q_i'|\exp\frac{i  H T}{\hbar}| y, q_i \rangle\nonumber
\end{align}
\\
Expanding each propagator as a path integral, noting
\\
\begin{align*}
\langle x, r_i |\exp{-\frac{i H T}{\hbar}}|x', r_i'\rangle =&
\int_{x'}^{x} \mathcal D[x(t)] \int_{r_i'}^{r_i} \mathcal
D[r_i(t)] \exp \frac{i}{\hbar} \mathcal S[x(t),r_i(t)] \\
\langle x, r_i |\exp{-\frac{i H T}{\hbar}}|x', r_i'\rangle =&
\int_{y'}^{y} \mathcal D[y(t)] \int_{q_i'}^{q_i} \mathcal
D[q_i(t)] \exp -\frac{i}{\hbar} \mathcal S[y(t),q_i(t)]
\end{align*}
\\
the density matrix at time $T$ becomes
\\
\begin{align*}
\rho(x, r_i; y, q_i; T) &= \int dx' dy' dr_i' dq_i' \int_{x'}^{x}
\mathcal D[x(t)] \int_{r_i'}^{r_i} \mathcal D[r_i(t)] \exp
\frac{i}{\hbar} \mathcal
S[x(t),r_i(t)] \\
& \times \langle x', r_i'|\rho(0)| y', q_i'\rangle \int_{y'}^{y}
\mathcal D[y(t)] \int_{q_i'}^{q_i} \mathcal D[q_i(t)] \exp
-\frac{i}{\hbar} \mathcal S[y(t),q_i(t)]\nonumber
\end{align*}
\\
However, suppose we're only interested in the dynamics of the
subsystem $x(t)$, regardless of the specific behaviour of the
harmonic oscillator subsystems. To eliminate these variables,
perform the trace over the $\{r_i\}$ variables to obtain the
so-called reduced density operator
\\
\begin{align*}
\tilde\rho(x; y; T) =& \int dr_i \int dx' dy' dr_i' dq_i'
\int_{x'}^{x} \mathcal D[x(t)] \int_{r_i'}^{r_i} \mathcal
D[r_i(t)] \exp \frac{i}{\hbar} \mathcal
S[x(t),r_i(t)] \\
& \times \langle x', r_i'|\rho(0)| y', q_i'\rangle \int_{y'}^{y}
\mathcal D[y(t)] \int_{q_i'}^{r_i} \mathcal D[q_i(t)] \exp
-\frac{i}{\hbar} \mathcal S[y(t),q_i(t)]
\end{align*}
\\
Assume that the $t=0$ density matrix is separable in the two
subsystems, i.e. that they are initially disentangled and
\\
\begin{equation}
\rho(x, r_i; y, q_i; 0) = \rho_x(x, y; 0) \rho_r(r_i,q_i; 0)
\end{equation}
\\
Further, assume that the simple harmonic oscillators are initially
in thermal equilibrium so that $\rho_r(r_i,q_i; t=0)$ is given
by\cite{feynman:1965}
\\
\begin{equation*}
\rho_r(r_i,q_i; 0) = \prod_i \sqrt{\frac{m\omega_i}{2\pi\hbar
\sinh\hbar\omega_i\beta}}\exp
-\frac{m\omega}{2\hbar\sinh\hbar\omega_i\beta} \left(
(r_i^2+q_i^2)\cosh \hbar\omega_i\beta - 2r_iq_i \right)
\end{equation*}

The reduced density matrix is then expressible as
\\
\begin{equation}
\tilde\rho(x; y; t=T) = \int dx' \int dy' J(x, y, T; x', y', 0)
\rho_x(x', y', 0)
\end{equation}
\\
where
\\
\begin{equation}
J(x, y, T; x', y', 0) = \int_{x'}^{x} \mathcal D[x(t)]
\int_{y'}^{y} \mathcal D[y(t)] \exp \frac{i}{\hbar}(\mathcal
S_x[x(t)]-\mathcal S_x[y(t)]) \mathcal F[x(t),y(t)]
\end{equation}
\\
is the propagator for the density operator and
\\
\begin{align}
F[x(t),y(t)] &= \int dr_i dr_i' dq_i' \rho_r(r_i',q_i',0)
\int_{r_i'}^{r_i} \mathcal D[r_i(t)] \int_{q_i'}^{r_i} \mathcal
D[q_i(t)] \\
& \times \exp \frac{i}{\hbar} \left(S_r[r_i(t)] + S_{int}
[r_i(t),x(t)] - S_r[q_i(t)] - S_{int} [q_i(t),y(t)]\right)
\nonumber
\end{align}
\\
is the influence functional \cite{feynman:1963}. Evaluating this
for the central coordinate $x(t)$ coupled linearly to a set of
environmental modes described by simple harmonic oscillators with
spectrum $\omega_i(t)$
\\
\begin{align}
F[x,y] = \exp -\frac{1}{\hbar} \int_0^T dt \int_0^t ds
\left(x(t)-y(t)\right) \left(\alpha(t-s)x(s) -
\alpha^*(t-s)y(s)\right)
\end{align}
\\
where
\begin{equation}
\alpha(t-s) =\sum_i \frac{C_i^2}{2m\omega_i}\left(\exp
-i\omega_i(t-s) + \frac{2\cos\omega_i(t-s)}{\exp\hbar
\omega_i\beta -1} \right)
\end{equation}
\\
where $C_i$ are the linear coupling parameters.

\subsection{Quantum Brownian motion}

Caldeira and Leggett \cite{leggett:1983} interpret the influence
functional result as quantized damped dynamics. The problem of
quantizing Brownian motion was not entirely understood. Their idea
of coupling to a bath of oscillators to achieve Brownian motion
(which, of course, from there is easily quantizable) was one of
many proposed in the 1980's and 90's. The classical equation of
motion for Brownian motion, the Langevin equation, is
\\
\begin{equation}\label{langevin}
m\ddot x + \eta \dot x + V'(x) = F(t)
\end{equation}
\\
where $m$ is the mass of the particle, $\eta$ is a damping
constant, $V(x)$ is the potential acting on the particle and
$F(t)$ is the fluctuating force. This force obeys
\\
\begin{align}\label{classforcecorr}
\langle F(t) \rangle =& 0 \nonumber \\
\langle F(t)F(t') \rangle =& 2\eta kT\delta(t-t')
\end{align}
\\
where $\langle \; \rangle$ denote statistical averaging.

With such a force, the propagator of the density matrix of system
$x$ is given by
\\
\begin{equation*}
J(x,y,t;x',y',0) = \int \mathcal D[x] \mathcal D[y] exp
\frac{i}{\hbar} \left(\mathcal S[x]-\mathcal S[y]+\int_0^t d\tau
(x(\tau)-y(\tau))F(\tau)\right)
\end{equation*}

Assuming that the fluctuating force $F(t)$ has the probability
distribution functional $P[F(\tau)]$ of different histories
$F(\tau)$, the averaged density matrix propagator becomes
\\
\begin{align}
J(x,y,t;x',y',0) = \int \mathcal D[x] \mathcal D[y] \mathcal
D[F]\;P[F(\tau)] & \exp \frac{i}{\hbar} \left(\mathcal
S[x]-\mathcal S[y] \right.\\
&\left.+\int_0^t d\tau (x(\tau)-y(\tau))F(\tau)\right) \nonumber
\end{align}
\\
We can perform the path integration over $F(\tau)$ if we assume
$P[F(\tau)]$ is a Gaussian distribution, yielding
\\
\begin{align}
J(x,y,t;x',y',0) = &\int  \mathcal D[x] \mathcal D[y] exp
\frac{i}{\hbar} \left(\mathcal S[x]-\mathcal S[y]\right) \\
& \times \exp -\frac{1}{\hbar^2} \int_0^t\int_0^{\tau} d\tau ds
(x(\tau)-y(\tau))A(\tau-s)(x(s)-y(s)) \nonumber
\end{align}
\\
where $A(\tau-s)$ is the correlation of forces, $\langle
F(\tau)F(s) \rangle$.

The real exponentiated term in the influence functional is
\\
\begin{equation}\label{FVrealpart}
\exp -\frac{1}{\hbar}\int_0^t\int_0^{\tau} d\tau ds
(x(\tau)-y(\tau))\alpha_R(\tau-s)(x(s)-y(s))
\end{equation}
\\
where
\\
\begin{equation}
\alpha_R(\tau-s) = \sum_i \frac{C_i^2}{2m\omega_i}
\coth\frac{\hbar\omega_i}{2k_BT}\cos\omega_i(\tau-s)
\end{equation}
\\
where $C_i$ denotes the coupling coefficient to the $i$th
environmental mode. Assuming instead a continuum of $k$ states
with density
\\
\begin{equation}
\rho_D(\omega)C^2(\omega) = \left\{%
\begin{array}{ll}
    \frac{2m\eta\omega^2}{\pi}, & \hbox{$\omega < \Omega$;} \\
    0, & \hbox{$\omega > \Omega$.} \\
\end{array}%
\right.
\end{equation}
\\
the influence functional result becomes in a high temperature
limit ($\coth\frac{\hbar\omega}{2kT} \to \frac{2kT}{\hbar\omega}$)
\\
\begin{equation}
\hbar \alpha_R(\tau-s) =\langle F(\tau)F(s) \rangle=2\eta kT
\frac{\sin\Omega(\tau-s)}{\pi(\tau-s)}
\end{equation}
\\
which tends to (\ref{classforcecorr}) in the limit $\Omega \to
\infty$.

The imaginary phase term in the influence functional is
manipulated to give an $x^2$ frequency shift which renormalizes
the external potential. In addition to this, there is a new action
term corresponding to a damping force
\\
\begin{equation}
\Delta \mathcal S = -\int_0^t dt M\gamma(x\dot x-y\dot y + x\dot
y- y\dot x)
\end{equation}
\\
Note that the forward and backward paths are interacting so that
the new effective action is coupled in $x(t)$ and $y(t)$.

The relaxation constant $\gamma$ is
\begin{equation}
\gamma=\frac{\eta}{2M}
\end{equation}
where the damping constant $\eta$ is dependent on the density of
states of the environmental modes. For our treatment where the
environmental modes are magnons, we know explicitly the magnon
density of states, going as $\frac{Q}{k^2+Q^2}$ (recall
$\omega(k)=ckQ$) rather than $\omega^2$ as assumed above, so that
our analysis does not simplify to a frequency independent damping
function.

Castro Neto and Caldeira \cite{caldeira:1993} consider the problem
of a central coordinate coupled linearly to a set of oscillators;
however, as opposed to the \citet{leggett:1983} problem, the
central system, $X(t)$, is a solution in the same medium as the
set of oscillators. Hence, as in our problem, there is no linear
coupling with position, but instead, we find a linear coupling
$C_i r_i \dot X(t)$ between oscillators $\{r_i\}$ and the
velocity.

They simplify their results by assuming a Born approximation.
Although they lose the resulting frequency dependent motion, they
do find that the damping coefficients and correlation integrals
now possess memory effects. We will discuss these issues after
results have been simplified for our vortex-magnon system.

\subsection{Semiclassical solution of perturbed
magnons}\label{section:pertprop}

Before evaluating the influence functional, we first need the
propagator of the magnon system perturbed by the vortex presence.
The effect of important perturbing terms have been discussed
already using perturbation theory. The one magnon coupling endows
the vortex with an effective mass and makes the vortex motion
dissipative by radiating magnons. The leading two magnon coupling
provides an overall zero-point energy shift to the vortex-magnon
system that is associated to the quantized vortex. Although the
two magnon couplings, or indeed any of the many magnon couplings,
may give more significant dissipation, we neglect these
contributions in this treatment. In the influence functional, the
forward and backward paths have cancelling zero point energy
shifts and hence we will ignore entirely the many magnon
couplings.

Treat the disturbance of a magnetic vortex centered at $\mathbf
X(t)$ with vorticity $q$ and polarization $p$ by the magnons via
the one magnon coupling, (\ref{firstordercoupling}),
\\
\begin{equation}
\mathcal L_{int} = -S \int \frac{d^2r}{a^2} \dot{\mathbf X} \cdot
\mathbf \nabla \phi_v \sin\theta_v \vartheta
\end{equation}
\\
We must evaluate the propagator for the system of magnons, again
in the $\varphi$ basis
\\
\begin{equation}
\langle\varphi | \exp{-\frac{i \mathcal H t}{\hbar}} |
\varphi'\rangle = \int_{\varphi'}^{\varphi} \mathcal D[\varphi]
\mathcal D[\vartheta] \exp \frac{i}{\hbar}\int_0^T dt (\mathcal
L_m + \mathcal L_{int})
\end{equation}
\\
where $\mathbf X(t)$ is considered now an externally controlled
parameter.

Introduce the plane wave decomposition (\ref{fouriertransform}) so
that the action becomes
\\
\begin{align}
\mathcal S_{m+int}[\varphi,\vartheta] = S \int \frac{a^2
d^2k}{(2\pi)^2} \int_0^T dt & \left(  \dot \varphi_{\mathbf k}
\vartheta_{-\mathbf k} - \frac{c}{2} \left( k^2 \varphi_{\mathbf
k} \varphi_{-\mathbf k} + Q^2 \vartheta_{\mathbf k}
\vartheta_{-\mathbf k} \right)
\right. \nonumber \\
& \left. - \int \frac{d^2r}{a^2} e^{ -i \mathbf k \cdot \mathbf r}
\dot{\mathbf X} \cdot \mathbf \nabla \phi_v \sin\theta_v
\vartheta_{\mathbf k} \right)
\end{align}
\\
The equations of motion are modified by a force term, that, for
simplicity, we denote as
\\
\begin{align}
f_{\mathbf k}[\mathbf X] =& \int \frac{d^2r}{a^2} e^{ i \mathbf k
\cdot \mathbf r} \dot{\mathbf X} \cdot \mathbf \nabla
\phi_v \sin\theta_v \nonumber \\
=& \frac{2i\pi q}{a^2} \frac{\dot{\mathbf X} \cdot \hat
\varphi_{\mathbf k}}{kQr_v} e^{ i \mathbf k \cdot \mathbf X}
\end{align}
\\
and become
\\
\begin{equation}
\left(%
\begin{array}{cc}
  ck^2 & \frac{\partial}{\partial t} \\
  -\frac{\partial}{\partial t} & cQ^2 \\
\end{array}%
\right)
\left(%
\begin{array}{c}
  \varphi_{\mathbf k}^{cl} \\
  \vartheta_{\mathbf k}^{cl} \\
\end{array}%
\right) = \left(%
\begin{array}{c}
  0 \\
  -f_{\mathbf k}[\mathbf X] \\
\end{array}%
\right)
\end{equation}
\\
The solution with boundary conditions $\varphi_{\mathbf k}(0) =
\varphi_{\mathbf k}'$ and $\varphi_{\mathbf k}(T) =
\varphi_{\mathbf k}$ is
\\
\begin{align}
\left(%
\begin{array}{c}
  \varphi_{\mathbf k}^{cl} \\
  \vartheta_{\mathbf k}^{cl} \\
\end{array}%
\right) =& \frac{\varphi_{\mathbf k} - \int_t^T ds \cos\omega_k
(T-s) f_{\mathbf k}[\mathbf X]}{\sin \omega_k T}
\left(%
\begin{array}{c}
  \sin\omega_k t \\
  \frac{k}{Q} \cos\omega_k t \\
\end{array}%
\right) \nonumber \\
&+ \frac{\varphi_{\mathbf k}' + \int_0^t ds
  \cos\omega_k s f_{\mathbf k}[\mathbf X]}{\sin \omega_k T}
\left(%
\begin{array}{c}
  \sin\omega_k (T-t) \\
  -\frac{k}{Q} \cos\omega_k (T-t) \\
\end{array}%
\right)
\end{align}

Substituting this solution into the action gives the classical
contribution
\\
\begin{align}
\mathcal S_{m+int}^{cl}(k) =& \frac{Sk}{2Q\sin\omega_k T} \left(
\left(\varphi_{\mathbf k}'\varphi_{-\mathbf k}' + \varphi_{\mathbf
k} \varphi_{-\mathbf k} \right) \cos\omega_kT - 2 \varphi_{\mathbf k}
\varphi_{-\mathbf k}' \right. \\
&- 2\varphi_{\mathbf k} \int_0^T dt \cos\omega_k t f_{-\mathbf
k}[\mathbf X(t)] + 2\varphi_{\mathbf k}'\int_0^T dt
\cos\omega_k(T-t)f_{-\mathbf k}[\mathbf X(t)] \nonumber \\
& \left. +2\int_0^T dt \int_0^t ds \cos \omega_k (T-t) \cos
\omega_k s f_{\mathbf k}[\mathbf X(t)] f_{-\mathbf k}[\mathbf
X(s)] \right) \nonumber
\end{align}
\\
The quantum fluctuations introduce a pre-factor given by solving
the relevant Jacobi equation
\begin{align}
\left(%
\begin{array}{cc}
  ck^2 & \frac{\partial}{\partial t} \\
  -\frac{\partial}{\partial t} & cQ^2 \\
\end{array}%
\right)
\left(%
\begin{array}{c}
  \varphi(t) \\
  \vartheta(t) \\
\end{array}%
\right) = 0
\end{align}
with initial conditions $\varphi_k(0) = 0$ and $S\vartheta(0)=1$.
The determinant is given as $i\varphi(T)$. Combining the
pre-factor $(\det)^{-1}$ with path integration measure factors,
give the overall result, (\ref{prefac}),
\begin{equation}
\sqrt{\frac{S k}{2 \pi i \hbar Q \sin \omega_kT}}
\end{equation}

The final propagator is
\\
\begin{equation}
\langle\varphi | \exp{-\frac{i \mathcal (H_{m+int}) t}{\hbar}} |
\varphi'\rangle = \prod_{\mathbf k} \sqrt{\frac{S k}{2 \pi i \hbar
Q \sin \omega_kT}} \exp \frac{i}{\hbar} \int \frac{a^2
d^2k}{(2\pi)^2} \mathcal S_{m+int}^{cl}(k)
\end{equation}

\subsection{Evaluating the influence functional}

Substituting the semiclassical solutions to the two path integrals
and for the thermal equilibrium density matrix, the problem is
reduced to three regular gaussian integrals, ignoring pre-factors,
which cancel anyway after all integrals when the density matrix is
properly normalized,
\\
\begin{align}
F[\mathbf X,\mathbf Y] &= \prod_{\mathbf k} \int d\varphi_{\mathbf
k} d\varphi_{\mathbf k}' d\tilde\varphi_{\mathbf k}' \\
& \exp \left[ -\frac{Sk}{2\hbar Q \sinh \hbar \omega_k \beta}
\Big(\left(\varphi_{\mathbf k}'\varphi_{-\mathbf k}' +\tilde
\varphi_{\mathbf k}'\tilde \varphi_{-\mathbf k}' \right) \cosh
\hbar \omega_k \beta - 2 \tilde \varphi_{\mathbf k}'
\varphi_{-\mathbf k}'   \Big)\right]
\nonumber \\
& \exp \bigg[ \frac{iSk}{2\hbar Q\sin\omega_k T} \Big(
\left(\varphi_{\mathbf k}'\varphi_{-\mathbf k}' + \varphi_{\mathbf
k} \varphi_{-\mathbf k} \right) \cos\omega_kT - 2 \varphi_{\mathbf
k} \varphi_{-\mathbf k}' \Big.\bigg.\nonumber \\
&- 2\varphi_{\mathbf k} \int_0^T dt \cos\omega_k t f_{-\mathbf
k}[\mathbf X(t)] + 2\varphi_{\mathbf k}'\int_0^T dt
\cos\omega_k (T-t)f_{-\mathbf k}[\mathbf X(t)] \nonumber \\
& \bigg.\Big. +2\int_0^T dt \int_0^t ds \cos \omega_k (T-t) \cos
\omega_k s f_{\mathbf k}[\mathbf X(t)] f_{-\mathbf k}[\mathbf
X(s)] \Big) \bigg]
\nonumber \\
& \exp \bigg[ -\frac{iSk}{2\hbar Q\sin\omega_k T} \Big(
\left(\tilde\varphi_{\mathbf k}'\tilde\varphi_{-\mathbf k}' +
\varphi_{\mathbf k}\varphi_{-\mathbf k} \right)\cos\omega_{\mathbf
k} T - 2\tilde\varphi_{\mathbf k}'\varphi_{-\mathbf k}\Big.
\bigg.\nonumber \\
&- 2\varphi_{\mathbf k} \int_0^T dt \cos\omega_k t f_{-\mathbf
k}[\mathbf Y(t)]  + 2\tilde\varphi_{\mathbf k}'\int_0^T dt
\cos\omega_{\mathbf k} (T-t)f_{-\mathbf
k}[\mathbf Y(t)] \nonumber \\
&  \bigg.\Big.+2\int_0^T dt \int_0^t ds \cos \omega_k (T-t) \cos
\omega_k s f_{\mathbf k}[\mathbf Y(t)] f_{-\mathbf k}[\mathbf
Y(s)] \Big) \bigg] \nonumber
\end{align}
\\
where the $\prod_{\mathbf k}$ applies to everything (and hence
implies integrals over $\mathbf k$ within exponentials).

Performing these integrals mimics very closely the calculations
for the analogous problem of a central coordinate $x(t)$ coupled
linearly to the position coordinate of a system of simple harmonic
oscillators \cite{feynman:1963}. In fact, the final expression is
the same, with the same substitution $m\omega \rightarrow Sk/Q$
found earlier in evaluating the magnon propagator (see section
\ref{section:magprop}),
\\
\begin{align}
F[\mathbf X,\mathbf Y] = \exp -\frac{1}{\hbar} \int \frac{a^2
 d^2k}{(2\pi)^2} &\int_0^T dt \int_0^t ds
\left(f_{\mathbf k}[\mathbf X(t)]-f_{\mathbf k}[\mathbf
Y(t)]\right)\\
&\left(\alpha_k(t-s)f_{-\mathbf k}[\mathbf X(s)] -
\alpha_k^*(t-s)f_{-\mathbf k}[\mathbf Y(s)]\right) \nonumber
\end{align}
\\
where
\\
\begin{equation}
\alpha_k(t-s) =  \frac{Sk}{2Q} \left(e^{-i\omega_k (t-s)} +
\frac{2\cos\omega_k (t-s)}{e^{\hbar \omega_k \beta} - 1} \right)
\end{equation}

The propagator of the density operator can be written as
\\
\begin{align}\label{propdensity}
J(\mathbf X, \mathbf Y, T; \mathbf X', &\mathbf Y', 0) =
\int_{\mathbf X}^{\mathbf X'} \mathcal D\mathbf X \int_{\mathbf
Y}^{\mathbf Y'} \mathcal D\mathbf X \exp \frac{i}{\hbar}
\Bigg[\mathcal S_v[\mathbf X]-\mathcal S_v[\mathbf Y] -  \\
 \int \frac{a^2 d^2k}{(2\pi)^2} \int_0^T dt &\int_0^t
ds \bigg( \left(f_{\mathbf k}[\mathbf X(t)] - f_{\mathbf
k}[\mathbf Y(t)]\right) \alpha^I_k(t-s)\left(f_{-\mathbf
k}[\mathbf
X(s)] + f_{-\mathbf k}[\mathbf Y(s)]\right)  \nonumber\\
 &-i  \left(f_{\mathbf k}[\mathbf X(t)]-f_{\mathbf k}[\mathbf
Y(t)]\right) \alpha^R_k(t-s) \left(f_{-\mathbf k}[\mathbf X(s)] -
f_{-\mathbf k}[\mathbf Y(s)]\right)\bigg)\Bigg]\nonumber
\end{align}
\\
where $\alpha^R_k(t-s)$ and $\alpha^I_k(t-s)$ are the real and
imaginary parts of $\alpha_k(t-s)$
\\
\begin{align}
\alpha^R_k(t-s) =& \frac{Sk}{2Q} \cos\omega_k (t-s) \coth
\frac{\hbar
\omega_k \beta}{2} \nonumber \\
\alpha^I_k(t-s) =&- \frac{Sk}{2Q} \sin \omega_k (t-s)
\end{align}

Ordinarily, we would extract a spectral function $\mathcal
J(\omega,T)$ from this result that gives the frequency waiting of
the functions $\alpha$. For example, the results of Caldeira and
Leggett can be re-expressed as
\begin{align}\label{spectralfunc}
\alpha^R_k(t-s) =& \int \frac{d\omega}{\pi} \mathcal J(\omega,T)
\cos\omega (t-s)
\coth \frac{\hbar \omega \beta}{2} \nonumber \\
\alpha^I_k(t-s) =& -\int \frac{d\omega}{\pi} \mathcal J(\omega,T)
\sin \omega (t-s)
\end{align}
In our case, however, we must first integrate over the angular
dependence of $\mathbf k$. This, however, gives the sum of two
terms with Bessel function factors of order 0 and 2, themselves
dependent on the wavenumber $k$ and the coordinate path $\mathbf
X(t)$. In order to define a spectral function, we would have to
disentangle the $t-s$ and $k$ behaviours, which, with the
additional $J_i(k|\mathbf X(t)- \mathbf X(s)|$ factors is rather
involved.

\subsection{Interpreting the imaginary
part}\label{section:imagpart}

The one magnon coupling treated perturbatively endows the vortex
with an effective mass and introduces dissipation. In the
influence formalism, we expect to obtain terms in the effective
action of the forward/backward paths interpretable as
particle-like inertial terms. Dissipation arises due to
fluctuating forces inflicted by scattered magnons on the vortex.
We expect the fluctuating forces to be accompanied by
corresponding damping forces.

Our one magnon term couples to the vortex velocity and not
position as treated by Caldeira and Leggett. This is because the
vortex is a solution itself of the system, so that all first order
variations vanish. The velocity term survives because the vortex
is to zeroth order a stationary solution. The potential
renormalization found earlier going like $x^2$ should here appear
as a shift $\sim \dot X(t)^2$, or an inertial term from which we
can deduce an effective vortex mass.

Substituting for $f_{\mathbf k}$ into the imaginary term yields
the phase, including the additional minus sign in
(\ref{propdensity}),
\\
\begin{align}
\Phi=-\frac{Sq^2}{2 a^2} & \int d^2k \int_0^T dt \int_0^t ds
\left(\dot{\mathbf X}(t)e^{ i \mathbf k \cdot \mathbf X(t)} -
\dot{\mathbf Y}(t)e^{ i \mathbf k \cdot \mathbf Y(t)}\right)\cdot
\hat \varphi_{\mathbf k} \frac{\sin
\omega_k(t-s)}{kQ^3r_v^2} \nonumber \\
& \left(\dot{\mathbf X}(s) e^{-i \mathbf k \cdot \mathbf X(s)} +
\dot{\mathbf Y}(s)e^{-i \mathbf k \cdot \mathbf Y(s)}\right)\cdot
\hat \varphi_{\mathbf k}
\end{align}
where we define the phase angles via
\begin{equation}
F = \exp \left(\frac{-i}{\hbar} (\Phi-i\Gamma)\right)
\end{equation}

Performing first the integral over $\chi_k$ from $0$ to $\pi$
(refer to an identical calculation in the perturbation calculation
of multi-vortex mass corrections in section
\ref{section:manyvortexmass}) to leading order yields, for the
$X^2$ term only for conciseness (the other factors have the same
form),
\begin{align}
\Phi=-\frac{\pi Sq^2}{2 a^2} & \int dk \int_0^T dt \int_0^t ds
\frac{\sin \omega_k(t-s)}{Q^3r_v^2} \Bigg( \dot{\mathbf X}(t)\cdot
\dot{\mathbf X}(s)
J_0(k|\mathbf X(t)-\mathbf X(s)|) \Bigg. \nonumber \\
&  + \dot X(t)\dot X(s)\left( (\hat{\mathbf X}(t)\cdot\hat{\mathbf
e}_{\Delta})
(\hat{\mathbf X}(s)\cdot\hat{\mathbf e}_{\Delta}) \right. \\
& \Bigg. \left. -(\hat{\mathbf X}(t)\times\hat{\mathbf
e}_{\Delta})\cdot (\hat{\mathbf X}(s)\times\hat{\mathbf
e}_{\Delta}) \right)J_2(k|\mathbf X(t)-\mathbf X(s)|) \Bigg)
\mbox{ + etc.} \nonumber
\end{align}
where $\hat{\mathbf e}_{\Delta}$ denotes the unit vector
connecting $\mathbf X(t)$ and $\mathbf X(s)$.

Integrate by parts in $t-s$ to get two terms, one with two time
derivatives in $X(s)$ and another with a single time derivative in
$X(s)$. Note, we ignore the derivatives of the Bessel functions
since the extra factor of $k$ makes these higher order corrections
and we assume the vortex curves slowly to ignore derivatives of
the unit vectors. The boundary terms from the integration by parts
are zero for $t=s$ and otherwise unimportant (they don't
contribute to the equations of motion, being just boundary
dependant). Finally, we have
\\
\begin{align}
& \Phi=- \frac{Sq^2\pi}{2 a^2}  \int dk\;k dt ds
\left(\frac{\sin\omega_0(t-s)}{kQ^3r_v^2} \ddot{\mathbf X}(s) -
\frac{c\cos\omega_0(t-s)}{Q^2r_v^2}
\dot{\mathbf X}(s)\right) \cdot \mathbf X(t)\nonumber \\
&  \times J_0(k|\mathbf X(t) - \mathbf X(s)|) +
\left(\frac{\sin\omega_0(t-s)}{kQ^3r_v^2} \ddot{X}(s) -
\frac{c\cos\omega_0(t-s)}{Q^2r_v^2}
\dot{X}(s)\right) \\
& \times \left[ \hat{\mathbf X}(s) \cdot \Big( \hat{\mathbf
e}_{\Delta} (\mathbf X(t) \cdot\hat{\mathbf e}_{\Delta}) -
\hat{\mathbf e}_{\bot} (\mathbf X(t) \cdot\hat{\mathbf e}_{\bot})
\Big) \right] J_2(k|\mathbf X(t) - \mathbf X(s)|) \mbox{ + etc.}
\nonumber
\end{align}
where $\hat{\mathbf e}_{\bot}$ is a unit vector perpendicular to
$\hat{\mathbf e}_{\Delta}$.

Split this integral into the sine and cosine portions, $\Phi =
\Phi_s + \Phi_c$. Consider first the sine integrals. Integrate by
parts again in $t-s$. The non-zero boundary terms are
\\
\begin{align}
\Phi_s^{BC}=& \frac{Sq^2\pi}{2 a^2} \int dk \int_0^T dt
\frac{1}{ckQ^4r_v^2} \left( \ddot{\mathbf X}(t)\cdot\mathbf
X(t) - \ddot{\mathbf Y}(t)\cdot\mathbf Y(t) \right) \nonumber \\
&= \frac{Sq^2\pi r_v^2}{2 a^2 c} \int_0^T dt \left( \ddot{\mathbf
X}(t)\cdot\mathbf X(t) - \ddot{\mathbf Y}(t)\cdot\mathbf Y(t)
\right)\ln \frac{R_S}{\sqrt{a^2+r_v^2}}
\end{align}
\\
or equivalently,
\begin{equation}
\Phi_s^{BC}=-\frac{Sq^2\pi r_v^2}{2 a^2 c} \int_0^T dt \left(
\dot{\mathbf X}^2(t) - \dot{\mathbf Y}^2(t) \right) \ln
\frac{R_S}{\sqrt{a^2+r_v^2}}
\end{equation}
where we've again split the sine integral into $\Phi_s =
\Phi_s^{BC} + \Phi_s^{int}$.

These provide an inertial mass term to the effective action of
each the forward and backward paths.

The remaining terms, ignoring Bessel function derivatives as
before,
\\
\begin{align}
 \Phi_s^{int}=&-\frac{Sq^2\pi}{2 a^2}  \int dk\;k \int_0^T dt
\int_0^t ds  \frac{\cos ckQ(t-s)}{ckQ^4r_v^2} \ddot{\mathbf
X}(s)\cdot \dot{\mathbf X}(t) \nonumber \\
&J_0(k|\mathbf X(t) - \mathbf X(s)|) \mbox{ + etc.} \nonumber \\
\approx & -\frac{2Sq^2\pi^2 r_v^2}{2 a^2 c} \int_0^T dt \int_0^t
ds  \frac{\cos ckQ(t-s)}{ckQ^4r_v^2} \ddot{\mathbf
X}(s)\cdot \dot{\mathbf X}(t) \nonumber \\
&J_0(k|\mathbf X(t) -\mathbf X(s)|) \mbox{ + etc.}
\end{align}
\\
This is much smaller than the log divergent boundary term. Note
that we've neglected the smaller still $J_2$ terms. By varying the
$\ddot X \dot X$ terms with respect to $X$, we would obtain a
small third order time derivative term, $\dddot X$, in the
equations of motion. In an attempt to explain their numerical
simulation results, Mertens et.
al.\cite{mertens:1997,mertens:1997,volkel:1994} artificially
introduce a third order term by expanding the energy functional
assuming both position and velocity as collective coordinates.
This, of course, is a misapplication of the collective coordinate
formalism, where a collective coordinate is meant to replace a
continuous symmetry that the soliton breaks. The freedom they
introduced by assuming velocity as a collective variable is not
actually available in the original problem.

Consider the cosine term next. We can re-express this damping in
terms of various damping functions
\\
\begin{align}\label{phidamping}
\Phi_c=\int_0^T dt \int_0^t ds & \Big( \gamma_{||}(t-s, |\mathbf
X(t) - \mathbf X(s)|)\dot{\mathbf X}(s)\cdot\mathbf
X(t) \Big.\\
&+ \gamma_{\Delta}(t-s, |\mathbf X(t) - \mathbf X(s)|) \dot{X}(s)
\hat{\mathbf X}(s) \cdot \hat{\mathbf e}_{\Delta} (\mathbf X(t)
\cdot\hat{\mathbf e}_{\Delta}) \nonumber \\
& \Big. + \gamma_{\bot}(t-s, |\mathbf X(t) - \mathbf X(s)|)
\dot{X}(s)\hat{\mathbf X}(s)\cdot \hat{\mathbf e}_{\bot} (\mathbf
X(t) \cdot\hat{\mathbf e}_{\bot}) \Big) \nonumber \mbox{ + etc.}
\end{align}
\\
where
\\
\begin{align}\label{gammafull}
\gamma_{||}(t-s, \Delta)=& \frac{S^2J\pi q^2}{2}\int dk \;k
\frac{\cos\omega_0(t-s)J_0(k\Delta)}{Q^2r_v^2} \nonumber \\
\gamma_{\Delta}(t-s,\Delta)=& \frac{S^2J\pi q^2}{2}\int dk
\;k \frac{\cos\omega_0(t-s)J_2(k\Delta)}{Q^2r_v^2}\nonumber\\
\gamma_{\bot}(t-s,\Delta)=&-\gamma_{\Delta}(t-s,\Delta)
\end{align}
Note we cannot perform the $k$ integrals in analytic form due to
the $\omega_0=ckQ$ argument in the cosine.

The damping forces depend on the previous motion of the vortex.
These memory effects appear as averages over Bessel
functions---this form is because the vortex exists in a 2D system.
The first damping term is of the regular form, that is, a force
acting in the opposite direction to the particle velocity. The
next damping, $\gamma_{\Delta}$, is the same as the first if the
vortex travels in a straight line, however, for a curved path, is
dependent on its change in direction. The last damping,
$\gamma_{\bot}$, contributes damping perpendicular to the
$\gamma_{\Delta}$ damping, which, in the case of a slowly curving
path, is transverse to the vortex motion.

Comparing with the dissipation results of \citet{sloncz:1984},
although we find frequency dependent dissipation (via the $kX(t)$
coupling in the Bessel functions), we do not see any of the same
small frequency behaviour predicted by Slonczewski. Likely, his
treatment considers a different source of dissipation than the
contribution considered here. As noted in section
\ref{section:vortmagdiss}, this dissipation arises due to the same
scattering processes that yield an inertial energy. In
Slonczewski's treatment, on the other hand, his inertial energy
calculation is for intermediate distance magnon scattering, while
his dissipation arises from far field scattering.

\subsection{Interpreting the real part}

In the paper of Caldeira and Leggett \cite{leggett:1983}, the real
part of the influence functional is interpreted as the correlation
of forces in the classical regime. The real phase of their
influence functional is
\\
\begin{equation}
\Gamma=\sum_k \frac{C_k^2}{2m\omega_k} \coth
\frac{\hbar\omega_k\beta}{2} \int_0^T dt \int_0^t ds
\left(x(t)-y(t)\right) \cos\omega_k(t-s) \left(x(s)-y(s)\right)
\end{equation}
\\
which they compare to the contribution of a normally distributed
classical fluctuating force $F(t)$ with correlation $\langle
F(t)F(s) \rangle = A(t - s)$
\\
\begin{equation}
\tilde \Gamma=\frac{1}{\hbar} \int_0^T dt \int_0^t ds
\left(x(t)-y(t)\right) A(t-s) \left(x(s)-y(s)\right)
\end{equation}
\\
Since these terms have the same form, the real part of the
influence functional must be interpretable as the correlation of
forces in the classical regime.

The real phase of the vortex influence functional is, after
substitution for $f_k$,
\\
\begin{align}
\Gamma=\frac{Sq^2}{2 a^2} \int d^2k \coth \frac{\hbar \omega_k
\beta}{2} & \int_0^T dt \int_0^t ds \left(\dot{\mathbf X}(t)e^{ i
\mathbf k \cdot \mathbf X(t)} - \dot{\mathbf Y}(t)e^{ i \mathbf
k\cdot \mathbf Y(t)} \right)\cdot
\hat \varphi_{\mathbf k} \nonumber \\
& \frac{\cos \omega_k(t-s)}{kQ^3r_v^2} \left(\dot{\mathbf
X}(s)e^{-i \mathbf k \cdot \mathbf X(s)} - \dot{\mathbf Y}(s)e^{-i
\mathbf k \cdot \mathbf Y(s)}\right) \cdot \hat \varphi_{\mathbf
k}
\end{align}
\\
The integral over $\varphi_k$ can be performed exactly as was done
for the imaginary part yielding Bessel function pre-factors
\\
\begin{align}
\Gamma=\frac{\pi Sq^2}{2a^2}& \int dk \;k \coth \frac{\hbar
\omega_k \beta}{2} \int_0^T dt \int_0^t ds \frac{\cos
\omega_k(t-s)}{kQ^3r_v^2}  \nonumber  \\
& \Bigg( \dot{\mathbf X}(t)\cdot \dot{\mathbf X}(s)J_0(k|\mathbf
X(t)-\mathbf X(s)|)+ \dot X(t)\dot X(s)\left( (\hat{\mathbf
X}(t)\cdot\hat{\mathbf e}_{\Delta})
(\hat{\mathbf X}(s)\cdot\hat{\mathbf e}_{\Delta}) \right. \Bigg.\nonumber \\
& \Bigg. \left. -(\hat{\mathbf X}(t)\times\hat{\mathbf
e}_{\Delta})\cdot (\hat{\mathbf X}(s)\times\hat{\mathbf
e}_{\Delta}) \right)J_2(k|\mathbf X(t)-\mathbf X(s)|) \Bigg)
\mbox{ + etc.}
\end{align}

Integrating by parts twice to cast this into a similar form to the
regular dissipation term of Caldeira and Leggett
\\
\begin{align}\label{gammareal}
\Gamma=\frac{\pi Sq^2}{2 a^2}& \int dk \; \omega_k^2 \coth
\frac{\hbar \omega_k \beta}{2} \int_0^T dt \int_0^t ds
\frac{\cos \omega_k(t-s)}{Q^3r_v^2}  \nonumber  \\
& \Bigg( \mathbf X(t)\cdot \mathbf X(s)J_0(k|\mathbf X(t)-\mathbf
X(s)|)+ \left( (\mathbf X(t)\cdot\hat{\mathbf e}_{\Delta})
(\mathbf X(s)\cdot\hat{\mathbf e}_{\Delta}) \right. \Bigg.\nonumber \\
& \Bigg. \left. -(\mathbf X(t)\times\hat{\mathbf e}_{\Delta})\cdot
(\mathbf X(s)\times\hat{\mathbf e}_{\Delta}) \right)J_2(k|\mathbf
X(t)-\mathbf X(s)|) \Bigg) \mbox{ + etc.}
\end{align}
\\
where, as usual, we neglect derivatives of the Bessel functions
since their derivatives provide higher order corrections. There
are additional boundary terms depending only on initial and final
positions that don't affect the vortex dynamics.

The real phase can be interpreted as the correlation of forces.
However, here the fluctuating forces are now vector forces and
there are correlations between various components of the
fluctuating forces. The appearance of the various Bessel
functions, arising because the vortex is an extended object in 2D,
differs from the treatment of Caldeira and Leggett because of a
different density of states of the environmental modes.

\section{Discussion of vortex effective dynamics}

\subsubsection{Markovian approximation}

We can apply the Markovian approximation as in Castro Neto and
Caldeira's treatment of solitons\cite{caldeira:1993}. That is,
approximate $\gamma(t) \to \gamma\frac{\delta(t)}{\omega_0}$ and
similarly in the force correlation integral, (\ref{gammareal}),
giving
\\
\begin{align}
\Gamma_M=  \frac{\tilde A(\beta)}{\hbar} \int_0^T dt  & (\mathbf
X(t)- \mathbf Y(t))\cdot(\mathbf X(t) - \mathbf Y(t))
\end{align}
\\
where
\\
\begin{equation}
\tilde A(\beta) = S^2J\pi q^2\hbar \int dk \;k
\frac{1}{Q^2r_v^2}\coth \frac{\hbar \omega_k \beta}{2}
\end{equation}
\\
Note, all $J_2$ terms disappear in this approximation.

In this limit, the longitudinal damping coefficient,
(\ref{phidamping}) becomes
\begin{align}\label{eta}
\eta =& \frac{S\pi q^2}{a^2}\int dk \frac{1}{Q^3r_v^2} \nonumber \\
=& \frac{S\pi q^2}{a^2}
\end{align}
\\
The classical fluctuation-dissipation theorem is now satisfied in
the high temperature limit ($\coth x  \to \frac 1 x$)
\begin{equation}
\tilde A(\beta) = 2 k_BT\eta
\end{equation}
where $T$ here denotes temperature.

This limit corresponds to the limit where the timescale of
interest is much greater than the correlation time of the magnons.

\subsection{Comparison with radiative
dissipation}\label{section:infldiss}

The dissipation found in the Markovian approximation can be
compared with the over-simplified calculation performed using
second order perturbation theory. There, assuming only the
emission of a magnon and no inter-magnon scattering, we found that
the dissipation rate was given by the integral
\\
\begin{equation}\label{radrate}
\gamma = 2\pi \int d^2k \frac{Sq^2}{2a^2kQ^3r_v^2}
\left(\dot{\mathbf X} \cdot \hat{\mathbf \chi}_k \right)^2
\delta\left(\hbar ckQ\right)
\end{equation}
where we evaluated this integral in section
\ref{section:vortmagdiss}. Comparing the $k$ dependance of this
integral with that of the damping coefficient in equation
(\ref{eta}), we find they differ only by the $\delta$-function.
The dissipation is now dependent on the entire magnon spectrum. In
the previous calculation, we made the simplifying assumption that
there were initially no magnons and hence only zero energy magnons
could be scattered.

In the Markovian limit, the effective damping force found in the
imaginary part of the influence functional phase gives roughly the
energy dissipation
\\
\begin{align}
E_{diss} \sim & \int dX \cdot \eta \dot{\mathbf X}\nonumber  \\
=& \int dt \eta \dot X^2\\
=& \int dt \frac{S\pi q^2}{a^2} \dot X^2 \nonumber \\
=& \frac{S\pi q^2 T}{a^2} \dot X^2\nonumber
\end{align}

Here, the full spectrum of magnons is excited, with probability of
finding a certain $k$ state weighted by its corresponding
Boltzmann factor. Thus, even assuming no vortex inertial energy,
we can find scattering between infinitesimally spaced $k$ states
throughout the spectrum.

\subsection{Extending results to many vortices}

The entire treatment can be repeated for a collection of vortices.
Assuming the vortices are well enough separated to neglect core
interactions, the unperturbed spin configuration is
\begin{align}
\phi_{tot} =& \sum_{i=1}^n q_i \chi(\mathbf X_i) \nonumber \\
\theta_{tot} =& \sum_{i=1}^n \theta_v(\mathbf r - \mathbf X_i)
\end{align}
where $\mathbf X_i$ denotes the center of the $i$th vortex. The
center coordinates are elevated to operators within the collective
coordinate formalism. Expanding the Lagrangian in terms of this
spin field, without magnon interactions, we find gyrotropic
momentum terms and inter-vortex potentials
\begin{equation}
\mathcal L_v^0 = \sum_i \left( -E_{v,i}+  \mathbf P_{gyro,i} \cdot
\dot{\mathbf X}_i  + 2S^2J\pi \sum_{j\neq i} q_iq_j \ln
\frac{X_{ij}}{r_v} \right)
\end{equation}
where $E_{v,i}$ is the unimportant rest energy of the vortices,
$\mathbf P_{gyro,i}=-\frac{\pi Sq_ip_i}{a^2} \mathbf X_i \times
\hat{\mathbf z}$ is the vector potential giving the gyrotropic
force, and the last term accounts for inter-vortex interactions.

The magnon interactions are treated to leading order only---the
zero-point energy shifts do not affecting dynamics, and higher
order dissipation is not treated here. There is a one magnon
coupling with the vortex velocities $\dot{\mathbf X}_i$ to the
magnons that is integrated over in the influence functional.

The resulting influence functional now has effective action terms
coupling not only the forward and backward paths of the same
vortex, but also the paths for different vortices. Without going
through all the details, the general results are presented. Were
we to neglect all inter-vortex terms, the influence functional
would simply be the product over each single vortex influence
functional.

Including inter-vortex terms to leading order now, the mass tensor
is exactly the same as that found using second order perturbation
theory (see section \ref{section:manyvortexmass}) and, in fact,
the calculations here are nearly identical to those. The mass
tensor is
\begin{align}\label{manyvortexmass}
M_{ij} =& \int d^2k  \frac{q_iq_j}{Ja^4 k^2 Q^4 r_v^2}
\left(\hat{\dot{\mathbf X}}_i \cdot \hat \chi_k\right)
\left(\hat{\dot{\mathbf X}}_j \cdot \hat \chi_k \right)
e^{i\mathbf k \cdot (\mathbf X_i-\mathbf X_j)} \nonumber \\
=& \frac{\pi q_iq_j r_v^2}{Ja^4} \left\{%
\begin{array}{ll}
    \ln \frac{R_S}{r_{ij}} + \frac{1}{2}
    \left( (\hat{\mathbf X}_i\cdot\hat{\mathbf e}_{ij}) (\hat{\mathbf
    X}_j\cdot\hat{\mathbf e}_{ij})   \right. & \\
    \left. -(\hat{\mathbf X}_i\times
    \hat{\mathbf e}_{ij}) \cdot (\hat{\mathbf
    X}_j\times\hat{\mathbf e}_{ij}) \right), & \hbox{$i\neq j$;} \\
    \ln \frac{R_S}{\sqrt{a^2+r_v^2}}, & \hbox{$i=j$.} \\
\end{array}%
\right.
\end{align}
\\
where $\hat{\mathbf e}_{ij} = \frac{\mathbf X_i-\mathbf
X_j}{|\mathbf X_i-\mathbf X_j|}$.

There are inter-vortex damping forces behaving essentially like
the single vortex damping forces: there exist forces longitudinal
and transverse to the motion of a vortex, however, acting on a
second vortex. The damping decreases as a function of vortex
separation as $\sim J_0(kr_{ij})$. This dissipation is thus quite
small when we assume well separated vortices, in keeping with
previous calculations (refer to the inter-vortex forces
calculation in section \ref{section:intervortex}).

Similarly, in the force correlation integral, we find that the
fluctuating forces acting on various vortices are
inter-correlated. This shouldn't be surprising at all: we have
damping terms intermingling the motion of vortex pairs so that we
should therefore expect that the fluctuating forces on these
vortices are inter-dependent.

The final effective density matrix propagator becomes
\\
\begin{align}\label{densmatprop}
J&(\mathbf X_i, \mathbf Y_i; \mathbf X_i', \mathbf Y_i') =\prod_i
\int_{\mathbf X_i,\mathbf Y_i}^{\mathbf X_i',\mathbf Y_i'}
\mathcal D[\mathbf X_i(t),\mathbf Y_i(t)] \exp
\frac{i}{\hbar}(\mathcal
S_v[\mathbf X_i(t)]-\mathcal S_v[\mathbf Y_i(t)])\nonumber \\
& \exp -\frac{1}{\hbar^2}\sum_i \int_0^T \int_0^t dt ds
\Bigg(\sum_{i,j} A_{ij}(t-s)\mathbf X_i(t)\cdot\mathbf X_j(s)
J_0(k|\mathbf X_i(t)-\mathbf X_j(s)|) \Bigg.\nonumber \\
&+ \sum_i A_{ii}(t-s)\Big( (\mathbf X_i(t)\cdot\hat{\mathbf
e}_{\Delta_i})
(\mathbf X_i(s)\cdot\hat{\mathbf e}_{\Delta_i}) \Big. \nonumber \\
& \Bigg. \Big. -(\mathbf X_i(t)\times\hat{\mathbf
e}_{\Delta_i})\cdot (\mathbf X_i(s)\times\hat{\mathbf
e}_{\Delta_i}) \Big)J_2(k|\mathbf X_i(t)-\mathbf X_i(s)|) \Bigg)
\mbox{ + etc.}
\end{align}
\\
where the force correlations as applied to vortices $i$ and $j$
are
\begin{equation}\label{flucforcecorr}
A_{ij}(t-s) = \frac{\hbar \pi Sq_iq_j}{2a^2} \int dk \frac{\coth
\frac{\hbar \omega_k \beta}{2}}{Q^3r_v^2} \omega_k^2 \cos
\omega_k(t-s)
\end{equation}

The vortex effective action has been redefined to include the
inertial mass and damping terms
\begin{align}
\mathcal S_v =&  \int dt \Bigg(\mathcal L_v^0 + \sum_{i,
j}\frac{1}{2} M_{ij} \dot{\mathbf X}_i\dot{\mathbf X}_j \Bigg.\nonumber \\
& - \int_0^t ds \bigg( \sum_{i, j}\gamma_{||}^{ij}\big(t-s,
|\mathbf X_j(t) - \mathbf X_i(s)|\big)\dot{\mathbf
X}_i(s)\cdot\mathbf X_j(t) \bigg. \\
& + \sum_{i}\gamma_{\Delta}^i\big(t-s, |\mathbf X_i(t) - \mathbf
X_i(s)|\big)\dot{X}_i(s) \big(\hat{\mathbf X}_i(s)\cdot
\hat{\mathbf e}_{\Delta_i}\big)
\big(\mathbf X_i(t)\cdot\hat{\mathbf e}_{\Delta_i}\big) \nonumber \\
& \Bigg.\bigg. + \gamma_{\bot}^i\big(t-s, |\mathbf X_i(t) -
\mathbf X_i(s)|\big) \dot{X}_i(s) \big(\hat{\mathbf X}_i(s)\cdot
\hat{\mathbf e}_{\bot_i}\big) \big(\mathbf X_i(t)
\cdot\hat{\mathbf e}_{\bot_i}\big) \bigg)\Bigg)\nonumber
\end{align}
\\
where the damping functions are from (\ref{gammafull})
\begin{align}
\gamma_{||}^{ij}(t-s, \Delta)=& \frac{S^2J\pi q_iq_j}{2}\int dk
\;k
\frac{\cos\omega_0(t-s)J_0(k\Delta)}{Q^2r_v^2} \nonumber \\
\gamma_{\Delta}^i(t-s,\Delta)=& \frac{S^2J\pi q_i^2}{2}\int dk
\;k \frac{\cos\omega_0(t-s)J_2(k\Delta)}{Q^2r_v^2}\nonumber\\
\gamma_{\bot}^i(t-s,\Delta)=&-\gamma_{\Delta}^i(t-s,\Delta)
\end{align}

Note in the limit of slow motion and large inter-vortex
separation, that is, $J_0\to1$ for same vortex terms and all
others are negligible, this effective action has the same form for
each vortex as found in the quantum Brownian motion described by
\citet{leggett:1983}, however, with inter-vortex terms introducing
Coulomb-like forces.

\subsection{Frequency dependent motion}\label{section:freqdepmot}

Perhaps a better way of understanding the role of the Bessel
function pre-factors is to decompose them according to the sum
rules
\begin{equation}
J_{\nu}(k|\mathbf x -\mathbf y|) = \sum_{m=-\infty}^{\infty}
J_m(kx)J_{\nu+m}(ky)e^{i(\nu+m)(\phi_x-\phi_y)}
\end{equation}

Denote $\mathbf X_{km}^i = J_m(kX_i)e^{im\phi_{i}}\mathbf X$. The
effective Lagrangian is transformed to
\begin{align}\label{freqeffaction}
\mathcal L_v =& \mathcal L_v^0 + \sum_m \int dk \frac{1}{2}
\sum_{i,j} M_{k}^{ij}\dot{\mathbf X}_{km}^i \cdot \dot{\mathbf
X}_{km}^j  +  \frac{1}{2}\sum_i M_{k}^{ii} \dot{\mathbf X}_{km}^i
\cdot \dot{\mathbf X}_{k,m+2}^i e^{i2\phi_{i}}\nonumber \\
&- \int_0^t ds \Bigg( \sum_{i,j} \gamma_{||k}^{ij}(t-s)
\dot{\mathbf X}_{km}^i(s) \cdot \mathbf
X_{km}^j(t) \Bigg. \\
& + \sum_i \gamma_{\Delta k}^i(t-s) e^{i2\phi_{is}}
\dot{X}_{km}^i(s) \big( \hat{\mathbf X}^i(s)\cdot \hat{\mathbf
e}_{\Delta_i}\big)
\big(\mathbf X_{k,m+2}^i(t)\cdot\hat{\mathbf e}_{\Delta_i}\big) \nonumber \\
&+ \Bigg. \gamma_{\bot k}^i(t-s)
e^{i2\phi_{is}}\dot{X}_{km}^i(s)\big( \hat{\mathbf X}^i(s)\cdot
\hat{\mathbf e}_{\bot_i}\big) \big(\mathbf X_{k,m+2}^i(t)
\cdot\hat{\mathbf e}_{\bot_i}\big) \Bigg)\nonumber
\end{align}
where $M_{k}^{ij} = \frac{\pi q_iq_j}{Ja^4 k Q^4 r_v^2}$,
recalling that the mass tensor can be expressed as an integral
over $k$ with Bessel function factors (refer to section
\ref{section:manyvortexmass}). The new damping function is defined
as
\begin{align}
\gamma_{||k}^{ij}(t-s)=& \frac{S^2J\pi q_iq_j}{2}
\frac{k}{Q^2r_v^2}\cos\omega_0(t-s) \nonumber \\
\gamma_{\Delta k}^i(t-s)=& \frac{S^2J\pi q_i^2}{2}
\frac{k}{Q^2r_v^2}\cos\omega_0(t-s) \nonumber\\
\gamma_{\bot k}^i(t-s)=&-\gamma_{\Delta k}^i(t-s,\Delta)
\end{align}

The real part of the influence functional can be re-expressed now
as
\begin{align}\label{freqflucforcecorr}
\Gamma=\frac{1}{\hbar} &\sum_{ijm}\int dk \int_0^T dt \int_0^t ds
A_k^{ij}(t-s) \big(\mathbf X_{km}^i(t) - \mathbf Y_{km}^i(t) \big)
\cdot \big(\mathbf X_{km}^j(s) - \mathbf
Y_{km}^j(s) \big)  \nonumber \\
&+ \delta_{ij} A_k^{ii}(t-s)e^{i2\phi_{it}} \mathbf
X_{km}^i(t)\cdot \bigg( \hat{\mathbf e}_{\Delta_i} \big(\mathbf
X_{k,m+2}^i(s)\cdot\hat{\mathbf e}_{\Delta_i}\big)
\bigg. \\
& \bigg. -\hat{\mathbf e}_{\bot_i} \cdot \big(\mathbf
X_{k,m+2}^i(s)\cdot\hat{\mathbf e}_{\bot_i}\big) \bigg) \mbox{ +
etc. in $XY$ and $Y^2$}\nonumber
\end{align}
where
\begin{equation}
A_k^{ij}(t-s)=\frac{\hbar \pi Sq_iq_j}{2a^2} \frac{\coth
\frac{\hbar \omega_k \beta}{2}}{Q^3r_v^2} \omega_k^2 \cos
\omega_k(t-s)
\end{equation}

Recall that the density matrix propagator is not simply the
product of non-interacting forward and backward paths. As in
(\ref{densmatprop}), we also have damping terms coupling the
forward and backward paths.

Thus, we find that the motion of the collection of vortices
behaves as interacting Brownian particles; however, with frequency
dependent damping and fluctuating forces. The formalism of
\citet{leggett:1983} can be applied to each frequency component,
with the added complexity of inter-vortex forces.

\subsection{Summary}
A collection of vortices are quantized by considering the small
perturbations about them. This amounts to including vortex-magnons
interactions. We considered two couplings in depth: a first order
coupling between the vortex velocity and the magnon spin field,
and a second order magnon coupling. All vortex-magnon couplings
create dissipation via magnon radiative processes. We considered
only the dissipation due to the first order coupling, first in
perturbation theory and later via the influence functional. Higher
order couplings also create dissipation, and may, in fact,
contribute more significantly\cite{stamp:1991,dube:1998}, however,
these weren't considered here.

The one magnon coupling creates an inertial energy endowing the
vortex with an effective mass. A collection of vortices are
strongly coupled: in addition to the usual inter-vortex forces,
there are inter-vortex inertial terms such as $\frac 1 2 M_{ij}
\dot{\mathbf X}_i\cdot \dot{\mathbf X}_j$ that are non-negligible.
The zero point energy shift from the two magnon coupling is log
divergent and, being due to the presence of the vortex, is
considered the quantized vortex's zero point energy. Note, we did
not calculate the full effect of this two magnon coupling, only
that portion independent of magnon populations. This shift was
calculated first by considering magnon scattering in Chapter
\ref{chapter:vortex} and next in this chapter by simply rewriting
the interaction in terms of magnon creation/annihilation
operators.

The influence functional reconfirms the effective mass
calculations and gives explicitly the damping forces and
corresponding fluctuating forces responsible for dissipation.
These act longitudinally and transverse to the vortex motion.
Again, a collection of vortices are coupled via the damping
forces: damping forces due to the motion of a first vortex act on
a second vortex. Damping forces depend on the entire history of
the vortex dynamics.

\chapter{Conclusions}

We study the dynamics of a collection of magnetic vortices in an
easy plane two dimensional insulating ferromagnet. The system is
approximated by a continuous spin field because we are only
interested in the low energy response. The vortices interact with
magnons via a variety of couplings. The effective dynamics bear
many similarities to that in the more complex superfluid and
superconducting vortex bearing systems.

We reviewed the derivations of the gyrotropic force and the
inter-vortex force by expanding the vortex action about a
stationary superposition of vortex solutions. We reviewed the
inertial mass derivation by calculating vortex profile distortions
when in motion and showed the equivalence of this method with
ordinary perturbation theory. We reviewed magnon phase shift
calculations and how these phase shifts modify the vortex zero
point energy. By rewriting the scattering potential in terms of
magnon creation and annihilation operators, we found an
equivalence of the phase shift calculations with the immediate
energy shift revealed in the second quantized form.

We suggest a new interpretation of the gyrotopic force as a
Lorentz-type force with the vortex vorticity behaving like charge,
$4\pi\epsilon_0 q$ (in SI units), in an effective perpendicular
magnetic field, $\mathbf B = \frac{S^2J}{4\epsilon_0 r_v}p_i \hat
z$, due to the vortex's own out-of-plane spins. We rewrite the
effective action term giving the gyrotropic force instead as a
vector potential shift in the vortex momentum. This momentum term
was then verified by direct integration of the operator generating
translations. The vector potential possesses gauge freedom,
allowed by the same freedom of gauge in the Berry's phase.

We next employed the Feynman-Vernon influence functional
formalism, assuming the vortex-magnon systems are initially
uncoupled with the magnons in thermal equilibrium (thus
introducing temperature). The systems interact and entangle. The
dynamics of the vortices were isolated by tracing over magnons.
The resulting effective vortex motion is acted upon by
longitudinal and transverse damping forces. Before now, no damping
force acting transverse to the vortex motion has been suggested in
a magnetic system.

The vortex is a stable solution of the easy plane ferromagnet. As
such, when we expand about it to quantize magnons in its presence,
we find no linear coupling between the two fields. However, the
vortex is a stationary solution, so that setting it into motion,
we find a first order coupling between the magnon field and the
vortex velocity.

This lowest order coupling, responsible for endowing the vortex
with an effective mass, is dissipative and yields effective
damping forces acting on a moving vortex. The damping forces are
accompanied by fluctuating forces that average to zero and with
time correlations such that the fluctuation-dissipation theorem is
satisfied in a generalized way.

We found both longitudinal and transverse damping forces dependent
on the prior motion of the vortex. For a collection of vortices,
the damping forces also act between vortices: the motion of a
first vortex causes a damping force to act on a second vortex.
Correspondingly, there are non-zero correlations between forces
acting on two different vortices.

The vortex dynamics were described by the propagator of the vortex
reduced density matrix. The forward and backward paths are
coupled, as already described for quantum Brownian motion by
\citet{leggett:1983}. The damping forces possess memory effects, a
common feature in general when describing a soliton as a quantum
Brownian particle\cite{caldeira:1993}. In our two dimensional
system, however, we found additional Bessel function factors.
These considerably complicate the extraction of a spectral
function describing the ensuing Brownian motion. By decomposing
the vortex motion in a basis of Bessel functions, we find that the
various frequency components behave as a coupled ensemble of
quantum Brownian particles.

\section{Open questions}

The analogy of a vortex as a charged particle in a magnetic field
can be extended. For instance, there should be excitations within
the gauge field giving the gyrotropic momentum. The magnetic field
is a result of the out-of-plane spins at the vortex center.
Perhaps, gauge fluctuations are related to vortex core flips.

Future work on magnetic vortex motion should check the relative
importance of higher order dissipative couplings. The basic motion
of a small collection of vortices can be examined now including
inertial and damping forces. For instance, one could verify the
claim of \citet{sloncz:1984} that damping forces acting on a
vortex pair only decay circular orbits inward and parallel ones
outward. The similarities with superfluid vortices should be
further examined by attempting to calculate the influence
functional of a superfluid vortex and, likewise, the Aharanov-Bohm
interference effects of magnons passing a moving vortex.

\appendix

\chapter{Some mechanics}\label{chapter:mechanics}

A classical system is describable by its Lagrangian, which is a
function of the system coordinates $q_i$ and velocities $\dot q_i$
\\
\begin{equation*}
\mathcal L(q_i,\dot q_i, t)
\end{equation*}
\\
The action of the system is defined by
\\
\begin{equation}
\mathcal S[q_i(t)] = \int_0^T dt L(q_i,\dot q_i, t)
\end{equation}
\\
The equations of motion of the system are given by the principle
of least action, otherwise known as Hamilton's principle, stating
that the system evolves from initial state $\{q_i(0)\}$ to final
state $\{q_i(T)\}$ via the path $q_i(t)$ that extremizes the
action, $\mathcal S$.

Given that $\mathcal L = \mathcal L(q_i,\dot q_i, t)$, extremizing
the action we arrive at the Euler-Lagrange equations
\\
\begin{equation}\label{ELeqns}
\frac{d}{dt} \frac{\partial \mathcal L}{\partial \dot q_i} -
\frac{\partial \mathcal L}{\partial q_i} = 0
\end{equation}

Alternatively, we can describe the system by its Hamiltonian. We
transform from the Lagrangian to the Hamiltion via a Legrendre
transformation
\\
\begin{equation}
H(q_i,p_i,t) = \sum_i p_i \dot q_i - \mathcal L(q_i,\dot q_i, t)
\end{equation}
\\
where we've defined the conjugate momenta $p_i$ defined by
\\
\begin{equation}\label{conjmom}
p_i = \frac{\partial \mathcal L}{\partial \dot q_i}
\end{equation}
\\
Hamilton's equations are a restatement of (\ref{ELeqns}) and
(\ref{conjmom})
\\
\begin{equation}\label{Hameqns}
\frac{\partial q_i}{\partial t} = \frac{\partial H}{\partial p_i};
\qquad \frac{\partial p_i}{\partial t} = -\frac{\partial
H}{\partial q_i}
\end{equation}
\\
For example, consider a particle of mass $m$, position $x$,
residing in a potential $V(x)$. The Hamiltonian is simply the
total energy of the system
\\
\begin{equation}
H(x,p, t) = \frac{p^2}{2m} + V(x)
\end{equation}
\\
where the conjugate momentum $p = m\dot x$, as usual. The
Lagrangian is then given as the difference in kinetic and
potential energies
\\
\begin{equation}
\mathcal L(x,\dot x, t)=\frac{1}{2}m\dot x^2 - V(x)
\end{equation}
\\
Application of either the Euler-Lagrange equation or Hamilton's
equations yields Newton's second law of motion, $F = m\ddot x$,
where the force $F = -\frac{\partial}{\partial x}V(x)$.

Define the Poisson bracket $\{\cdot,\cdot\}_{q,p}$ for a system
with coordinate $q$ and conjugate momentum $p$ via
\\
\begin{equation}\label{poisson}
\{A,B\}_{q,p} = \frac{\partial A}{\partial q} \frac{\partial
B}{\partial p} - \frac{\partial A}{\partial p}\frac{\partial
B}{\partial q}
\end{equation}
\\
Note that $\{q,p\}_{q,p}=1$. The above can be easily generalized
to a field theory by substituting $q\to \tilde\phi(x)$ and $p \to
\tilde\pi(x)$ and replacing all simple derivatives by functional
derivatives.

Going over to quantum mechanics, to quantize the motion of the
system, we impose the commutation relations
\\
\begin{equation}
[q,p] = i\hbar
\end{equation}
\\
In 1925, P. A. M. Dirac \cite{dirac:1925} observed that proper
quantum mechanical relations followed under the substitution
\\
\begin{equation*}
\{\cdot,\cdot\}_{q,p} \to \frac{1}{i\hbar}[\cdot,\cdot]
\end{equation*}

In a spin system, using coordinate $\phi$  and conjugate momentum
$S\cos\theta$ we can verify directly the classical version of
$[S_i, S_j]= i\hbar \varepsilon_{ijk}S_k$, that is,
\\
\begin{equation}
\{S_i, S_j\}_{\phi,S\cos\theta}= \varepsilon_{ijk}S_k
\end{equation}
\\
where we define $S = S(\sin\theta\cos\phi, \sin\theta\sin\phi,
\cos\theta)$. However, spin being an essentially quantum concept,
we must bear in mind that when speaking of spin directions given
by $(\phi,\theta)$, we mean the spin state of highest probability
to be found in that direction.

\section{Imaginary time path integral}\label{section:imagtime}

Consider a system in thermal equilibrium at temperature $\tau$. If
we decompose the system Hamiltonian into a set of eigenstates
$\xi_n(x)$ with eigenenergies $E_n$, then the probability of
observing the system in eigenstate $n$ is proportional to
$e^{-\frac{E_n}{k_B\tau}}$ where $k_B$ is Boltzmann's constant
\cite{feynman:1972}. The density matrix for this system is
\\
\begin{equation}\label{rhosum_n}
\rho(x',x) = \sum_n \xi_n(x') \xi_n^*(x) e^{-\beta E_n}
\end{equation}
\\
where $\beta=(k_B\tau)^{-1}$.

Compare this with the quantum propagator decomposed into this same
basis:
\\
\begin{align}
K(x',T;x,0) =& \langle x' | \exp -\frac{iHT}{\hbar} |
x \rangle \nonumber \\
=& \sum_n \langle x' | \xi_n \rangle \exp -\frac{i E_nT}{\hbar}
\langle \xi_n | x \rangle \\
=& \sum_n \xi_n(x') \xi_n^*(x) \exp -\frac{i E_nT}{\hbar}
\nonumber
\end{align}
\\
We see that the density matrix is formally identical to the
propagator corresponding to an imaginary time interval $T =
-i\beta\hbar$. In fact, if we consider the equation of motion of
the density matrix found by taking the derivative of
(\ref{rhosum_n}) with respect to $\beta$ \cite{feynman:1965}
\\
\begin{equation}
-\frac{\partial\rho}{\partial\beta} = \sum_n E_n \xi_n(x')
\xi_n^*(x) e^{-\beta E_n}
\end{equation}
\\
Recall that $E_n \xi_n(x') = H \xi_n(x')$. If we understand
$H_{x'}$ to act only on $x'$, we can write
\\
\begin{equation}\label{rhoeom}
-\frac{\partial\rho(x',x)}{\partial\beta} = \mathcal H_{x'}
\rho(x',x)
\end{equation}
\\
We know how to evaluate the propagator as a path integral for
simple Hamiltonians involving only the system coordinates and
their conjugate momenta. For example, for the Hamiltonian
\\
\begin{equation}
H = -\frac{\hbar^2}{2m} \frac{d^2}{dx^2} + V(x)
\end{equation}
\\
the solution over an infinitesimal time period $\epsilon$ is
\\
\begin{equation}
K(x',\epsilon;x,0) = \sqrt{\frac{m}{2\pi i\hbar \epsilon}} \exp
\frac{i}{\hbar}\left( \frac{m}{2}\frac{(x'-x)^2}{\epsilon} -
\epsilon V\left( \frac{x'+x}{2} \right) \right)
\end{equation}
\\
which can be verified by direct substitution into
\\
\begin{equation}
-\frac{\hbar}{i} \frac{\partial K(x',T;x,0)}{\partial T} = H_{x'}
K(x',T;x,0)
\end{equation}

Now, under an infinitesimal interval in the density matrix
$i\epsilon$ the solution is given by $\epsilon \rightarrow
-i\epsilon$
\\
\begin{equation}
\rho(x',x; \beta = \epsilon/\hbar) = \sqrt{\frac{m}{2\pi \hbar
\epsilon}} \exp -\frac{1}{\hbar}\left(
\frac{m}{2}\frac{(x'-x)^2}{\epsilon} + \epsilon V\left(
\frac{x'+x}{2} \right) \right)
\end{equation}
\\
which can be verified by direct substitution into (\ref{rhoeom}).

Stringing many of these solutions together for successive
intervals of time according to
\\
\begin{equation}
\rho(x',x; \beta') = \int dx'' \rho(x',x'',\beta') \rho(x'',x;
\beta)
\end{equation}
\\
for intermediate $x''$ at $\beta$, we obtain a path integral
formulation of the density matrix which is simply an imaginary
time version of the propagator path integral, that is, with the
substitution $T \rightarrow -i\beta\hbar$.

\chapter{Quantization of classical
solutions}\label{appendix:quant}

Suppose we have a particle described by position $x$ residing in a
potential $V(x)$. Classically, the particle follows a path $x(t)$
that satisfies Newton's second law of motion. In quantum theory,
the particle is no longer described by its position $x$, but by
its wavefunction $\psi(x)$ giving a probability distribution of
finding the particle at position $x$. If the energy is conserved,
the wavefunction can be decomposed into energy eigenstates,
$\psi_n(x)$, obeying Schroedinger's equation
\\
\begin{equation}
H\psi_n = E_n\psi_n(x)
\end{equation}
\\
where $H$ is the Hamiltonian of the system, quantized by elevating
the position and momentum variables to operators.

As preparation for a description of the quantization of a soliton
solution, consider some of the finer points of quantization of
classical particle solutions.

\begin{figure}
\centering
\includegraphics[width=6cm]{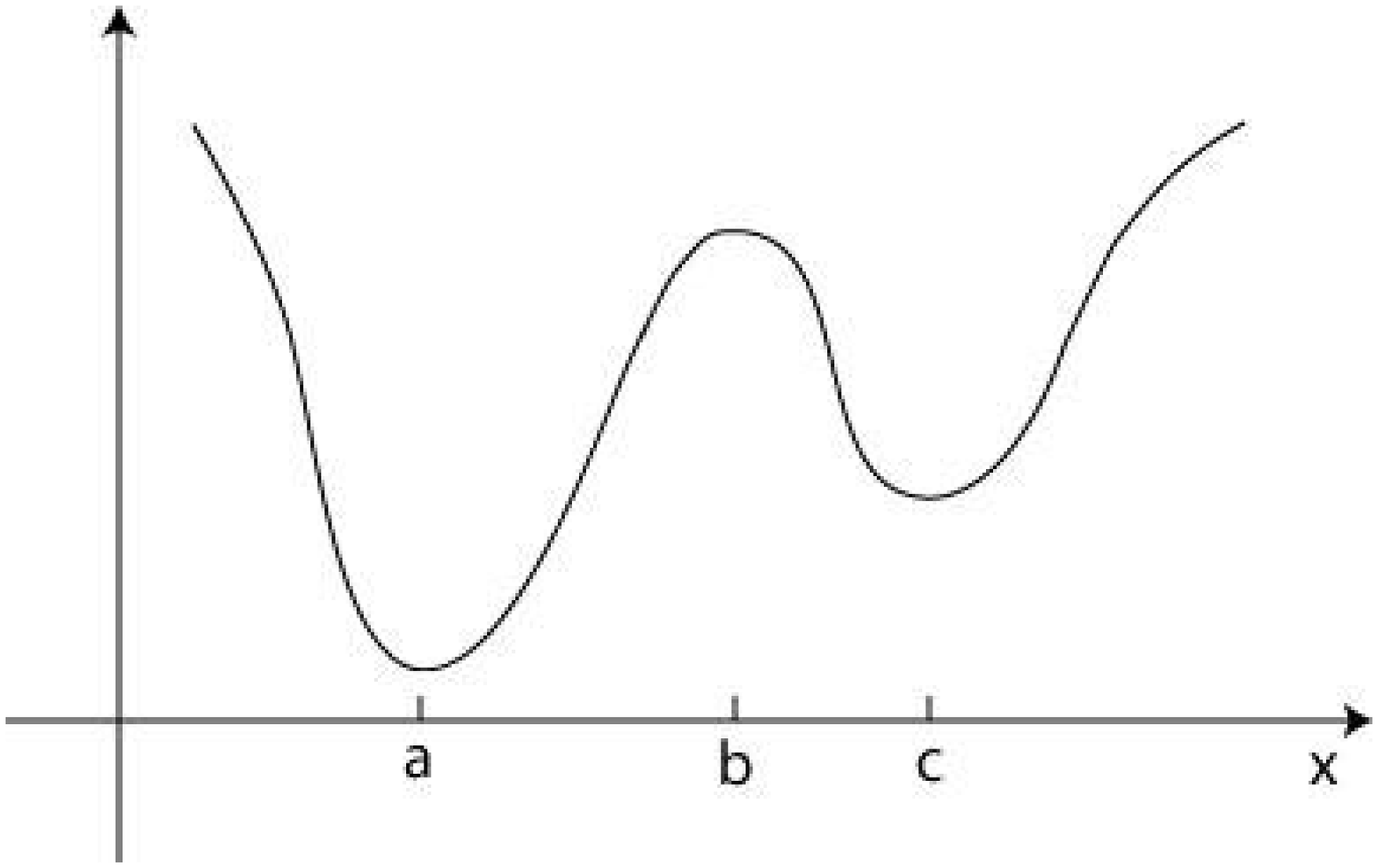}
\caption{An illustrative potential of a one dimensional
particle.}\label{figure:1dpotential}
\end{figure}

For the potential shown in Figure \ref{figure:1dpotential}, there
are three extrema and hence three stationary classical solutions.
The absolute minimum, $x=a$ is the classical ground state, having
the lowest attainable energy.

In quantum mechanics, according to the uncertainty principle, a
solution is not allowed to have zero momentum and a fixed
position. Thus, even in its ground state there are fluctuations.
Expanding $V(x)$ in a Taylor series, to lowest order the potential
is harmonic about the minimum and we have simple harmonic
excitations with frequency $\omega^2 = V''(x=a)$ and energies
\\
\begin{equation}
E_n = V(a) + (n+\frac{1}{2})\hbar \omega
\end{equation}
\\
The ground state energy becomes $E_0 = V(a) + \frac{1}{2}\hbar
\omega$. The additional $\frac{1}{2}\hbar \omega$ is the
zero-point energy due to quantum fluctuations.

The solution $x=c$ is a second stable classical solution. Quantum
mechanically, there are again fluctuations about this solution
that give a similar excitation spectrum. In this case, however,
since this is an excited state, there are possibly tunneling
processes that relax the state to its ground state about $x=a$. In
a field theory, this stable excited state is the analogue of a
soliton solution (while the tunneling processes are analogous to
instantons). However, for the magnetic solitonic solutions
considered here, these excited states belong to separate
topological sectors of the solution space so that there is
effectively an infinite energy barrier to the ground state.

The classical solution $x=b$ is unstable and would thus correspond
to an imaginary frequency. There are hence no set of quantum
levels formed about it.

Another interesting analogy to consider is the case of a constant
potential, $V(x) = V$. In that case, there is no clear choice of
minimum about which to expand and, should we attempt to, we would
find everywhere $\omega=0$.  Of course, in quantum mechanics, the
proper solutions to consider are the plane waves $e^{ikx}$ with
energies
\\
\begin{equation}
E_n = V+\frac{1}{2m} (\hbar k_n)^2
\end{equation}
\\
where $\hbar k_n = p_n$ are the momenta of these states. In field
theory, we find a zero frequency mode, or Goldstone mode, for
every broken continuous symmetry. Further, for each of these
broken symmetries, we find a corresponding conserved momentum,
analogous to the conserved $p_n$ in the particle case.

\section{Quantizing soliton solutions}\label{section:solitonzpe}

In field theory, quantizing a soliton follows analogously to the
regular quantization of a classical solution. The language is
changed somewhat however. For instance, the ground state of the
particle, $x=a$, is quantized to a hierarchy of simple harmonic
excitations. In field theory, we call the absolute potential
minimum the ground state, or the vacuum state. The hierarchy of
perturbative excitations are interpreted as mesons or
quasiparticles. In our system, these are the magnons.

When we expand about the solitonic excited state (analogous to the
second minimum, $x=c$), generally the quasiparticles are modified
by the soliton presence. In the simple particle case, this
corresponds to the general case where $V''(a)\neq V''(c)$. In the
particle case, the hierarchy of simple harmonic states are
interpreted as excited states about the minima.

In a field theory, the quasiparticles are generally extended
states and, in the presence of a soliton, are shifted but still
extended. In some cases, the soliton can trap a few quasiparticle
modes. These bound modes are interpreted as soliton excited
states. The remaining, extended states are interpreted as
unshifted quasiparticles, while all energy shifts due to the
soliton are attributed to the zero-point energy of the quantized
soliton.

The soliton acts perturbatively on the extended states, it itself
being localized in space, as a scattering center. Asymptotically
far from the soliton center, the quasiparticles are simply phase
shifted. Suppose that the relative phase shift between the
incoming and outgoing waves is $\delta(k)$, a function of the
wavevector $k$.

By enforcing periodic boundary conditions\symbolfootnote[2]{Or
alternatively, we could enforce fixed boundary conditions forcing
$k$ to be $\pi$-periodic rather than $2\pi$-periodic. In that
case, the asymptotic waveform must be modified from a plane wave
to a cosine wavefunction and we find that the phase shift is also
changed by a factor of 2. Thus, either set of boundary conditions
is equivalent.} on both the unperturbed wavevector $k$ and the
scattered wavevector $q$
\\
\begin{align}
L k_n =& 2n\pi \nonumber  \\
L q_n - \delta(q_n) =& 2n\pi
\end{align}
\\
we fix the allowed $k$ and $q$ values. In the $L\to \infty$ limit,
these allowed values merge to a continuum and the sum over
$k$-states is replaced by an integral
\\
\begin{equation*}
\sum_k \to  \frac{L}{2\pi}\int_{-\infty}^{\infty}dk
\end{equation*}

The energy correction to the soliton solution, taken as the
modification to the zero point energy of the vacuum, is thus,
noting that $\omega(q)= \omega(k+\frac{\delta}{L})$,
\\
\begin{align}
\Delta E =& \frac{1}{2} \hbar \sum_k \omega(q) - \omega(k)\nonumber \\
=& \frac{1}{2} \hbar \frac{L}{2\pi}\int_{-\infty}^{\infty}
dk \frac{\partial \omega(k)}{\partial k} \frac{\delta(q)}{L} \\
=& \frac{1}{4\pi} \hbar \int_{-\infty}^{\infty} dk \frac{\partial
\omega(k)}{\partial k} \delta(q)\nonumber
\end{align}
\\
found by expanding to first order in $\delta$.

In addition to small corrections to the quasiparticle continuum,
the soliton might bind discrete levels in the quasiparticle
spectrum. Those with $\omega=0$ are due to a continuous symmetry
broken by the soliton solution. These modes are dealt with using
collective coordinates. There can also be $\omega\neq 0$ discrete
modes. These are interpreted as soliton excited states. For an
example of these, see the quantization of the quantum kink of the
$\phi^4$ theory \cite{raj}---the magnetic vortex does not have any
such excited states.

\subsection{In a path integral formalism}
Using path integrals, we can find the excitation spectrum of a
system by taking the trace of the system's quantum propagator. We
first review the simple case of a regular particle in an external
potential and then generalize to field theory.

\subsubsection{Semiclassical approximation for a single particle}
The propagator of a single particle starting in position $q_a$ at
time 0 and ending in position $q_b$ at time $T$ is
\\
\begin{equation}
    K(q_b,T;q_a,0) = <q_b|e^{-iHT/\hbar}|q_a>
\end{equation}
\\
where $H(q,p) = \frac{1}{2}p^2+V(x)$, and, for simplicity, we've
set $m=1$.

Next, we set $q_a = q_b = q_0$ and integrate over the endpoint of
the periodic orbit

\begin{align}
    G(T)=& \int dq_0 <q_0|e^{-iHT/\hbar}|q_0> \nonumber \\
        =& \int dq_0 \sum_n
        <q_0|\phi_n>e^{-iE_nT/\hbar}<\phi_n|q_0> \\
        =& \sum_n e^{-iE_nT/\hbar} \nonumber
\end{align}
\\
where $\{\phi_n\}$ denote a complete orthonormal set of
eigenstates of $H$. This yields an expression giving the
excitation spectrum of the Hamiltonian.

For a particle in a potential $V(x)$ with a minimum at $x = x_0$,
the classical solution is simply $q_{cl} = x_0$. Expanding the
potential in a power series about this solution
\\
\begin{equation}
    V(x) = V(q_{cl}) + V'(q_{cl})(q-q_{cl}) +\frac{1}{2} V''(q_{cl}) (q-q_{cl})^2
    + \mathcal O(\Delta x^3)
\end{equation}
\\
the second term is zero since $q_{cl}$ is a minimum of $V(x)$. The
action expanded about this classical solution, $q(t) \to q_{cl} +
q'(t)$, is now
\\
\begin{equation}
\mathcal{S}[q(t)] =-V(x_0)+ \int_0^T dt \frac{1}{2} \dot{q'}^2 -
\frac{1}{2} \omega^2 q'^2
\end{equation}
\\
where $\omega^2 = V''(x_0)$, assumed positive (i.e. the classical
solution is stable). Note, at this point, the boundary conditions
of the periodic path are still not generally satisfied so that the
new perturbed solution must now satisfy $q'(0) = q'(T) = q_0 -
q_{cl}$.

The semiclassical approximation amounts to neglecting the
$\mathcal O(\Delta x^3)$ and higher order terms. But the terms in
$q'$ are just the action of a simple harmonic oscillator. To
evaluate the path integral
\\
\begin{equation}
G_{SHO}(T) = \int dq_0 \int \mathcal{D}[q'(t)]
e^{\frac{i}{\hbar}\mathcal{S}[q'(t)]}
\end{equation}
\\
we expand again about the simple harmonic oscillator classical
solution satisfying the appropriate boundary conditions. You may
ask why we didn't immediately go from the beginning action and
expand $V$ in a Taylor series and approximate there. Although that
would have proceeded identically, the additional step helps
clarify what to do when expanding in a field theory admitting
classical soliton solutions. The classical solution is now
\\
\begin{equation}
q_{cl}' = A\cos\omega t+B\sin\omega t
\end{equation}
\\
where the boundary conditions give
\\
\begin{align}\label{SHO_IC}
A =& q_0-q_{cl} \nonumber \\
A\cos\omega T+B\sin\omega T =& q_0-q_{cl}
\end{align}
\\
Evaluating this second classical contribution to the action gives
\\
\begin{equation}\label{SHO_Scl}
\mathcal{S}[q_{cl}'] = -2\omega (q_0 - q_{cl})^2
\frac{\sin^2\omega T/2}{\sin\omega T}
\end{equation}

The complete path integral becomes, setting $y(t) = q'(t) -
q_{cl}'$ and noting $y(t)$ now has the boundary conditions $y(0) =
y(T) = 0$,
\\
\begin{equation*}
G(T) = \int dq_0 e^{-\frac{i}{\hbar}V(x_0)-\frac{i}{\hbar} 2\omega
(q_0 - q_cl)^2 \frac{\sin^2\omega T/2}{\sin\omega T}} \int
\mathcal{D}[y(t)] e^{\frac{i}{2\hbar} \int_0^T dt
y(-\frac{\partial^2}{\partial t^2} - \omega^2) y}
\end{equation*}

Solving for the determinant of the remaining action $-\frac{1}{2}
\int_0^T dt y(\frac{\partial^2}{\partial t^2} + \omega^2) y$ we
solve the relevant Jacobi equation \cite{schulman}
$(\frac{\partial^2}{\partial t^2} + \omega^2) y =0$ with initial
conditions $y(0) = 0$ and $ y'(0) = 1$. This gives the prefactor
\\
\begin{equation}
\sqrt{\frac{\omega}{2\pi i\hbar\sin\omega T}}
\end{equation}

Evaluating the $q_0$ integral, the final result is
\\
\begin{align}
    G(T)=& \frac{1}{2i\sin \omega T/2} \nonumber \\
        = & e^{-i\omega T/2} \frac{1}{1 - e^{-i\omega T}}\\
        = & \sum_{n=0}^{\infty} e^{-i(n+\frac{1}{2}) \omega T -iT
        V(x_0)} \nonumber
\end{align}
\\
giving the excitation spectrum $E_n = \hbar \omega
(n+\frac{1}{2})$ as expected.

\subsubsection{Semiclassical approximation in field theory}

This follows almost identically to the single particle case, with
just a few technical points needing clarification. Suppose we have
a field theory in 1+1 dimensions with the Lagrangian density
\\
\begin{equation}
    \mathcal{L}(x,t) = \frac{1}{2} (\partial_\mu \phi)^2 - U[\phi]
\end{equation}
\\
Assume $\phi_{cl}(x)$ is a stationary extremum of this system.
Expanding the action about this solution, $\phi \to \phi' +
\phi_{cl}$
\\
\begin{equation}
\mathcal{S} = S_{cl} + \frac{1}{2} \int dx \int dt (\partial_\mu
\phi')^2 - \frac{\partial^2 U(\phi_{cl})}{\partial \phi^2} \phi'^2
\end{equation}
\\
Next, we integrate by parts to replace $(\partial_\mu \phi')^2 \to
\phi'(-\frac{\partial^2}{\partial t^2} +
\frac{\partial^2}{\partial x^2})\phi'$.

Assuming now that $\phi'(x,t)$ is separable, i.e. $\phi'(x,t) =
f(x)g(t)$, we solve for the eigenvalues of the spatial portion
\\
\begin{equation}
    \left( -\frac{\partial^2}{\partial x^2}+\frac{\partial^2
    U(\phi_{cl})}{\partial \phi^2} \right) f_r(x) = \omega_r^2 f_r(x)
\end{equation}
\\
Assume that the $f_r(x)$ eigenfunctions form an orthonormal basis.
Expressing the general solution $\phi'(x,t) = \sum_r f_r(x)
g_r(t)$ so that the integration measure becomes $\prod_r
\mathcal{D}[g_r(t)]$, the action becomes
\\
\begin{align}
\mathcal{S} =& \frac{1}{2}\int dx \int dt \sum_r f_r(x)g_r(t)
\sum_{r'}(-
\frac{\partial^2}{\partial t^2}-\omega_{r'}^2)f_{r'}(x)g_{r'}(t) \nonumber \\
=& \sum_r \frac{1}{2} \int dt g_r(t) (- \frac{\partial^2}{\partial
t^2}-\omega_{r}^2)g_r(t)
\end{align}
\\
by the orthonormality of the $f_r(x)$. Thus, the problem has
separated into a product on $r$ of equivalent single particle
problems
\\
\begin{equation}
    G(T) = e^{\frac{i}{\hbar}S_{cl}} \prod_r \left( \int \mathcal{D}[g_r(t)]
    e^{\frac{i}{2\hbar} \int dt g_r(t) (-\frac{\partial^2}{\partial t^2}-
    \omega_{r}^2)g_r(t)} \right)
\end{equation}
\\
which we know how to solve from the previous section.

The only remaining manipulation required is to note that
\\
\begin{equation}
    \prod_r \sum_{n_r}e^{-iT\omega_r(n_r+\frac{1}{2})} =
    \sum_{\{n_r\}}e^{-\frac{iT}{\hbar} \sum_r \hbar\omega_r(
    n_r+\frac{1}{2})}
\end{equation}
\\
where $\{n_r\}$ denotes a set of integers $n_r$.

\subsection{Collective coordinates}\label{section:collcoord}

Suppose the soliton exists in a system with translational
symmetry. The soliton itself is a localized entity, and hence
breaks this symmetry. The soliton must choose arbitrarily what
coordinate to center on. This is an example of spontaneously
broken symmetry.

This symmetry introduces to the quasiparticle spectrum a zero
frequency mode associated with the soliton. While to first order
presenting no problems, should we continue in the perturbative
expansion, the energy denominators would develop artificial
singularities.

In perturbing about the soliton solution, rather than as done
previously via
\\
\begin{align}
\phi =& \phi_0 + \sum_{n=0}^{\infty} a_n(t) \psi_n (x) \nonumber \\
=& \phi_0 + a_0(t) \frac{d\phi_0}{dx} + \sum_{n=1}^{\infty}a_n(t)
\psi_n (x)
\end{align}
where the $n=0$ mode is the translation mode, rewrite the
expansion as
\\
\begin{equation}
\phi = \phi_0(x-X(t))+\sum_{n=1}^{\infty}a_n(t) \psi_n (x)
\end{equation}
\\
where $X(t)$ is the collective coordinate associated to the
translational invariance. This is completely equivalent if we
expand $\phi_0(x-X(t))$ to first order in $X(t)$ and identify
$a_0(t)$ with $-X(t)$.

Rewriting the Lagrangian in terms of this expansion, the potential
terms, being translationally invariant by assumption, does not
depend on $X(t)$. The kinetic term depends only on $\dot X(t)$.

We can introduce conjugate momenta to $X(t)$ and to the $a_n(t)$,
denote these $P$ and $\pi_n$, and transform to the classical
Hamiltonian
\\
\begin{equation}
H = P\dot X(t) + \sum_{n=1}^{\infty} \pi_n \dot a_n(t) - L
\end{equation}

Quantizing the soliton now follows exactly as quantization of a
regular particle: we impose commutation relations on the various
degrees of freedom
\\
\begin{align}
[X,P] =& i\hbar \nonumber \\
[a_n,\pi_n] =& i\hbar
\end{align}
\\
The quantized quasiparticles have a zero-point energy shifted by
$\sum \frac{1}{2} \hbar \delta\omega_n$ that is attributed instead
to the quantized soliton. That is, if the vacuum quasiparticle
zero-point energy is $\sum \frac{1}{2} \hbar \omega_n$, while in
the presence of a soliton becomes $\sum \frac{1}{2} \hbar
(\omega_n+\delta\omega_n)$, the soliton is said to have the
zero-point energy $\sum \frac{1}{2} \hbar \delta\omega_n$ while
the quasiparticles are considered unchanged\cite{raj}.

\chapter{Spin path integrals}\label{section:SPI} Consider a spin
system $\Omega(t)$ with Hamiltonian $H$. The propagator for this
spin to evolve from state $\Omega_i$ at time $t=0$ to $\Omega_f$
at time $t=T$ is
\\
\begin{equation}
K(\Omega_f,T;\Omega_i,0) = \langle \Omega_f | \exp
-\frac{i}{\hbar}HT | \Omega_i \rangle
\end{equation}
\\
Inserting $N-1$ resolutions of the identity \cite{auerbach}
\\
\begin{equation}
\frac{2s+1}{4\pi} \int d\Omega |\Omega\rangle\langle \Omega | = 1
\end{equation}
\\
where a lower case $s$ denotes the dimensionless spin (whereas,
$S = \hbar s$), gives
\\
\begin{align*}
K(\Omega_f,T;\Omega_i,0) = \prod_{k=1}^{N-1}
\left(\frac{2s+1}{4\pi} \int d\Omega_k \right) &\langle \Omega_N| 
\exp -\frac{i}{\hbar}H\frac{T}{N}
|\Omega_{N-1}\rangle \\
& \langle \Omega_{N-1} | \cdots|\Omega_{1}\rangle\langle
\Omega_{1} | \exp -\frac{i}{\hbar}H\frac{T}{N}|\Omega_{0}\rangle
\end{align*}
\\
where $k=0$ denotes the initial state and $k=N$ the final state.
Define $\epsilon = T/N$.

Expand the exponential
\\
\begin{equation*}
\langle \Omega_{k+1} |\exp -\frac{i}{\hbar}
H\frac{T}{N}|\Omega_{k}\rangle =\langle \Omega_{k+1}
|\Omega_{k}\rangle \left(1-\frac{i}{\hbar}\epsilon \frac{\langle
\Omega_{k+1} |H|\Omega_{k}\rangle}{\langle \Omega_{k+1}
|\Omega_{k}\rangle} + \mathcal O(\epsilon^2)\right)
\end{equation*}
\\
Keeping terms to linear order in $\epsilon$, the $H$ term can be
approximated at equal times: define $H(t_k) = \langle \Omega_{k+1}
|H|\Omega_{k}\rangle$. Re-exponentiate the bracketed term to $\exp
-\frac{i}{\hbar}\epsilon H(t_k)$.

The overlap of two coherent states, $\Omega_{k}$ and
$\Omega_{k+1}$ is\cite{auerbach}
\\
\begin{equation}
\langle \Omega_{k+1} |\Omega_{k}\rangle  = \left( \frac{1+ \hat
\Omega_{k+1} \cdot \hat \Omega_{k}}{2} \right)^s e^{-is\psi}
\end{equation}
\\
where
\\
\begin{equation}
\psi = 2\tan^{-1} \left( \tan
\left(\frac{\phi_{k+1}-\phi_k}{2}\right) \frac{\cos
\frac{1}{2}(\theta_{k+1}+\theta_k)}{\cos
\frac{1}{2}(\theta_{k+1}-\theta_k)} \right) + \xi_{k+1} - \xi_k
\end{equation}
\\
and where $\xi$ is a gauge dependent phase that we can ignore. The
pre-factor is $1$ to leading order and the phase can be
approximated such that
\\
\begin{equation}
\langle \Omega_{k+1} |\Omega_{k}\rangle  = \exp \left( -is\epsilon
\frac{\phi_{k+1}-\phi_k}{2} \cos\theta_k \right)
\end{equation}

All together, letting $N\to\infty$, we find the spin path integral
\\
\begin{equation}
K(\Omega_f,T;\Omega_i,0) = \int \mathcal D[\Omega(t)] \exp \left(
-is\int_0^T dt \dot \phi(t) \cos\theta(t) - H(t) \right)
\end{equation}
\\
Note that there are no spurious boundary terms as there are in the
stereographic representation using $z$ and $z^*$, as found, for
example, by \citet{solari:1987}.

\section{The semiclassical approximation}
Evaluation of spin path integral is non-trivial as evidenced by
the series of papers suggesting various corrections. Klauder
\cite{klauder:1979} discussed the spin path integral in terms of
conjugate variables $\phi$ and $S\cos\theta$ and first addressed
the semiclassical approximation applied to the spin path integral.
He claimed that to evaluate properly the trace of the propagator
obtaining the excitation spectrum, real valued periodic orbits are
required. However, simple counting of degrees of freedom, given
two equations of motion (one for $\phi$ and another for
$S\cos\theta$) with two initial and two final conditions, results
in an overdetermined system. In fact, we are also trying to
simultaneously specify both $x$ and $p$ at each boundary,
disallowed by the familiar uncertainty principle.

Kuratsuji and Mizobuchi \cite{kuratsuji:1981} note this
overdeterminacy and claim only one of $\{x_i,x_f\}$ or
$\{p_i,p_f\}$ needs specifying, the other being fixed by the
equations of motion.

Solari \cite{solari:1987} finds an additional pre-factor
\\
\begin{equation}
\exp \frac{i}{2}\int_0^T dt \tilde A(t)
\end{equation}
\\
where $\tilde A(t)$ is a time-dependent operator appearing in the
action $z\tilde A(t)z^*$ where $z$ is the spin coherent state in
the stereographic projection. We won't worry about this correction
since in our treatment there is no such term in the action.

Various authors \cite{shibata:2001} have even claimed that the
spin path integral can only be properly evaluated discretely.
However, a continuous version is reliable with the proper
additional phase of Solari, as argued by Stone et.
al.\cite{stone:2000}.

Below, we review the usage of the Jacobi equation for evaluating
the path integral of a regular particle, then generalizing to the
path integral over a field. Finally, we derive the analogous
Jacobi equation for a spin path integral, following closely the
work of Kuratsuji \cite{kuratsuji:1981}.

\subsection{Coherent state path integral}

In the classical limit, a spin coherent state $|\Omega(t)\rangle$
can be interpreted simply as a spin lying along the direction
$\Omega(t)$.

The spin path integral, including the trace over periodic orbits,
can be written as\symbolfootnote[2]{For the moment considering a
single spin -- the generalization to a field of spins follows
identically to the treatment in Appendix \ref{appendix:quant}.}
\\
\begin{equation}
G(T) = \int_{\Omega(0) = \Omega(T)} \mathcal{D}[\Omega(t)]
e^{\frac{i}{\hbar} \int_0^T dt -S\dot{\phi}\cos\theta -
H[\Omega(t)]}
\end{equation}
\\
This step is analogous to the quantum perturbations about a
soliton solution. We now are solving for the quantum propagator
for these perturbations.

Let $\phi_{cl}(t)$ and $\theta_{cl}(t)$ be a classical solution of
this action (analogous to the simple harmonic oscillator solutions
of the single particle case). Attempting to impose periodic
boundary conditions results, in general, in an over-determined
system of equations. Instead, we set only $\phi_{cl}(0) =
\phi_{cl}(T)$ allowing the equations of motion to fix boundary
conditions for $\theta_{cl}(t)$.

Expanding $\phi = \phi_{cl}(t)+x(t)$ and $S\theta =
S\theta_{cl}(t) +y(t)$, the action becomes to second order
variations (neglecting higher orders in keeping with the
semiclassical approximation)
\\
\begin{equation}
\mathcal{S} = \mathcal{S}_{cl} - \int dt \left(\dot{x}y +
\frac{1}{2} \left( A(t) x^2 + 2B(t) xy +C(t) y^2 \right)\right)
\end{equation}
\\
where $A(t) = \frac{\partial^2H}{\partial \phi^2}$, $B(t) =
\frac{\partial^2H}{\partial \phi \partial S\cos\theta}$ and $C(t)
= \frac{\partial^2H}{\partial (S\cos\theta)^2}$.

In the discrete version\symbolfootnote[3]{\label{bcterm}In
arriving at this expression, note that in the discrete version
there is actually an average of $y(t) \to \frac{y_k+y_{k-1}}{2}$
which under careful analysis gives boundary terms as found by
\citet{solari:1987}. We neglect these terms and approximate
$\frac{y_k+y_{k-1}}{2} \approx y_k$.}, introducing the small
timestep $\epsilon$, we complete the square in $y_k$ to obtain
\\
\begin{align}
\mathcal{S} = \mathcal{S}_{cl} -\frac{1}{2}\sum_{k=1}^{N} \epsilon
C_k & \left( y_k + \frac{B_k}{C_k}x_k y_k + \frac{(x_k -
x_{k-1})}{\epsilon C_k} \right)^2 \\
& + \epsilon x_k^2 \left(A_k - \frac{B_k^2}{C_k}+\frac{d}{dt}
\left(\frac{B}{C}\right)_k\right) - \frac{(x_k -
x_{k-1})^2}{\epsilon C_k} \nonumber
\end{align}
\\
Notice we use the extra integration over the periodic orbit
coordinate $\theta_N$ to integrate over all $N$ $y_k$'s; whereas,
we only use $N-1$ integrations over the $x_k$'s. Impose the
boundary conditions $x_0 = x_N = 0$.

For the general case, where we do not have the additional
integration over boundary conditions, we must introduce this
additional integration as an averaging over the final coordinate.
This doesn't change the physics since this final coordinate is
necessarily fixed by the equations of motion anyway.

The $N$ Gaussian integrals over $y_k$ give the pre-factors
$\prod_{k=1}^{N}\sqrt{\frac{2\pi}{i\epsilon C_k}}$. The complete
expression becomes
\\
\begin{align}
G(T) = \lim_{N\to\infty}\left(\frac{1}{2\pi}\right)^N & \int
d\phi_0 \left( \prod_{k=1}^{N-1}dx_k\right) \left(
\prod_{k=1}^{N} \sqrt{\frac{2\pi}{\epsilon C_k}}\right) \nonumber \\
& \exp{\frac{i}{2}\sum_{k=1}{N}
\frac{x_k^2-2x_kx_{k-1}+x_{k-1}^2}{\epsilon C_k} - \epsilon x_k^2
a_k}
\end{align}
\\
where $a_k =A_k - \frac{B_k^2}{C_k}+\frac{d}{dt}
\left(\frac{B}{C}\right)_k$.

The problem becomes that of solving for the determinant of the
$(N-1)\times (N-1)$ matrix
\\
\begin{equation*}
    -i\left(%
\begin{array}{ccccc}
  \tilde{a}_1 & -\frac{1}{\epsilon C_2} &  &  &  \\
  -\frac{1}{\epsilon C_2} & \tilde{a}_2 &  &  &  \\
   &  & \ddots &  &  \\
   &  &  & \tilde{a}_{N-2} & -\frac{1}{\epsilon C_{N-1}} \\
   &  &  & -\frac{1}{\epsilon C_{N-1}} & \tilde{a}_{N-1} \\
\end{array}%
\right)
\end{equation*}
\\
where $\tilde{a}_k = -\epsilon a_k +\frac{1}{\epsilon
C_k}+\frac{1}{\epsilon C_{k+1}}$.

Re-express the product of pre-factors from the $y_k$ Gaussian
integrals as
\\
\begin{equation*}
    \prod_{k=1}^{N}\sqrt{\frac{2\pi}{i\epsilon C_k}} = \left(\det\left(%
\begin{array}{cccc}
  iC_2 \epsilon &  &  &  \\
   & iC_3 \epsilon &  &  \\
   &  & \ddots &  \\
   &  &  & iC_N \epsilon \\
\end{array}%
\right)i\epsilon C_1\right)^{1/2}
\end{equation*}
\\
and noting that $\det(AB) = \det(A)\det(B)$, we multiply the two
matrices to yield
\\
\begin{equation*}
    iC_1 \epsilon \det\left(%
\begin{array}{ccccc}
  C_2 \epsilon \tilde{a}_1 & -1  \\
  -\frac{C_3}{C_2} & C_3 \epsilon \tilde{a}_2  \\
   &  & \ddots  \\
   &  &  & C_{N-1} \epsilon \tilde{a}_{N-2} & -1 \\
   &  &  & -\frac{C_N}{C_{N-1}} & C_{N} \epsilon \tilde{a}_{N-1} \\
\end{array}%
\right)
\end{equation*}

Denote the determinant of the submatrix ending in the $k$'th row
and column by $D_k$. We can then write down the recursion relation
\\
\begin{eqnarray*}
    D_k &=& \epsilon C_{k+1}\tilde{a}_k D_{k-1} -
    \frac{C_{k+1}}{C_k} D_{k-2}\\
        &=& \left(1+\frac{C_{k+1}}{C_k}-\epsilon^2C_{k+1}\left( A_k -
        \frac{B_k^2}{C_k}+\frac{d}{dt}\left(\frac{B}{C}\right)_k \right)
        \right)D_{k-1}-\frac{C_{k+1}}{C_k} D_{k-2}
\end{eqnarray*}
\\
Letting $D_k$ be a function of $k\epsilon$, this can be rewritten
\\
\begin{align*}
\frac{D_k - 2D_{k-1} + D_{k-2}}{\epsilon^2} =& \frac{(C_{k+1}-C_k)
(D_{k-1}-D_{k-2})}{C_k\epsilon^2}\\
& -C_{k+1}\left( A_k - \frac{B_k^2}{C_k} +\frac{d}{dt}
\left(\frac{B}{C}\right)_k \right)D_{k-1}
\end{align*}
\\
or in a continuum limit
\\
\begin{equation}
    \frac{d^2D}{dt^2} = \frac{1}{C}\frac{dC}{dt}\frac{dD}{dt} -
    CD\left(A - \frac{B^2}{C} +\frac{d}{dt}\left(\frac{B}{C}\right)\right)
\end{equation}
\\
The initial conditions on $D$ can be found directly from the first
and second submatrix determinants
\\
\begin{eqnarray*}
   D_1 &=& iC_1 \epsilon C_2 \epsilon \tilde{a}_1 \\
   \frac{D_2 - D_1}{\epsilon} &=& iC_1 \left(C_3\epsilon\tilde{a}_2
   C_2\epsilon\tilde{a}_1 - \frac{C_3}{C_2} - C_2\epsilon\tilde{a}_1
   \right)
\end{eqnarray*}
\\
giving, in the limit of $\epsilon \to 0$, $D(0) = 0$ and
$\dot{D}(0) =iC(0)$. But this is equivalent to the system of
equations from the original formulation
\\
\begin{eqnarray}
    \frac{dy}{dt} &=& Ax+By \nonumber \\
    \frac{dx}{dt} &=& -Bx-Cy
\end{eqnarray}
\\
with initial conditions $x(0)=0$ and $y(0) = -1$ after eliminating
$y(t)$ and setting $ix(t)=D(t)$. Thus the required determinant is
$ix(T)$.

\subsection{Spectrum of a ferromagnetic plane of spins}

The action of a ferromagnetic plane of spins with easy plane
anisotropy in a continuum limit is
\\
\begin{equation}\label{ferro_S}
\mathcal{S} = S \int \frac{d^2r}{a^2} \int dt \dot{\phi}\theta -
\frac{c}{2} \left(-\phi \nabla^2 \phi -\theta\nabla^2\theta +
\frac{\theta^2}{r_v^2} \right)
\end{equation}
\\
where the various constants are as defined in Chapter
\ref{chapter:magnons}.

Choose a set of spatial eigenfunctions such that $\nabla^2 \to
-k^2$ and $\int \frac{d^2r}{a^2} f_{k'}f_k = \delta^2(k'-k)$.

Thus, the integration measure becomes a product over $k$ states,
now decoupled, leaving within the time integral of the action
(note $S$ was factored out into the integration measure)
\\
\begin{equation}
    \frac{1}{2} \left(%
\begin{array}{cc}
  \phi & \theta \\
\end{array}%
\right)\left(%
\begin{array}{cc}
  -c\nabla^2 & -\frac{\partial}{\partial t} \\
  \frac{\partial}{\partial t} & -c\left(\nabla^2-\frac{1}{r_v^2}\right) \\
\end{array}%
\right)\left(%
\begin{array}{c}
  \phi \\
  \theta \\
\end{array}%
\right)
\end{equation}
\\
where $\omega = ckQ$. The periodic classical solutions can be
written
\\
\begin{equation}
    \left(%
\begin{array}{c}
  \phi_k(t) \\
  \theta_k(t) \\
\end{array}%
\right) = A\left(%
\begin{array}{c}
  \cos\omega_k t \\
  -\frac{k}{Q}\sin\omega_k t \\
\end{array}%
\right) +B\left(%
\begin{array}{c}
  \sin\omega_k t \\
  \frac{k}{Q}\cos\omega_k t \\
\end{array}%
\right)
\end{equation}
\\
with the periodicity condition on $\phi_k(t)$ imposing identical
conditions on $A$ and $B$ as in (\ref{SHO_IC}), with $q_0-q_{cl}
\to \phi_{k0}$. Note that the periodicity condition was previously
$\phi(x,0) = \phi(x,T) = \phi_0$; however, after the
transformation to diagonalize the equations in $k$, each
coefficient $\phi_{k0}$ must now be periodic and integrated over.

The classical action for these periodic orbits becomes
\\
\begin{equation}
    \mathcal{S}_{cl} = -2\frac{k}{Q} \phi_{ko}^2
    \frac{\sin^2\omega_k
    T/2}{\sin\omega_k T}
\end{equation}

The perturbed action has the same form as the original linearized
action above, (\ref{ferro_S}). Calling the small perturbations in
$\phi$, $x$, and those in $\theta$, $y$, we need a solution such
that $x(0) = 0$ and $y(0) = 1$ (the change of sign here arises
from linearizing $\cos\theta \to -\theta$ in the Berry phase
term). This corresponds to
\\
\begin{equation}
    \left(%
\begin{array}{c}
  x(t) \\
  y(t) \\
\end{array}%
\right) = \left(%
\begin{array}{c}
  \frac{Q}{k}\sin\omega_k t \\
  \cos\omega_k t \\
\end{array}%
\right)
\end{equation}
\\
and the determinant evaluates to $ix(T) = i\frac{Q}{k}\sin\omega_k
T$.

\bibliographystyle{plainnat}
\bibliography{mybiblio}


\end{document}